\documentclass[
 reprint,
superscriptaddress,
 amsmath,amssymb,
 aps,
pre,
floatfix,
]{revtex4-2}

\usepackage[colorlinks=true,urlcolor=blue,linkcolor=blue,citecolor=blue]{hyperref}
\usepackage[normalem]{ulem}
\newcommand{\customref}[2]{\hyperref[#1]{\ref*{#1}#2}}
\usepackage{footmisc}

\usepackage{graphicx}
\usepackage{dcolumn}
\usepackage{bm}
\usepackage{hyperref}
\usepackage{color}
\usepackage[table,dvipsnames]{xcolor}

\usepackage[braket, qm]{qcircuit}
\usepackage{booktabs}
\usepackage{multirow}
\usepackage{float}

\begin{document}

\preprint{APS/123-QED}

\title{Quantum Circuit Discovery for Fault-Tolerant Logical State Preparation with Reinforcement Learning}

\author{Remmy Zen}
\email{remmy.zen@mpl.mpg.de}
\affiliation{Max Planck Institute for the Science of Light, Staudtstra{\ss}e 2, 91058 Erlangen, Germany}
\author{Jan Olle}
\affiliation{Max Planck Institute for the Science of Light, Staudtstra{\ss}e 2, 91058 Erlangen, Germany}
\author{Luis Colmenarez}
\affiliation{Institute for Quantum Information, RWTH Aachen University, 52056 Aachen, Germany}
\affiliation{Peter Gr\"{u}nberg Institute, Theoretical Nanoelectronics, Forschungszentrum J\"{u}lich, 52425 J\"{u}lich, Germany}
\author{Matteo Puviani}
\affiliation{Max Planck Institute for the Science of Light, Staudtstra{\ss}e 2, 91058 Erlangen, Germany}
\author{Markus M\"{u}ller}
\affiliation{Institute for Quantum Information, RWTH Aachen University, 52056 Aachen, Germany}
\affiliation{Peter Gr\"{u}nberg Institute, Theoretical Nanoelectronics, Forschungszentrum J\"{u}lich, 52425 J\"{u}lich, Germany}
\author{Florian Marquardt}
\affiliation{Max Planck Institute for the Science of Light, Staudtstra{\ss}e 2, 91058 Erlangen, Germany}
\affiliation{Department of Physics, Friedrich-Alexander Universit\"{a}t Erlangen-N\"{u}rnberg, Staudtstra{\ss}e 5, 91058 Erlangen, Germany}

\date{\today}

\begin{abstract}
The realization of large-scale quantum computers requires not only quantum error correction (QEC) but also fault-tolerant operations to handle errors that propagate into harmful errors. Recently, flag-based protocols have been introduced that use ancillary qubits to flag harmful errors. However, there is no clear recipe for finding a fault-tolerant quantum circuit with flag-based protocols, especially when we consider hardware constraints, such as qubit connectivity and available gate set. In this work, we propose and explore reinforcement learning (RL) to automatically discover compact and hardware-adapted fault-tolerant quantum circuits. We show that in the task of fault-tolerant logical state preparation, RL discovers circuits with fewer gates and ancillary qubits than published results without and with hardware constraints of up to $15$ physical qubits. Furthermore, RL allows for straightforward exploration of different qubit connectivities and the use of transfer learning to accelerate the discovery. More generally, our work opens the door towards the use of RL for the discovery of fault-tolerant quantum circuits for addressing tasks beyond state preparation, including magic state preparation, logical gate synthesis, or syndrome measurement.
\end{abstract}

\maketitle

\section{Introduction} 

Quantum systems are highly fragile due to their susceptibility to errors caused by decoherence. Furthermore, quantum operations are imperfect and error-prone. Therefore, in order to harness quantum systems for computation, the error rates must be significantly reduced. Quantum error correction (QEC) is essential to protect quantum information from these errors, allowing us to perform complex and reliable computations~\cite{preskill2018quantum,gottesman2010introduction}. The basic idea behind QEC is to encode logical qubits into multiple noisy physical qubits in such a way that we can detect and correct errors without destroying the logical state. Although the implementation of QEC is a challenging task, recently we have seen several experimental breakthroughs of QEC with different quantum computing platforms~\cite{postler2022demonstration,cong2022hardware,krinner2022realizing,ryan-anderson_implementing_2022,abobeih2022fault,zhao2022realization,wang2023fault}, first quantum circuits carried out on up to $48$ logical qubits~\cite{bluvstein2023logical,mayer_benchmarking_2024} and crossing the break-even point of beneficial QEC~\cite{sivak2023real,acharya_suppressing_2023,da_silva_demonstration_2024}. 

QEC operations are often expressed using a sequence of quantum gates that forms a quantum circuit. However, these gates are faulty and multi-qubit gates proliferate the errors, compromising the scalability of QEC. In general, the more gates, the more errors are introduced, making QEC less effective~\cite{aharonov_fault-tolerant_2008,knill_resilient_1998,kitaev_fault-tolerant_2003}. Therefore, we want to minimize the number of possible faulty operations that can lead to harmful errors: this is achieved by designing \textit{fault-tolerant} (FT) circuits~\cite{campbell2017roads}. In FT circuits, all faults (gates, measurements, errors, resets) that our QEC code cannot correct become less likely to occur below a specific threshold as the distance of the code increases (see Sec. \ref{sec:bg-ft} for more details). In consequence, only by using FT schemes we can ensure systematic improvement in correction as the size of the code scales. Therefore, FT is of paramount importance in making scalable quantum computers~\cite{gottesman2010introduction,aharonov_fault-tolerant_2008,knill_resilient_1998,kitaev_fault-tolerant_2003}.  Several classes of FT protocols have been proposed~\cite{campbell2017roads,sohn_introduction_2018,chao2020flag,heusen_measurement-free_2023,ryan-anderson_implementing_2022,preskill1998reliable}. Among the first ones was Shor-type error correction \cite{shor_fault-tolerant_1996} which relies on additional GHZ states and repeated measurements to check for errors. Another scheme is Steane-type error correction, which uses additional logical qubits to detect errors~\cite{steane1997active,huang_comparing_2023,postler_demonstration_2023}. Both approaches suffer from a large qubit overhead. Recently, \textit{flag fault-tolerant error correction}~\cite{yoder2017surface,chao2020flag,chamberland2018flag,chao_quantum_2018,chamberland_very_2020,chamberland_topological_2020} was introduced as a way to achieve fault-tolerant protocols with a minimal number of ancilla qubits, e.g.~sometimes requiring only one extra qubit. For instance, in the specific case of preparing a state fault-tolerantly, a flag fault-tolerant protocol uses a \textit{verification circuit} after the encoding circuit that utilizes a few extra ancilla qubits, known as \textit{flag qubits}, to flag harmful errors while keeping the logical state intact. There are already examples of flag verification circuits in state preparation on several QEC codes~\cite{goto_minimizing_2016,paetznick_fault-tolerant_2013,butt_fault-tolerant_2023,goto2023measurement,chamberland2019fault}.  They have also been shown to be effective in reducing logical error rates in experimental realizations~\cite{postler2022demonstration,ryan-anderson_implementing_2022,ryan2021realization,abobeih2022fault,hilder2022fault,pogorelov_experimental_2024}.  

Despite their success, flag-based protocols are typically handcrafted. Furthermore, flag-based protocols have so far been implemented in devices with all-to-all qubit connectivity. A \textit{transpilation} process~\cite{qiskittranspiler,hua2023qasmtrans,younis2022quantum} can be applied to the circuit to respect the qubit connectivity and gate set, but it will generally make it non-FT. In other words, the automatic \textit{compilation}~\cite{maronese2022quantum,schmid2023computational,sivarajah2020t,paraskevopoulos2023spinq,kreppel_quantum_2023}  of  FT circuits has not been widely explored yet. 

\begin{figure}[t]
	\includegraphics[width=.49\textwidth]{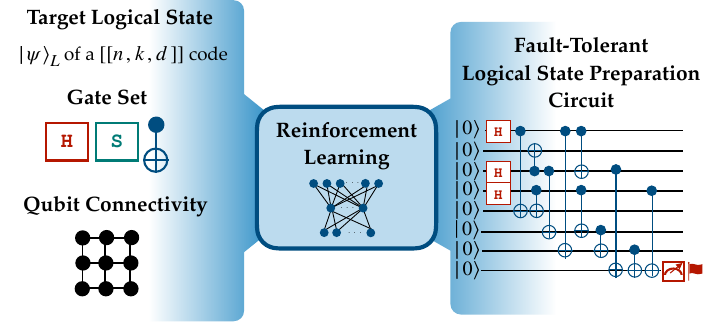}
	\caption{\label{fig:overview-intro} Discovery of fault-tolerant logical state preparation circuit with reinforcement learning (RL). Given the target logical state $|\psi\rangle_L$ of a specified $[[n,k,d]]$ code, a gate set, and a qubit connectivity, we use RL to automatically discover circuits for preparing $|\psi\rangle_L$ fault-tolerantly with flag qubits.  }
\end{figure}

In this work, we present a novel approach based on reinforcement learning (RL) to automatically discover fault-tolerant quantum circuits for QEC. RL is an approach in which an agent learns to make decisions by interacting with an environment in order to maximize its reward through guided trial and error. We apply our method to the task of logical state preparation. Specifically, our approach is based on the automatic discovery of quantum circuits that fault-tolerantly prepare the logical states of a given QEC code under hardware constraints. That is, we can constrain the qubit connectivity and the gate set based on the quantum platforms of interest, as shown in Fig.~\ref{fig:overview-intro}. The RL agent needs to find an optimal strategy by applying a discrete gate at each step, guided by reward signals.

Recently, reinforcement learning (RL)~\cite{sutton2018reinforcement} has emerged as a useful tool for solving various problems in quantum technologies~\cite{krenn2023artificial}.  It has been applied to quantum error correction~\cite{foesel2018reinforcement,nautrup2019optimizing,andreasson2019quantum,sweke2020reinforcement,olle2023simultaneous,puviani2023boosting}, quantum control and state preparation~\cite{zhang2019does,bukov2018reinforcement,porotti2022deep,haug2020classifying,sivak2022model,giordano2022reinforcement}, and quantum compilation~\cite{moro2021quantum,zhang2020topological,chen2022efficient,he2021variational,gong2023no,trenkwalder2023compilation,preti2023hybrid} among many others. It has been shown that RL stands out in quantum state preparation when the gates applied by the agent are discrete~\cite{zhang2019does}. Furthermore, RL is suitable for our task because it can be formulated as a goal-oriented task that is specified in the reward signals. In addition, we find that RL is capable of efficiently navigating large and complex quantum circuit spaces. Finally, RL enables efficient and flexible automated discovery through transfer learning, i.e., reusing trained RL agents for similar but different quantum circuit problem instances. This is not possible, for example, in the method proposed in Ref.~\cite{shutty_decoding_2022}, where finding a fault-tolerant quantum circuit is framed as a Satisfiability Modulo Theory (SMT) problem.

We test our method on several QEC codes, including the 5-qubit perfect code, the 7-qubit Steane code, the 9-qubit Shor code, and the 15-qubit Reed-Muller code. Our first RL approach is to separate the task of finding a FT logical qubit encoding protocol into a logical state preparation task followed by a verification circuit synthesis task. Individually, the RL method for each task already produces quantum circuits that have better or similar performance to existing circuits. More interestingly, by integrating the logical state preparation and verification circuit synthesis tasks, a single RL agent can directly prepare FT logical states and is able to outperform all other available approaches. Thus, this work establishes RL as a viable approach for FT quantum circuit synthesis tasks that go beyond the preparation of logical states.

The paper is organized as follows. In Sec.~\ref{sec:background}, we give a brief background on FT QEC and RL. In Sec.~\ref{sec:rl-framework}, we describe our general reinforcement learning framework for fault-tolerant logical state preparation.  The preparation of a FT logical state can be divided into two successive tasks: the preparation of the logical state, described in Sec.~\ref{sec:logical-state}, followed by the synthesis of the verification circuit, described in Sec.~\ref{sec:verification-circuit}. In Sec.~\ref{sec:ft-logical-state}, we go beyond the separation of tasks and present our main integrated approach, where we directly prepare FT logical states. In Sec.\ref{sec:conclusion} we summarize our work and discuss further extensions.

\section{Background}
\label{sec:background}

\subsection{Quantum Error Correction}
\label{sec:bg-qec}

Here, we briefly review basic concepts from stabilizer quantum error correcting (QEC) codes and introduce the notation that will be used in this paper. Readers familiar with these concepts are invited to skip to Sec.~\ref{sec:bg-circuit}.

The main idea of quantum error correction is to introduce redundancy by encoding  $k$ logical qubits into $n>k$ noisy physical qubits. In this work, we focus on a specific type of QEC codes called \textit{stabilizer codes}~\cite{gottesman1997stabilizer}. Given the Pauli group of $n$ qubits,  the set of stabilizers $S$ is a subgroup such that all elements of $S$ commute with each other and $-I \notin S$. If $S$ is generated by the set $G = \langle g_1, \dots, g_{n-k}\rangle$, then the code space corresponds to the joint +1 subspace of all generators $g_i$, hosting logical quantum states $|\psi\rangle$, for which $g_i|\psi\rangle = |\psi\rangle$ for all generators. 
Within the code space, each code word can be transformed into one another using the \textit{logical operators} $Z^{i}_L,X^{i}_L$, with $i=1,...,k$, where $Z^{i}_L$ and $X^{i}_L$ commute with all elements of the stabilizer group and satisfy $[Z_L^i,X_L^j]=2Z_L^iX_L^j\delta_{ij}$, where $\delta_{ij}$ is the Kronecker delta. For instance, for the case of a QEC code hosting a single logical qubit, $k=1$, $Z^1_L|0\rangle_L = |0\rangle_L$, $Z^1_L|1\rangle_L = -|1\rangle_L$ and $X^1_L|0\rangle_L = |1\rangle_L$. Thus, once $S$ is chosen, the choice of logical operators fixes the codewords $|0\rangle_L$ and $|1\rangle_L$ and all their linear combinations.

The \textit{weight} of a Pauli operator is the number of non-identity components within that operator. The minimum weight among all possible choices of logical operators defines the  \textit{distance} $d$ of the QEC code. A QEC code is able to detect $d-1$ errors and correct $\lfloor (d-1) / 2 \rfloor$ errors. A distance $d$ QEC code encoding $k$ logical qubits into $n$ physical qubits is denoted as $[[n,k,d]]$. 

A QEC code can be defined solely by its stabilizer generators $g_i$. When the stabilizer generators consist of either $X$ or $Z$ Pauli matrices, such that they can be related to two independent classical codes $C_X$ and $C_Z$ for the $X$ and $Z$ stabilizers, they are called \textit{Calderbank-Shor-Steane (CSS) codes}~\cite{steane1996multiple,calderbank_good_1996}. Due to their simplicity and connection to classical codes, CSS codes are at the frontier of theoretical and practical implementations of QEC \cite{postler2022demonstration,ryan2021realization,acharya_suppressing_2023}. Two famous examples of CSS codes are the surface/toric code~\cite{kitaev_quantum_1997,kitaev_fault-tolerant_2003} and the color code~\cite{bombin2015gauge,bombin_topological_2006}. In our work we consider the search for FT and non-FT encoding circuits for several CSS codes (including color codes) and the 5-qubit code, which is a non-CSS code (see Appendix \ref{app:stabilizer-generators} for the code definitions).

\subsection{Logical Qubit Encoding Circuit}
\label{sec:bg-circuit}

Once a QEC code and the logical operators are chosen, the next step is to find a way to encode the desired logical states. For stabilizer codes, one approach is to measure the stabilizers and apply conditional local operations that bring the state back into the code space \cite{krinner2022realizing,abobeih2022fault}. This approach relies on stabilizer measurements, which has the disadvantage of being susceptible to measurement errors, and forces repeated measurements according to the code distance to ensure FT, resulting in a large gate count.   An alternative approach is to find a unitary circuit $U$ that encodes such information using the given code \cite{ryan2021realization,postler2022demonstration,cong2022hardware,ryan-anderson_implementing_2022}. For instance, encoding a logical zero $|0\rangle_L$ implies finding a circuit that performs the task $|0\rangle_L = U|0\rangle^{\otimes n}$. Unlike stabilizer measurement encodings, unitary encodings avoid repeated stabilizer measurements, potentially reducing the number of gates. Importantly, even after choosing a QEC code and codeword, there is no unique recipe for finding an encoding unitary $U$.

NISQ devices often have specific constraints, such as limited qubit connectivity and \textit{native gate set} availability. To fulfill these constraints, a transpilation process is commonly applied to the circuit. The whole procedure typically involves mapping the qubits in the circuit to physical qubits, routing the qubits based on the connectivity by inserting swap gates, decomposing gates into native gates, and optimizing the final circuit~\cite{qiskittranspiler,hua2023qasmtrans,younis2022quantum}. Since the procedure involves inserting and decomposing gates, this process will, in general, increase the size of the circuit.

Due to their simplicity and relevance, we restrict ourselves to logical Pauli eigenstates only. Thus, we can focus only on Clifford circuits, where the logical state $|\psi\rangle = U|0\rangle^{\otimes n}$ is always determined by its \textit{stabilizer tableau} \cite{aaronson2004improved}. In particular, a tableau of a single logical codeword contains the $n-k$ stabilizer generators and the $k$ logical operators, which can be represented as a binary matrix that scales quadratically with respect to $n$. Appendix~\ref{app:tableau} shows more details on the tableau representation. While different tableaus may represent the same state, their canonical form~\cite{gottesman1997stabilizer} remains the same. A \textit{canonical tableau} can be obtained by applying Gaussian elimination to the tableau~\cite{aaronson2004improved}. This means that different encoding circuits preparing the same logical state will have the same representation. We will use this representation later as an input to the RL agent. The canonical representation helps the RL agent to learn more effectively and efficiently by reducing complexity and ensuring consistency of the input space. 

It has been proven that Clifford circuits can be efficiently simulated using classical computers~\cite{aaronson2004improved}. Despite its simplicity, finding a compact circuit is still not trivial~\cite{bravyi20226,kliuchnikov2013optimization}. Several methods have been proposed to prepare arbitrary stabilizer states~\cite{aaronson2004improved,bravyi2021clifford,schneider2023sat,winder2023architecture,niemann2014efficient}.  Some methods have also been developed specifically for the preparation of logical states of stabilizer QEC codes~\cite{amaro2020scalable,nielsen_quantum_2010,rengaswamy2020logical,mondal2023systematic,tandeitnik2022evolving}. However, these techniques generally do not include any hardware constraints, nor do they output fault-tolerant quantum circuits, the latter of which we will focus on next.

\subsection{Fault-Tolerant State Preparation}
\label{sec:bg-ft}
 
In practice, quantum gates are faulty and thus introduce errors in state preparation. A simple but effective model for gate failures is to consider the perfect gate to be applied only with probability $1-p$, where $p$ is the probability that a fault occurs when the gate is applied. In this work, we consider any fault consisting of bit flips ($X$ Pauli), phase flips ($Z$ Pauli), or both ($Y$ Pauli). Therefore, single qubit gates have $3$  error generators $\mathcal{E}=\{\sigma_k\} / I$ and two qubit gates have $15$ error generators $\mathcal{E}=\{\sigma_k \otimes \sigma_m\}/(I \otimes I)$, where $k,m=0,1,2,3$ denotes the Pauli matrices including the identity. More formally, the errors introduced by the gates are modeled by introducing a depolarizing channel after the gates: 
\begin{equation}\label{eq:gate_error}
G \rho G^{\dagger} =  (1-p) G \rho G^{\dagger} +\sum_{E \in \mathcal{E}} \dfrac{p}{|\mathcal{E}|} EG \rho G^{\dagger} E,
\end{equation}
where $G$ is the ideal gate, $p$ is the probability of having a gate error, and $|\mathcal{E}|$ is the number of elements in the set of all error generators $\mathcal{E}$. This is the standard modeling of gate errors, often referred to as \textit{circuit-level noise} \cite{nielsen_quantum_2010,ryan-anderson_implementing_2022,zhao2022realization,postler2022demonstration}. 

An error $E$ can be propagated through the circuit in such a way that a unitary $U_E = \tilde{E} U$ can always be written as the error-free $U$ followed by the propagated error $\tilde{E}$. For Clifford unitaries, $\tilde{E}$ remains a single Pauli error obtained by propagating $E$ through the individual gates one by one \cite{nielsen_quantum_2010}. There are two classes of errors that we consider according to their propagated version: (i) $\tilde{E}$ is a member of the stabilizer group, thus acting trivially on the stabilizer states, or its weight is small enough that it can be removed by QEC, in which case, we say the error is \textit{tolerable}. (ii) Its weight is large enough that it cannot be corrected, causing a logical failure after a QEC cycle. We call such errors \textit{harmful}. 

In practice, any circuit that is not carefully designed will have components whose failures lead to harmful errors. For example, even a single gate failure with probability $p$ can lead to a logical error.  Therefore, increasing the code distance of the QEC code would not suppress the logical error rate, because there would always be uncorrectable events with probability $p$. Formally, for a code of distance $d$ able to correct errors of weight $t=\lfloor (d-1) / 2 \rfloor$, the logical error rate in a FT architecture $p_L$ scales as $p_L\sim p^{t+1}$ for $p$ below the threshold \cite{aharonov_fault-tolerant_2008,knill_resilient_1998,shor_fault-tolerant_1996}, which ensures an ever decreasing logical error rate when increasing the size (number of physical qubits) of the QEC code. In contrast, if a harmful error occurs with probability $p$, the expected gain from QEC is lost, no matter how large $d$ is. In other words, all error events with probability $p^{\alpha}$, $\alpha<(d+1)/2$ should be \textit{tolerable} in the sense that they are corrected after QEC cycles. A circuit or component that fulfills the latter condition is called \textit{fault-tolerant} (FT) \cite{chao2020flag,chamberland2018flag,steane_overhead_2003,chamberland_triangular_2020,knill_quantum_2005,tansuwannont_achieving_2022,chao_quantum_2018,heusen_measurement-free_2023,shutty_decoding_2022}. As an example, let us consider a code with $d=3$ that corrects any single qubit error. Some gate failures in this code can produce weight-two harmful errors. Therefore, $p_L\sim p$ if the circuit design is non-FT. If the encoding circuit is made fault-tolerant, then all errors coming from a single gate failure become tolerable, and only errors coming from two gate failures are harmful, hence $p_L\sim p^2$.
 
There is no unique way to render a circuit FT~\cite{shutty_decoding_2022,chamberland_triangular_2020,chamberland2018flag,heusen_measurement-free_2023,postler2022demonstration,bombin_fault-tolerant_2022,egan_fault-tolerant_2021}. Recently, \textit{flag verification circuits} ~\cite{chao2020flag,chamberland2018flag,chao_quantum_2018} have been proposed for turning non-FT circuits into FT ones. The flag verification procedure is based on coupling additional ancilla flag qubits to the main register in such a way that the error-free state is unperturbed and the flag qubit(s) always have the same measurement outcome. When faults that lead to harmful errors occur in any component of the circuit, the flag qubit(s) are triggered, i.e., flip the measurement outcome of the flag ancilla qubit(s). Thus, harmful errors are flagged out by the flag qubit, allowing us to do post-selection on the ancilla measurement outcomes. One can then apply a repeat-until-success mechanism (rejecting outcomes with triggered flag qubits and accepting outcomes without triggered flag qubits) and apply QEC to remove the tolerable errors.

\subsection{Reinforcement Learning}

Reinforcement learning (RL)~\cite{sutton2018reinforcement} aims to train an agent to take an optimal set of actions in an environment (here a simulation of our physical system). This goal is achieved by maximizing the expected returns or the cumulative rewards via a guided trial-and-error approach. In this work, we focus on model-free reinforcement learning, where the agent does not know about the model of the environment. Formally, an RL agent observes the state of the RL environment $s_t$, applies a discrete action $a_t$ at time step $t$ that changes the state of the environment from $s_t$ to $s_{t+1}$, and receives an instantaneous reward $r_t$. An \textit{episode} is a trajectory of states and actions $\tau=(s_0, a_0, s_1, a_1, \dots, s_T)$ from the initial state $s_0$ to the terminal state $s_T$. An RL agent learns a \textit{policy} function $\pi_{\theta}$ parameterized by $\theta$, which maps each state of the environment to a probability distribution over all possible actions.  $\pi_{\theta}(a_t | s_t)$ gives the probability of applying action $a_t$ for a given state $s_t$ of the environment. The RL agent is trained to maximize the expected returns (cumulative reward) over multiple episodes $\mathcal{J(\theta)} = \mathbb{E}_{\tau \sim \pi_{\theta}}[\sum_{t=0}^{T} r_t]$. 

Policy gradient methods~\cite{sutton1999policy} optimize the objective function $\mathcal{J(\theta)}$ with gradient ascent. In this work, we use a deep reinforcement learning algorithm where a deep neural network is used to compute $\pi_{\theta}$, where $\theta$ corresponds to the weights and biases of a neural network. We use a state-of-the-art variant of policy gradient methods called Proximal Policy Optimization (PPO)~\cite{schulman2017proximal}. In PPO, we use two networks: an actor and a critic network. The former determines the action taken by the agent, while the latter measures the quality of the action taken by the agent. Both networks take the representation of the observation as input. The actor network outputs the probability of taking each discrete action, while the value network outputs a value that corresponds to the expectation value of the cumulative reward. During training, the parameter $\theta$ of the networks is updated in such a way that the objective is satisfied.  

\section{Reinforcement Learning Framework for Quantum Circuit Discovery}

\label{sec:rl-framework}

\begin{figure}[htb]
	\includegraphics[width=.45\textwidth]{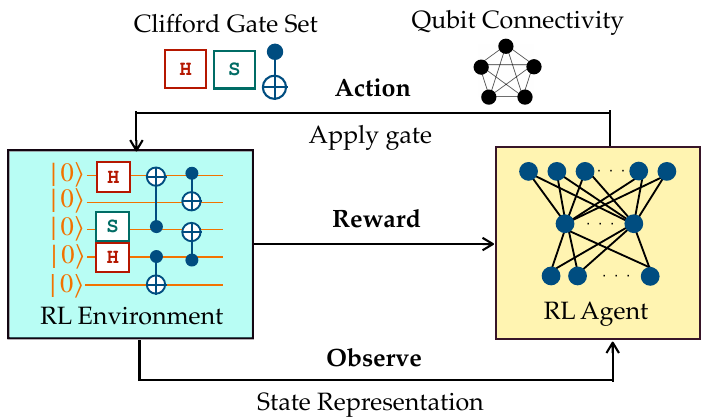}
	\caption{\label{fig:overview-rl} The general RL framework in this work. The circuit is the environment, where its state is represented by its stabilizer canonical tableau. At each step, the RL agent observes the environment and applies a discrete Clifford gate as an action from the specified available gate set (e.g., the Hadamard gate $H$, the phase gate $S$, and the CNOT gate), taking into account qubit connectivity constraints. Subsequently, the agent receives a reward depending on the given task and the quality of the proposed circuit. }
\end{figure}

Here we first introduce the general RL framework as shown in Fig.~\ref{fig:overview-rl}. In this work, an RL agent is trained to output circuits (suggesting a sequence of gates) for a given task (i.e. logical state preparation, verification circuit synthesis, or integrated fault-tolerant logical state preparation). At each step, the RL agent observes the state of a quantum circuit and applies a discrete Clifford gate to the quantum circuit as an action. A trajectory stops when the number of gates is greater than a preset maximum number (counting as a failure) or when it reaches the success criteria defined by the task. We assume that all physical qubits in the circuit are initialized in the $|0\rangle$ state. The hardware constraints, such as the set of available Clifford gates and the qubit connectivity of the device considered, determine the set of possible actions that the agent can take. The reward is then given according to how well the quantum circuit proposed by the agent fulfills the task, which will be explained in the following sections. 

We must then choose a representation of the RL agent's observation. The most common representation is to directly observe the quantum circuit~\cite{fosel2021quantum,ostaszewski2021reinforcement} or to observe the state vector of the state that the circuit represents~\cite{porotti2022deep,gabor2022applicability,zhang2019does,haug2020classifying}. However, multiple quantum circuits could represent the same state, and the state vector representation scales exponentially with $n$. Since we are focusing on stabilizer codes, we can use the stabilizer tableau of the circuit as a representation of the state of the environment. Even better, we can use the canonical tableau as the representation, so that different circuits producing the same output state have the same representation. This representation scales quadratically with $n$. 

Although the state representation is polynomial in  $n$, a brute-force search of the circuit scales exponentially with the number of gates $L$. Suppose we choose a gate set consisting of $G_{1}$ one-qubit gates and $G_{2}$ two-qubit gates with all-to-all qubit connectivity. At each step, the agent must decide over $nG_{1} + (n^2-n)G_{2}$ possible actions, which scales quadratically with $n$. Furthermore, if we assume that a circuit has $L$ gates, then the space of all possible solutions grows exponentially as $(nG_{1} + (n^2-n)G_{2})^{L}$,  making search algorithms infeasible.

As a side note, instead of using a discrete Clifford gate set, one can also use a continuous gate set with a parameterized circuit and use a variational approach as in Ref.~\cite{xu2021variational}. However, in a variational approach, the state can no longer be efficiently described within the stabilizer formalism. Furthermore, one has to design an ansatz and optimize the parameters, which generally does not scale well due to barren plateaus~\cite{mcclean2018barren}.

We use the \textsc{PureJaxRL} library~\cite{lu2022discovered} for the implementation of the PPO algorithm, which is written with the \textsc{Jax}~\cite{jax2018github} library to allow very fast parallel training on a GPU. We then implement the environment for each task using \textsc{Jax}. This means that the simulation of the Clifford circuits and the computation of the rewards run very fast in parallel on the GPU. Therefore, we train multiple agents in parallel and each agent is trained on multiple environments also in parallel.  The code is available online~\footnote{\url{https://github.com/remmyzen/rlftqc}\label{githublink}}. The details of the hyperparameters used and the training process are described in Appendix~\ref{app:training-details}.

\section{Logical State Preparation}
\label{sec:logical-state}

\begin{figure}[tb]
	\includegraphics[width=.45\textwidth]{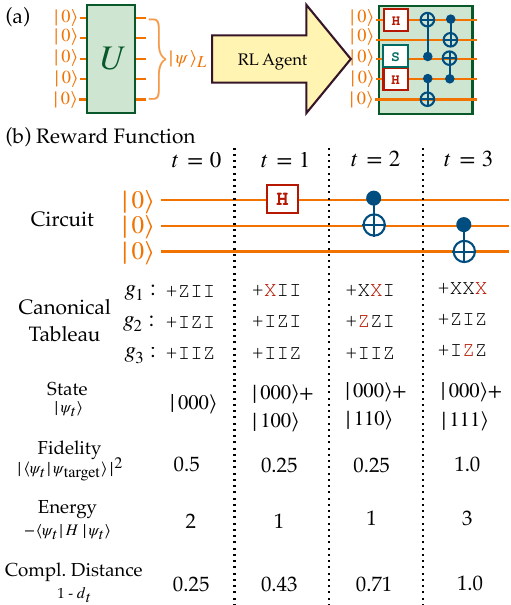}
	\caption{\label{fig:reward} Description and reward function for the logical state preparation task. (a) The logical state preparation task outputs a circuit $U$ that prepares a target logical state $|\psi\rangle_L$ of a $[[n,k,d]]$ code.  (b) The preparation of the state $| \psi_{\rm{target}}\rangle = |000\rangle + |111\rangle$ (normalization factors are not shown for simplicity) from the initial state $| \psi_{0}\rangle = |000\rangle$. We show the value of the three possible functions at each time step $t$ for the reward: fidelity $|\langle \psi_{t} | \psi_{\rm{target}}\rangle|^2$, energy $\sum_i \langle \psi_{t} | H | \psi_{t} \rangle$ used in~\cite{xu2021variational}, and our proposed complementary tableau distance $1 - d_t$. In this case, the proposed complementary tableau distance is monotonically increasing, which is easier for RL algorithms to learn compared to the other functions.   }
	%See main text for more details.
\end{figure}

\subsection{Task Description and Reward Function}

The goal of the logical state preparation task is to find a circuit $U$ that prepares the target stabilizer state (see Fig.~\ref{fig:reward}(a)).  The task requirement is the canonical tableau $T_{\rm{target}}$ of the target stabilizer state $|\psi_{\rm{target}}\rangle$.    Note that although in this paper we focus on preparing logical states of a stabilizer code, this task is general enough to prepare any stabilizer state.

In any RL application, it is of utmost importance to design a good reward function according to the goal. A natural choice of the reward function is the \textit{fidelity} of the state~\cite{zhang2019does,haug2020classifying,bukov2018reinforcement,baba2023deep,mackeprang2020reinforcement,gabor2022applicability}. For a given state at time step $t$, $|\psi_{\rm{t}}\rangle$, the fidelity can be computed as $|\langle \psi_{\rm{t}} | \psi_{\rm{target}}\rangle|^2$, however it suffers from the sparse reward problem~\cite{sutton2018reinforcement}. As an illustration, consider preparing $|\psi_{\rm{target}}\rangle = |111\rangle$ from an initial state of $ |\psi_{\rm{0}}\rangle = |000\rangle$. The agent would have to apply the Pauli $X$ gate to every qubit that changes the state from $|000\rangle$ to $|100\rangle$ to $|110\rangle$ to $|111\rangle$. However, the fidelity value only changes on the last step, since $|\langle 000 | 111\rangle|^2 = |\langle 100 | 111\rangle|^2 = |\langle 110 | 111\rangle|^2 =  0$.  This makes the RL agent harder to train because it does not get immediate feedback.

Since we are preparing a stabilizer state, there are seemingly better rewards that we can use, but there are still drawbacks. In Ref.~\cite{xu2021variational}, finding a logical state of a stabilizer code is framed as finding the ground state of a Hamiltonian $H = - \sum_{i=1}^{n-k} g_i - \sum_{j=1}^{k} O^j_L$, where $g$ are the generators  of the target state and $O_L$ are the logical operators. This then allows one to compute the \textit{energy} as $E = \sum_i \langle \psi_{\rm{t}} | H | \psi_{\rm{t}} \rangle$. If $|\psi_{\rm{t}}\rangle  = |\psi_{\rm{target}}\rangle$, then the ground state energy $E_0 = -n$.  Ref.~\cite{xu2021variational} used $E$ as a cost function for the variational optimization of a parameterized circuit. In our case, we use $-E$ instead,  since we want to maximize the cumulative reward.  Although the computation of this function scales linearly with $n$, it still suffers from the sparse reward problem. One can see that there are only $2n$ possible discrete energy values ranging from $-n$ to $n$.  

We introduce another measure that does not suffer from the sparse reward problem, and the computation of its value does not scale exponentially. We refer to it as the tableau distance $d_{\rm{t}}$, which is the distance between the tableau describing the output state of the currently proposed quantum circuit and the tableau of the target state. We convert the tableaus into binary vectors and measure the binary distance $d_{\rm{t}}$ between the two. Here, we use the Jaccard distance (see discussion in Appendix~\ref{app:distance-comparison}). We normalize the $d_{\rm{t}}$ so that it ranges from $0$ to $1$ and we use the \textit{complementary tableau distance} $1 - d_{\rm{t}}$ since the training of RL maximizes the cumulative reward.

Fig.~\ref{fig:reward}(b) illustrates how the three possible functions (fidelity, energy, and complementary tableau distance) for the reward change for preparing $|000\rangle + |111\rangle$ (normalization factors are not shown for simplicity) from $|000\rangle$. We see that in this case,  unlike the other functions, the proposed complementary tableau distance $1 - d_t$ always increases when gates are applied, giving good feedback to the RL agent.  We can also see that from $t=1$ to $t=2$, applying the correct gate does not change the fidelity and energy functions, which is not a good feedback to the RL agent. This is not the case for $1 - d_t$.  Our empirical numerical experiments also showed that using our proposed complementary tableau distance function as a reward leads to faster convergence of the training of the RL agent compared to using the rewards based on the fidelity and the energy.

Finally, one can give the reward only at the last time step (e.g. $r_t = 1 - d_{L}$ at $t = L$, otherwise $r_t = 0$). However, this means that the agent does not receive immediate feedback after performing an action. Instead, we use the reward shaping technique~\cite{sutton2018reinforcement} by giving a small intermediate value at each step so that the training converges faster. Therefore, at each time step $t$, we give the difference of the complementary tableau distance  between $t$ and $t-1$, or more formally,
\begin{equation}
\label{eq:logical-state-reward}
r_t =  d_{\rm{t-1}} - d_{\rm{t}}. 
\end{equation}

In this case, the cumulative reward $\sum_{t=0}^{L} r_t$ is still $1 - d_{L}$.  A trajectory stops when the complementary tableau distance is greater than a threshold $\epsilon$ close to $1$ (success) or the number of gates is greater than a threshold $L$ (failure). 

As a side note, one might notice that the reward function does not have a term that minimizes the number of gates. This is intrinsically embedded in the RL formulation, which we explain in more detail in Appendix~\ref{app:reward-explanation}. Additionally,  it is straightforward to extend the reward function to consider different objectives or constraints. For example, to minimize the number of two-qubit gates, we could add a term that gives a higher cost for two-qubit gates than for single-qubit gates.

\subsection{Results}

\label{sec:result_logical_state}

We apply our approach to prepare logical states of different QEC codes. Our goal is not only to demonstrate the generality of our approach by benchmarking it as broadly as possible but also to address the ongoing and timely challenge of identifying optimized circuits. We are interested in the preparation of logical states of the following QEC codes. The first code that we consider is the smallest complete error-correcting code, the $[[5,1,3]]$ perfect code~\cite{laflamme1996perfect}, which has been realized experimentally, for example in~\cite{ryan-anderson_implementing_2022,abobeih2022fault}. This code is non-CSS. We then consider several CSS codes. The first quantum error correction code, the $[[9,1,3]]$ Shor code~\cite{shor1995scheme}, which has been realized experimentally, for example in~\cite{nguyen2021demonstration,zhang2022loss}. We also consider 2D and 3D color codes~\cite{bombin_topological_2006,bombin2015gauge}. The $[[7,1,3]]$ Steane code~\cite{steane1996multiple} is the smallest CSS and triangular 2D color code that has been realized experimentally, for example in~\cite{nigg2014quantum,ryan2021realization,ryan-anderson_implementing_2022,postler2022demonstration,bluvstein2022quantum}. We also explore the distance $5$ 2D color code, which is the $[[17,1,5]]$ code~\cite{bombin_topological_2006}. Finally, we consider the smallest error-correcting 3D color code, the $[[15,1,3]]$ Reed-Muller code~\cite{anderson2014fault,bombin2015gauge}. The stabilizer generators of these codes are listed in Appendix~\ref{app:stabilizer-generators} for completeness.

\begin{figure}[tb]
	\includegraphics[width=.45\textwidth]{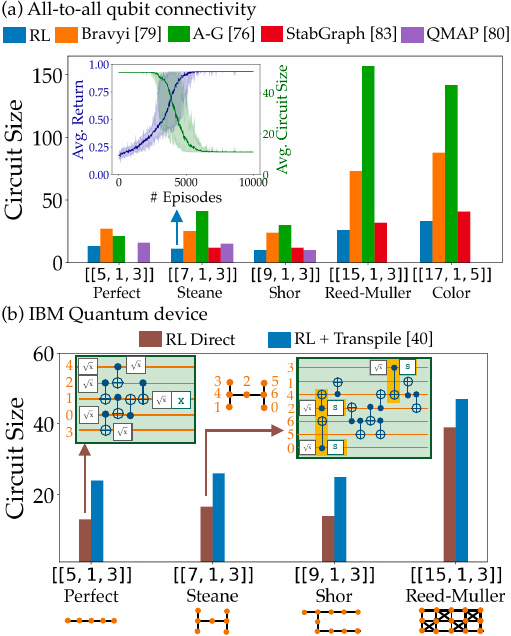}
	\caption{ Results for the logical state preparation task. (a)~The minimum circuit size of different methods for logical state preparation of different QEC codes with all-to-all qubit connectivity and $H$, $S$, and $\text{CNOT}$ gates. StabGraph~\cite{amaro2020scalable} does not work for non-CSS codes such as the $[[5,1,3]]$ perfect code. QMAP~\cite{schneider2023sat} could not prepare the state of the $[[15,1,3]]$ and the $[[17,1,5]]$ code in the allotted maximum time of $12$ hours. The inset shows an example of the training progress for preparing the $|0\rangle_L$ state of the $[[7,1,3]]$ Steane code. (b) Comparison of circuit size from an RL agent that includes the connectivity and gate set during training (RL Direct) with respect to RL-prepared circuits for all-to-all qubit connectivity that have been transpiled with \textsc{Qiskit}~\cite{qiskittranspiler} (RL + Transpile). Results shown for various IBM Quantum device connectivities~\cite{manila,jakarta,guadalupe,tokyo} using $\text{CNOT}$, $\sqrt{X}$, $X$, and $S = R_z(\pi / 2)$ gates. The inset shows examples of RL-prepared circuits for the $|0\rangle_L$ state of the $[[5,1,3]]$ perfect and the $[[7,1,3]]$ Steane code.}
	
	\label{fig:result-logical-state}
\end{figure}

In all of the codes mentioned above, we can choose $Z^{\otimes n}$ as the logical $Z_L$ operator. This means that we prepare the $|0\rangle_L$ states of these codes, except for the $[[9,1,3]]$ Shor code, where the same choice corresponds to the $|+\rangle_L$ state. The preparation of other logical states can be achieved by changing the target logical operator accordingly. As an evaluation metric, we measure the \textit{circuit size}, which corresponds to the number of gates in the circuit. 

We first discuss the preparation of logical states on a device with all-to-all qubit connectivity and a gate set consisting of the gates $H$, $S$, and $\text{CNOT}$, which we refer to as the \textit{standard gate set}. This connectivity and gate set is realistic, for example, in trapped-ion-based quantum computers~\cite{cirac1995quantum}.

We compare our RL method with four different Clifford circuit synthesis methods, where one provides the tableau and the respective methods automatically synthesize a Clifford circuit. Two of them are available in the \textsc{Qiskit}~\cite{qiskit} library, based on the algorithm provided by Bravyi et al.~\cite{bravyi2021clifford} and Aaronson-Gottesman~\cite{aaronson2004improved}. We also compare with StabGraph~\cite{amaro2020scalable}, which works only for CSS codes and uses graph states, and QMAP~\cite{schneider2023sat}, which converts the problem into a Boolean satisfiability (SAT) problem and solves it with a SAT solver. For QMAP, we use the MAX-SAT algorithm, use the depth as the optimization target, and then minimize the number of gates. We find that using the number of gates as the optimization target of the MAX-SAT algorithm is very slow even for small $n$.

Fig.~\ref{fig:result-logical-state}(a) shows the comparison of the smallest circuit size between different methods for preparing logical states of different codes. We see that the RL method always prepared a smaller circuit size compared to the other methods. StabGraph~\cite{amaro2020scalable} is specialized in preparing logical states of CSS codes, therefore it does not work for the $[[5,1,3]]$ perfect code.  QMAP~\cite{schneider2023sat} also did not finish the logical state preparation for $n > 10$ in the allotted maximum time of $12$ hours.  The inset of Fig.~\ref{fig:result-logical-state}(a) shows the training progress for the preparation of the $|0\rangle_L$ for the $[[7,1,3]]$ Steane code.  The shaded area indicates the minimum and maximum values over $10$ agents trained in parallel, where each agent saw $16$ environments in parallel. The entire training takes approximately $100$ seconds on a single NVIDIA Quadro RTX 6000 GPU and produces $10$ circuits. On average, the 10 agents converge after seeing about $6000$ episodes. In Appendix~\ref{app:lsp-circuits}, we show some examples of circuits prepared by the RL agent and discuss some of the strategies that the RL agent learned. For example, we see that in some cases the agent would first try to get the correct tableau without worrying about the sign, and then uses $Z$ gates (two $S$ gates) to fix the sign.

So far, the agent is used only once after training to generate circuits for a specific logical state. However, an advantage of the deep RL method is that one can reuse the agent trained for one task and retrain it for another task. This is commonly referred to as \textit{transfer learning}~\cite{taylor2009transfer,zhu2023transfer}. For example, one can take the agent that prepares the $|0\rangle_L$ state and reuse it to train another agent that prepares the $|+\rangle_L$ state and the $|+i\rangle_L$ state more efficiently. We show these results in Appendix~\ref{app:logical-state-transfer}.

We now show that the RL method is robust enough to adapt to different realistic qubit connectivities and gate sets from different hardware platforms by constraining the actions that the RL agent can take. We illustrate this by focusing on several IBM Quantum devices. The IBM Quantum devices have $\text{CNOT}$, $X$, $\sqrt{X}$, and  $R_Z$ gates (parameterized rotation along the $z$-axis) as their native gate set. Instead of using an arbitrary $R_Z$ gate, we choose to include the $S$ gate, which can be translated into a $R_Z(\pi / 2)$ gate.

We prepare the $|0\rangle_L$ state of the $[[5,1,3]]$ perfect code on the IBMQ Manila~\cite{manila} connectivity, the $|0\rangle_L$ state of the $[[7,1,3]]$ Steane code on the IBMQ Jakarta~\cite{jakarta} connectivity, the $|+\rangle_L$ state of the $[[9,1,3]]$ Shor code on the IBMQ Guadalupe~\cite{guadalupe} connectivity, and the $|0\rangle_L$ state of the $[[15,1,3]]$ quantum Reed-Muller code on the IBMQ Tokyo~\cite{tokyo} connectivity. These connectivities are shown at the bottom of Fig.~\ref{fig:result-logical-state}(b). We trained several agents and take the circuit with the minimum circuit size.

We refer to the RL method that directly restricts the connectivity and gate set in the training as \textit{RL Direct}. We compare it to the \textit{RL + Transpile} method, where we take the RL-prepared circuit for all-to-all qubit connectivity and transpile it with the \textsc{Qiskit} transpiler~\cite{qiskittranspiler}. Fig.~\ref{fig:result-logical-state}(b) shows the comparison of circuit size between the two methods. We see that circuits from the RL Direct method always have a smaller circuit size as compared to circuits obtained with the RL + Transpile method. This shows that restricting the actions of the RL agent based on the hardware constraint during the training is better than transpiling a circuit from all-to-all qubit connectivity.

The inset of Fig.~\ref{fig:result-logical-state}(b) shows examples of a circuit prepared by the RL agent for the $|0\rangle_L$ of the $[[5,1,3]]$ perfect code on the IBMQ Manila connectivity and the $[[7,1,3]]$ code on the IBMQ Jakarta connectivity.  Interestingly, we see that in the circuit for the $[[7,1,3]]$ code, the agent learns a new gate sequence $\sqrt{X}, \text{CNOT}$, and $ S$ (shaded in yellow in the figure). This gate sequence is equivalent to a $H$ gate followed by a $\text{CNOT}$ gate. The agent discovers this gate sequence because the $H$ gate is not available as a native gate on IBMQ devices.  Appendix~\ref{app:lsp-circuits-ibm} shows more examples of logical state preparation circuits on IBMQ devices.

In terms of efficiency, the training of the RL agent to prepare the $|0\rangle_L$ of the $[[7,1,3]]$ Steane code for the IBMQ Jakarta connectivity takes approximately $200$ seconds on a single NVIDIA Quadro RTX 6000 GPU. One could argue that this is much slower than transpiling a circuit for all-to-all qubit connectivity. However, as we have shown, the resulting circuit size is smaller and the training only needs to be done once. Furthermore, the training can be accelerated through transfer learning. In Appendix~\ref{app:logical-state-transfer-connectivity}, we show a technique where the agent trained to prepare a logical state for all-to-all qubit connectivity can be reused and retrained to prepare the same state with different connectivity.

In summary, we have shown that RL can prepare logical states of different QEC codes with smaller circuit sizes than other methods in all-to-all qubit connectivity. We also show that by directly incorporating the hardware constraint by restricting the connectivity and gate set in the training is better than transpiling a circuit for all-to-all qubit connectivity. Furthermore, we can reuse a trained RL agent to speed up the training of the RL agent for different but similar problems.

\section{Verification Circuit Synthesis}
\label{sec:verification-circuit}

\subsection{Task Description and Reward Function}

\begin{figure}[htb]
	\includegraphics[width=.5\textwidth]{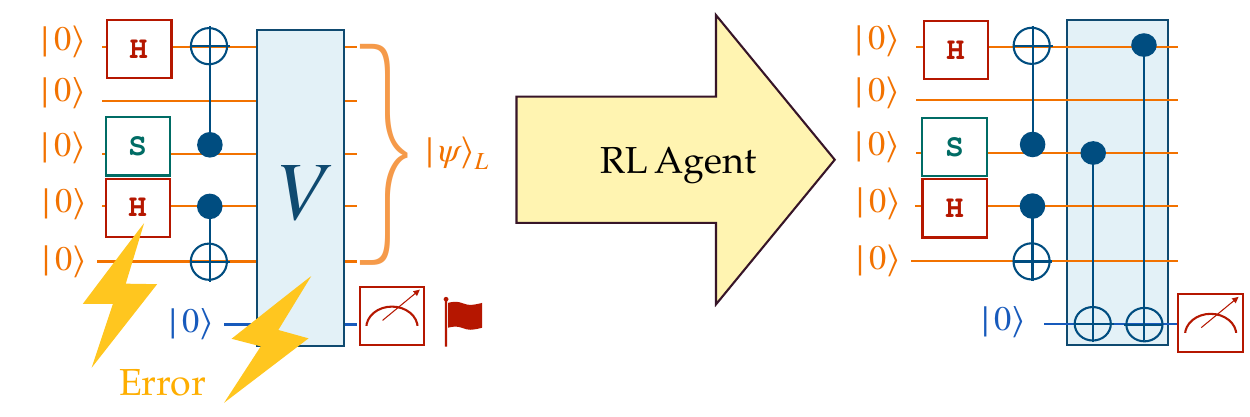}
	\caption{\label{fig:overview-vcs}  The verification circuit synthesis task prepares a circuit $V$ that uses flag qubits to flag harmful errors,  thereby rendering a state preparation fault-tolerant. }
\end{figure}

The goal of the verification circuit synthesis task is to synthesize a circuit $V$ and use the ancilla flag qubits to flag harmful errors and thereby render the encoding protocol fault-tolerant (see Fig.~\ref{fig:overview-vcs}). The task requirement is the sequence of gates that form the circuit $U$ to prepare the target logical state $|\psi\rangle_L$ and the number of the ancilla flag qubits $n_A$. It is possible that several circuits represent the same unitary $U$, but the propagated error would be different. The ancilla flag qubits are initialized in state $|0\rangle$ and are always placed last in the qubit ordering.  For a given circuit, it is usually not known a priori how many ancilla qubits are needed to flag all of the harmful errors. Therefore, it is possible that the agent cannot find a solution for a given number of ancillas. On the other hand, it is also possible that the agent will not use some ancillas if $n_A$ is larger than needed.

There are underlying principles for designing FT flag verification circuits \cite{chao2020flag,chamberland2018flag}, although they do not provide a direct recipe for FT circuit design. For instance, in the case of $d=3$ CSS codes, state verification is equivalent to measuring some logical operators \cite{goto_minimizing_2016}. However, for higher distances or non-CSS codes, a straight-up measurement of one of the logical operators does not necessarily lead to an FT verification. Our goal is to use reinforcement learning to automatically find verification circuits under very general constraints.

We consider three criteria that must be met for this task. The first and most important criterion is to ensure that all harmful errors are flagged. While applying gates to the ancilla, it is possible that the state will change. Therefore, preserving the logical state is the second criterion. Finally, we do not want that the data (non-ancilla) qubits are entangled with the flag qubits, since this will destroy the logical state when we measure the flag. Thus, the third criterion is that the final state is a separable or product state of the data qubits and the flag qubits such that $VU|00\dots0\rangle = |\psi\rangle_L|\phi\rangle_F$, where $|\phi\rangle_F$ is the state of the flag qubits. In summary, the RL agent must flag all harmful errors while preserving the logical state and keeping it disentangled from the flag qubits.

For the first criterion, we reward the agent based on the number of harmful errors that are flagged. We first apply circuit-level noise to the circuit $U$ and obtain the set of all possible error operators $\mathcal{E}$ in the circuit. When the agent applies a gate, which is faulty, we update the set $\mathcal{E}$ by propagating errors from the applied gate and also the old errors. The set $\mathcal{E}$ may grow with new errors or shrink because some errors may become obsolete.

At each time step $t$, we compute $f_t$ ($f$ for flag) given as,
\begin{equation}
    f_t = \sum_{E \in \mathcal{E}}\left\{
\begin{array}{ll}
      0 & \text{if $E$ is $I$ and a flag is triggered,}\\
      1 & \text{if $E$ is tolerable,}\\
      1 & \text{if $E$ is harmful and a flag is triggered,}\\
      0 & \text{if $E$ is harmful and flags are not triggered.} \\
\end{array} 
\right. 
\end{equation}

The first term is used to prevent the agent from choosing a naive strategy like always flagging the ancilla (e.g. applying an $X$ gate to the flag qubits).

We then normalize $f_t$ by dividing it by the total number of errors $|\mathcal{E}|$. Note that some errors may initially have a large weight, but can be reduced by multiplication with a member of the stabilizer group. For instance, an error with weight $4$ that is a member of the stabilizer group will have weight $0$ and become a tolerable error. Therefore, to consider whether an error is tolerable or harmful, we compute the minimum weight of each error by multiplying it by all members of the stabilizer group when computing the reward. For instance, for the 5-qubit code, we check all $2^{5}=32$ elements of the stabilizer group, including the state-dependent logical operator.

One might notice that if an error $E$ is tolerable and the flag is triggered, the agent still gets a reward. This is inevitable, since it is not possible to both flag all of the harmful errors and with the same circuit construction unflag all of the tolerable errors. We can consider flagged tolerable errors as ``unlucky'' cases - note that this does not compromise FT, of course. One could add additional terms in the reward function to minimize this. 

In principle, accurate FT circuit design requires consideration of every type of error that can occur in the circuit. However, CSS codes provide a further simplification in the design of fault-tolerant schemes. For instance, multiple $Z$ errors usually lead to logical failures of the type $Z_L|\Psi\rangle$ (assuming $Z_L$ consists only of $Z$ operators). Therefore, by restricting to $|0\rangle_L$, we make multiple $Z$ errors tolerable, since $Z_L|0\rangle_L=(+1)|0\rangle_L$. Thus, only $X$ and $Y$ errors at the end of the encoding circuit are potentially harmful. The same applies to the preparation of $X_L|+\rangle_L=|+\rangle_L$, if $X_L$ consists only of $X$ operators, it is sufficient to consider $Z$ and $Y$ errors. Synthesis of FT encoding circuits for codewords $|0\rangle_L$ and $|+\rangle_L$ of CSS codes is then easier in the sense that either $X$ or $Z$ is not harmful by construction. However, this is not true for non-CSS codes, because the stabilizers have both $X$ and $Z$ Paulis, so the logical operators consist of both $X$ and $Z$ operators.

For the second criterion, we need to make sure that the circuit preserves the logical state $|\psi\rangle_L$. Here we can use the three possible functions discussed in Sec.~\ref{sec:logical-state}. In this case, we reuse our proposed complementary tableau distance to measure the distance between the canonical tableau of the target logical state and the current error-free canonical tableau of the data qubits.

For the third criterion, we directly enforce the state to be a separable state of the data and ancilla flag qubits. This is necessary in order not to change the error-free logical state after the measurement of the ancilla qubits.  In the stabilizer formalism, the latter is achieved by targeting the stabilizer generators of the ancilla in the current error-free canonical tableau to be $Z$ in the location of the ancilla and $I$ in the others.  We can extract the canonical tableau of the ancilla qubits by taking the submatrix of the canonical tableau where the rows are $n$ to $n + n_{A}$. Therefore, we define a value $p_t$ ($p$ for product state) that measures the complementary tableau distance of the current error-free canonical tableau with the target tableau according to the above criteria. For an illustration of the reward calculation, see Appendix~\ref{app:reward-computation}.

We again use the reward shaping technique, which gives the reward function,
\begin{equation} 
\label{eq:vcs-reward}
r_t = \mu_{f}(f_t - f_{t-1}) + \mu_{d}(d_{t-1} - d_{t}) + \mu_{p}(p_t - p_{t-1}) ,
\end{equation}
where $\mu$ defines the weight for each individual reward.  A trajectory stops when all of the harmful errors are flagged, the prepared state is the logical state, and the data qubits and flag qubits are a product state (success) or the number of gates is greater than a threshold $L$ (failure).

\subsection{Results}

\label{sec:result_verification_circuit}

Let us take non-FT state preparation circuits from the literature and use the RL method to synthesize the verification circuits. We then compare them with known verification circuits. 

We use three metrics to compare different verification circuits. (i) First, we compare the number of two-qubit gates (one-qubit gates do not propagate errors) and the number of flag qubits.  (ii) The second metric is the \textit{acceptance rate}. A state outcome is accepted if, after running the circuit with noise, the flag qubits are not triggered. To determine the acceptance rate numerically, we simulate $10^{7}$ noisy circuit trajectories for each varying error probability $p$ by adding circuit-level noise using the \textsc{Stim}~\cite{gidney2021stim} library and count the number of accepted state outcomes.  (iii) The final metric is the \textit{logical error rate} $p_L$. When a state outcome is accepted, we perform a perfect round of error correction on the data qubits. We can then check if the decoded state is correct, otherwise, a logical error has occurred. As discussed in Sec.~\ref{sec:bg-ft},  $p_L$ of a fault-tolerant circuit should scale proportional to $p^2$ for distance-$3$ codes, while it scales as $p$ for non-fault-tolerant circuits.

In our numerical experiments, we choose to use the standard gate set ($H$, $S$, and $\text{CNOT}$) combined with the $\text{CZ}$ gate. The training of the agent starts with one flag qubit, and if the training does not converge, the number of flag qubits is incremented by one until a solution is found. We have also found empirically that prohibiting the agent from applying gates between the data qubits helps to speed up training convergence. In Appendix~\ref{app:vcs-varymiu}, we show how different values of the weights $\mu$ in the reward affect the acceptance and logical error rates.

\begin{figure}[htb]
	\includegraphics[width=.48\textwidth]{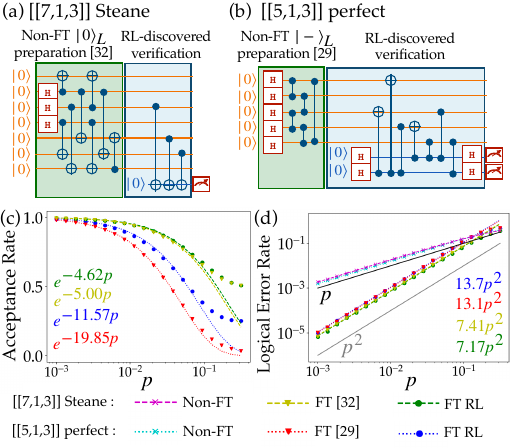}
	\caption{ Results for the verification circuit synthesis task. Examples of RL-discovered verification circuits (shaded in blue) for a given non-fault tolerant (FT) preparation (shaded in green) of (a) the $|0\rangle_L$ state of the $[[7,1,3]]$ Steane code from Ref.~\cite{goto_minimizing_2016} and (b) the $|-\rangle_L$ state of the $[[5,1,3]]$ perfect code from Ref.~\cite{chao_quantum_2018}. In Ref.~\cite{goto_minimizing_2016}, the verification circuit uses $1$ flag qubit and $3$ two-qubit gates, which is the same as the circuit discovered by the RL agent. In Ref.~\cite{chao_quantum_2018}, the verification circuit uses $6$ flag qubits (or $2$ flag qubits with $2$ qubit resets) and $15$ two-qubit gates, while the RL-discovered circuit in (b) uses only $2$ flag qubits and $7$ two-qubit gates.  Comparison of the acceptance rate (c) and logical error rate (d) with different simulated error probability $p$ for the circuits shown in (a) and (b) compared to non-FT circuits and circuits in~\cite{goto_minimizing_2016} and ~\cite{chao_quantum_2018}. }

	\label{fig:result-flag-fully-connected}
\end{figure}

First, we synthesize the verification circuit for CSS codes. As discussed in Sec.~\ref{sec:bg-ft}, CSS codes have the favorable property that some errors are tolerable. We illustrate this by synthesizing the verification circuit for the $|0\rangle_L$ preparation of the $[[7,1,3]]$ Steane code  with $Z_L = Z^{\otimes 7}$.  The circuit was proposed in~\cite{goto_minimizing_2016} (part of the circuit in Fig~\ref{fig:result-flag-fully-connected}a shaded in green) and has been experimentally realized in~\cite{postler2022demonstration,ryan2021realization,ryan-anderson_implementing_2022,hilder2022fault}. The RL agent discovers verification circuits with the same number of flag qubits and two-qubit gates as the one in~\cite{goto_minimizing_2016}. Part of the circuit in Fig~\ref{fig:result-flag-fully-connected}a, shaded in blue, shows an example of the verification circuit discovered by the RL agent. We show other discovered circuits in Appendix~\ref{app:vcs-circuits}. We observe that the RL agent learns to measure the stabilizer-equivalent logical $Z$ operator $IIZIZZI$ without being explicitly told. Although the discovered circuit has the same number of flag qubits and two-qubit gates, we see in Fig.~\ref{fig:result-flag-fully-connected}(c),(d) that the acceptance rate and the logical error rate of the RL-discovered circuit are marginally better than the verification circuit proposed in ~\cite{goto_minimizing_2016}.

We now move on to the synthesis of verification circuits for non-CSS codes. We choose to synthesize the verification circuit for the $|-\rangle_L$ preparation of the $[[5,1,3]]$ perfect code with $X_L = XXXXX$ proposed in~\cite{chao_quantum_2018} and experimentally realized in~\cite{ryan-anderson_implementing_2022,abobeih2022fault}. The blue-shaded part of the circuit in Fig~\ref{fig:result-flag-fully-connected}b shows an example of the verification circuit discovered by RL. The RL agent learns to measure the stabilizer $IIZXZ$ in the first ancilla and the stabilizer $XIXZZ$ in the second ancilla. In Fig. ~\ref{fig:result-flag-fully-connected}(c), we see that the RL-discovered circuit has a higher acceptance rate compared to the circuit in~\cite{chao_quantum_2018} due to the smaller circuit size and fewer flag qubits. Nevertheless, in Fig. ~\ref{fig:result-flag-fully-connected}(d), we see that the logical error rate is slightly worse than the circuit in~\cite{chao_quantum_2018}. However, the RL agent also discovers circuits with a lower logical error rate than the circuit in~\cite{chao_quantum_2018}, at the expense of requiring $3$ flag qubits. We show this circuit in Appendix~\ref{app:vcs-circuits}.

In this work, we consider only $d=3$ codes. The discovery of verification circuits for larger codes is challenging, but we could in principle extend our RL approach to higher distance codes. We illustrate the discussion with $d=5$ codes. In this case, the logical error rate $p_L$ must be such that $p_L \sim p^3$, so that errors from two gate failures must be propagated. Therefore, we only need to change the way the error is propagated when training the RL agent. If $L$ is the number of two-qubit gates, we need to propagate $O(L)$ errors for $d=3$ and $O(L^2)$ errors for $d=5$. In addition, $L$ is also generally higher for $d=5$ codes than for $d=3$ codes. Our RL approach can be further improved by restricting the action space or by designing a better reward function, which we leave for future work.

In summary, we have shown that the RL method can be used to discover verification circuits for given non-FT logical state preparation circuits. We even show a case where the RL method discovers a better circuit than the existing circuit in the literature. Furthermore, interestingly the RL method can also discover variants of verification circuits with different tradeoffs in terms of logical error rates, acceptance rates, and the number of flag qubits.

\section{Integrated Fault-Tolerant Logical State Preparation}
\label{sec:ft-logical-state}

\subsection{Task Description and Reward Function}

\begin{figure}[htb]
	\includegraphics[width=.5\textwidth]{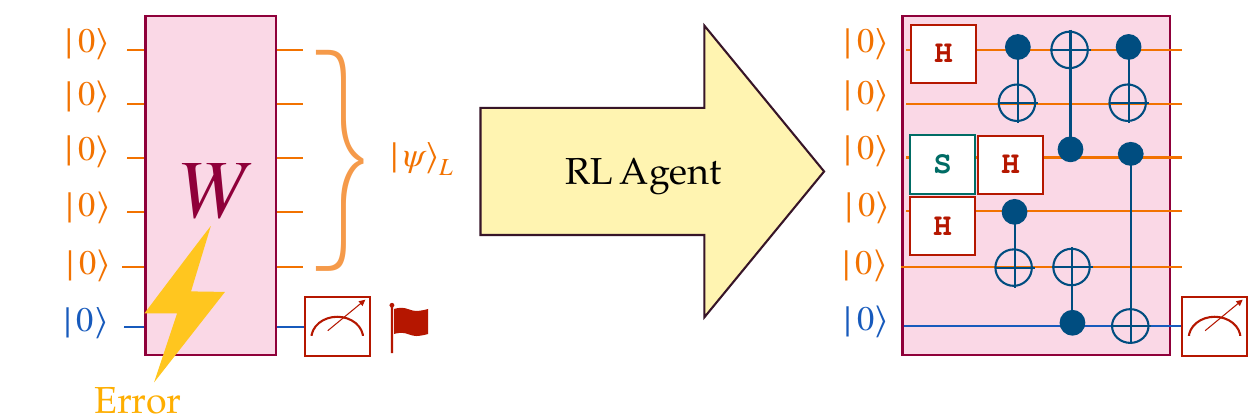}
	\caption{\label{fig:overview-ft-lsp} The integrated fault-tolerant logical state preparation task outputs a circuit $W$ that directly prepares $|\psi\rangle_L$ of a $[[n,k,d]]$ code in a fault-tolerant way. }
\end{figure}

Individually, we have shown that RL methods are able to achieve competitive results for the tasks of logical state preparation and verification circuit synthesis. Here, we go beyond the separation of the tasks and present our main approach that integrates them to directly prepare logical states in a fault-tolerant manner. 

We expect that this integration will allow the RL agent to devise a more effective strategy compared to separating the task for two main reasons.  First, it will take error propagation into account when preparing the logical state. In addition, we expect the agent to perform better when preparing a fault-tolerant logical state under limited qubit connectivity. When we consider the two goals separately, instead, the RL agent does not take into account which data qubits are connected to the flag qubits when preparing the logical state.

The goal is to find a circuit $W$ that prepares a logical state in a fault-tolerant way (see Fig.~\ref{fig:overview-ft-lsp}). The task requirement is the tableau of the target stabilizer state $s_{target}$ and the number of available flag qubits $n_A$. Note that, with respect to the previous two tasks, it is possible though not necessary that the circuit $W$ found by the RL agent is also decomposable into the state preparation circuit $U$ and the verification circuit $V$. 

The reward used for this task is the same as in Eq.~(\ref{eq:vcs-reward}). However, in this case, the RL agent starts from scratch, so the set of error operators $\mathcal{E}$ is initially empty and grows as the agent performs actions by adding gates to the circuit construction attempt.

\subsection{Results}

\label{sec:result_ft_logical_state}

\begin{figure*}[htb]
	\includegraphics[width=.95\textwidth]{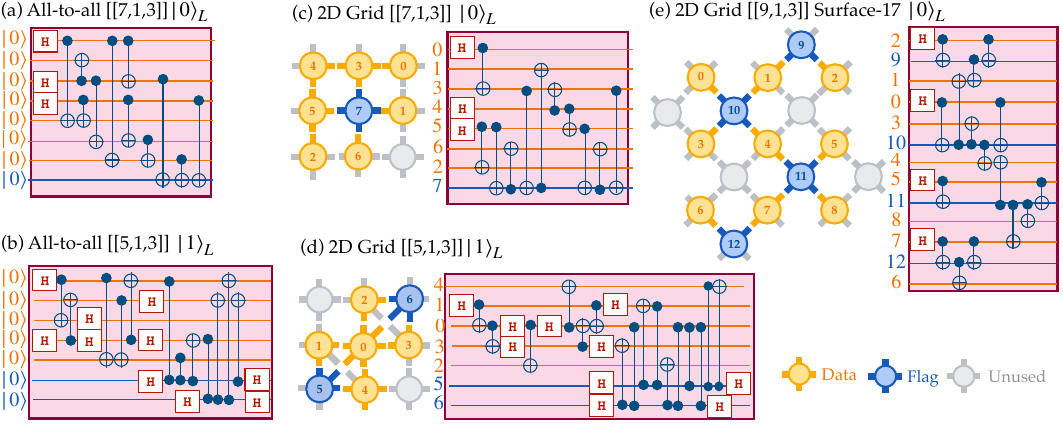}
	\caption{ \label{fig:result-ft-logical} Results for the RL-based integrated fault-tolerant logical state preparation (IFT-LSP).  An example of a fault-tolerant circuit prepared by an RL agent for (a) the $|0\rangle_L$ state of the $[[7,1,3]]$ Steane code and (b) the $|1\rangle_L$ state of the $[[5,1,3]]$ perfect code in all-to-all qubit connectivity. Parts (c) and (d) show learned fault-tolerant circuits for the same logical state preparation task on a 2D grid connectivity based on Google Sycamore~\cite{arute2019quantum} (c) and IBMQ Tokyo~\cite{tokyo} (d) devices. In (e), we show, for the first time, flag-based fault-tolerant $|0\rangle_L$ state preparation of the $[[9,1,3]]$ Surface-17 code on a 2D grid connectivity and qubit placement taken from Refs.~\cite{krinner2022realizing}.  Note that the flag qubits (in blue) are measured, but for simplicity, the measurement is not shown. Unused qubits or connections (in gray) mean that they are available for use by the RL agent, but are not used in the solution found by the agent.  }
\end{figure*}

\begin{table}[t]
\caption{\label{tab:comparison} The comparison of fault-tolerant logical state preparation circuits on all-to-all qubit connectivity between our two RL approaches and existing circuits. We show the minimum number of two-qubit gates and the number of flag qubits in parentheses. Bold text indicates methods with the lowest number of two-qubit gates. The first RL approach is the LSP+VCS, where we separate the task by first performing the logical state preparation (LSP in Sec.~\ref{sec:logical-state}) followed by the verification circuit synthesis (VCS in Sec.~\ref{sec:verification-circuit}). The second RL approach is our main approach, which is the integrated fault-tolerant logical state preparation (IFT-LSP in Sec.~\ref{sec:ft-logical-state}). We see that IFT-LSP always finds circuits with less or a similar number of two-qubit gates than LSP+VCS or existing circuits.}

\begin{tabular}{@{}ccccc@{}}
\toprule
\textbf{Code}  & \textbf{State}           & \textbf{LSP+VCS} & \textbf{IFT-LSP} & \textbf{Existing} \\ \midrule\midrule
$[[5,1,3]]$ & $|1\rangle_L$    & 14 (2)             &\textbf{12 (2)}         & -                   \\
perfect & $|-\rangle_L$   & \textbf{12 (2)}        & \textbf{12 (2)}     & 20 (6)~\cite{chao_quantum_2018}              \\\midrule
$[[7,1,3]]$  & $|0\rangle_L$   & 11 (1)             & 11 (1)          & 11 (1)~\cite{goto_minimizing_2016}              \\
Steane & $|+\rangle_L$  & 11 (1)             & 11 (1)          & -                   \\\midrule
$[[9,1,3]]$  & $|0\rangle_L$  & 6 (0)              & 6 (0)           & -                   \\
Shor & $|+\rangle_L$ & 11 (1)             & 11 (1)          & -                   \\\midrule
$[[9,1,3]]$ & $|0\rangle_L$   & 11 (1)              & 11 (1)           & -$^{\text{a}}$           \\
Surface-17 & $|+\rangle_L$  & 11 (1)             & 11 (1)          & -                   \\\midrule
$[[15,1,3]]$ & $|0\rangle_L$  & 29 (2)             & \textbf{25 (1)}          & \textbf{25 (1)}~\cite{butt_fault-tolerant_2023}              \\
 Reed-Muller & $|+\rangle_L$ & \textbf{31 (1)}         & \textbf{31 (1)}      & 32 (1)~\cite{butt_fault-tolerant_2023}               \\ \bottomrule
 \hline\hline
\end{tabular}

 \footnotesize \begin{flushleft}$^{\text{a}}$Ref.~\cite{goto2023measurement} shows the FT preparation of $|0\rangle_L$ state of the $[[9,1,3]]$ surface-17 code with 8 two-qubit gates and 0 flag qubits. However, the connectivity of qubits is different, namely a $3\times3$ grid without ancilla qubits. See also footnote~\cite{Note2}.\end{flushleft}
\end{table}

\subsubsection{All-to-all qubit connectivity}

We first compare our two RL approaches to prepare a logical state in a fault-tolerant manner on all-to-all qubit connectivity. The first approach separates the task into \textit{logical state preparation} (LSP) followed by \textit{verification circuit synthesis} (VCS), which we refer to as LSP+VCS. The second one, instead, is our main approach that directly prepares the fault-tolerant logical state, which we refer to as \textit{integrated fault-tolerant logical state preparation} (IFT-LSP). 

We will discuss the preparation of the following logical states. For the CSS codes considered (i.e.~the $[[7,1,3]]$ Steane code, $[[9,1,3]]$ Shor code, and $[[15,1,3]]$ Reed-Muller code), we prepare the $|0\rangle_L$ state with $Z_L = Z^{\otimes n}$ and the $|+\rangle_L$ state with $X_L = X^{\otimes n}$. For the non-CSS code (i.e.~the $[[5,1,3]]$ perfect code), we prepare the $|1\rangle_L$ state with $Z_L =ZZZZZ$ and the $|-\rangle_L$ state with $X_L = XXXXX$. For this task, we also consider the $[[9,1,3]]$ Surface-17 code~\cite{kitaev_fault-tolerant_2003}, which has been realized experimentally, for example in~\cite{krinner2022realizing,acharya_suppressing_2023,zhao2022realization}.

In our numerical experiments, we again use the standard gate set combined with the $\text{CZ}$ gate. The training of the agent starts with one flag qubit, and if the training does not converge, the number of flag qubits is incremented by one until a solution is found.

We compare the minimum number of two-qubit gates and the number of ancillas needed to prepare fault-tolerant logical states with the two RL approaches (LSP+VCS and IFT-LSP) and existing circuits in Table~\ref{tab:comparison}. IFT-LSP is better than LSP+VCS at preparing two states: $|1\rangle_L$ of $[[5,1,3]]$ perfect code and $|0\rangle_L$ of $[[15,1,3]]$ Reed-Muller code.  This is most likely because the LSP does not take error propagation into account when preparing the state. Compared to existing circuits in the literature, our RL approaches find a smaller number of two-qubit gates in two states: $|-\rangle_L$ of $[[5,1,3]]$ perfect code and $|+\rangle_L$ of $[[15,1,3]]$ Reed-Muller code. The first case is already shown in Fig.~\ref{fig:result-flag-fully-connected}(b), while the second case needs one two-qubit gate less than the existing one. The circuits are shown in Appendix~\ref{app:ftlsp-circuits}. In terms of efficiency, both RL approaches are comparable. For example, to prepare the $|0\rangle_L$ of the $[[7,1,3]]$ Steane code, IFT-LSP needs about $150$ seconds, while LSP+VCS needs about $180$ seconds on a single NVIDIA Quadro RTX 6000 GPU.

Fig.~\ref{fig:result-ft-logical}(a) and (b) show an example circuit for the fault-tolerant preparation of the $|0\rangle_L$ state of the $[[7,1,3]]$ Steane code and the $|1\rangle_L$ state of the $[[5,1,3]]$ perfect code discovered by IFT-LSP, respectively (see Appendix~\ref{app:ftlsp-circuits} for other examples of RL-discovered circuits). We can see that to prepare the $|0\rangle_L$ state of the $[[7,1,3]]$ Steane code, the RL agent measures the stabilizer-equivalent logical $Z$ operator $IIZZIIZ$. When preparing the $|1\rangle_L$ state of the $[[5,1,3]]$ perfect code, the agent measures the stabilizer-equivalent logical $Z$ operator $ZXIXZ$ via the first ancilla and $XXIZI$ via the second ancilla. 

As a side note, one can prepare other states by changing the logical operators. Alternatively, we can also apply logical gates to a prepared state. For example, it is known that the logical $H$ gate in the $[[7,1,3]]$ Steane code is transversal (applying $H$ to each physical qubit), so it is still fault-tolerant. Thus, one can prepare $|+\rangle_L$ by applying logical $H$ to the prepared $|0\rangle_L$. However, this is not obvious for example in the $[[5,1,3]]$ perfect code. In Ref.~\cite{ryan-anderson_implementing_2022}, the authors always prepare the $|-\rangle_L$ fault-tolerantly first, and then rotate the logical basis state to another state. With our RL approach, instead, we can automatically discover fault-tolerant preparation circuits for other states.

\subsubsection{Restricted qubit connectivity}

We now move to a more general and practically relevant case where we show fault-tolerant logical state preparation on a device with restricted qubit connectivity. There are some handcrafted recipes for specific codes, such as encoding the $|0\rangle_L$ state of the $[[9,1,3]]$ surface-17 code in a 1D array~\cite{goto2023measurement} and encoding a magic state of the $[[4,1,2]]$ code in an IBMQ device~\cite{gupta_encoding_2024}. Here, we want to use RL instead to automatically discover such circuits.

Note that transpiling a fault-tolerant circuit prepared for all-to-all qubit connectivity generally does not work, since it does not guarantee that the transpiled circuit is fault-tolerant. Additionally, we find that separating the task (LSP+VCS) fails in some cases. The first case is when a data qubit is connected only to the ancillas. In this scenario, one would have to use the ancilla as a ``bridge'' to the corresponding data qubit. The second case is when the VCS fails because the LSP does not take the position of the ancilla into account when preparing the logical state. In contrast, our main approach (IFT-LSP) works in these conditions. We discuss these two cases in more detail and give examples in Appendix~\ref{app:ftlsp-fail-cases}. 

We illustrate our main approach by preparing fault-tolerant logical states on a 2D grid, which is common in quantum chips based on superconducting qubits (e.g. Google Sycamore~\cite{arute2019quantum}, IBM Quantum devices, Rigetti Ankaa). We first demonstrate the fault-tolerant preparation of the $|0\rangle_L$ for the $[[7,1,3]]$ Steane code on a $3\times3$ grid based on the Google Sycamore device. Fig.~\ref{fig:result-ft-logical}(c) shows an example of an RL-discovered circuit. Impressively, the RL agent discovers a circuit with the same number of two-qubit gates and flag qubits as in the all-to-all qubit connectivity shown in  Fig.~\ref{fig:result-ft-logical}(a). Fig.~\ref{fig:result-ft-logical}(d) shows an example of the fault-tolerant preparation of the $|1\rangle_L$ for the $[[5,1,3]]$ perfect code on a 2D grid based on the IBMQ Tokyo~\cite{tokyo} device. Compared to the circuit on all-to-all qubit connectivity, it has the same number of flag qubits and requires only $4$ additional two-qubit gates. The RL approach also manages to discover circuits for the preparation of other logical states, including for the $[[9,1,3]]$ Shor code, which we show in Appendix~\ref{app:ftlsp-circuits-connectivity}.

\begin{figure}[tb]
	\includegraphics[width=.45\textwidth]{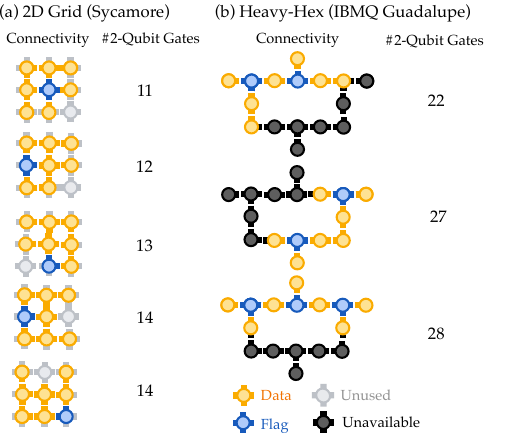}
	\caption{\label{fig:ft-logical-connectivity} Exploration of different qubit connectivity and placement for the integrated fault-tolerant $|0\rangle_L$ state preparation of the $[[7,1,3]]$ Steane code. We show the number of two-qubit gates in some of the RL-discovered circuits with (a) 2D grid (based on the Google Sycamore device) and (b) heavy-hex layout (based on the IBMQ Guadalupe~\cite{guadalupe} device). Unused qubits and connectivities (in gray) mean that the qubits were given to the RL agent to be used as flag qubits, but were not used. Unavailable qubits and connectivities (in black) mean that the qubits are not set as available to the RL agent.}
\end{figure}

\begin{figure}[tb]
	\includegraphics[width=.45\textwidth]{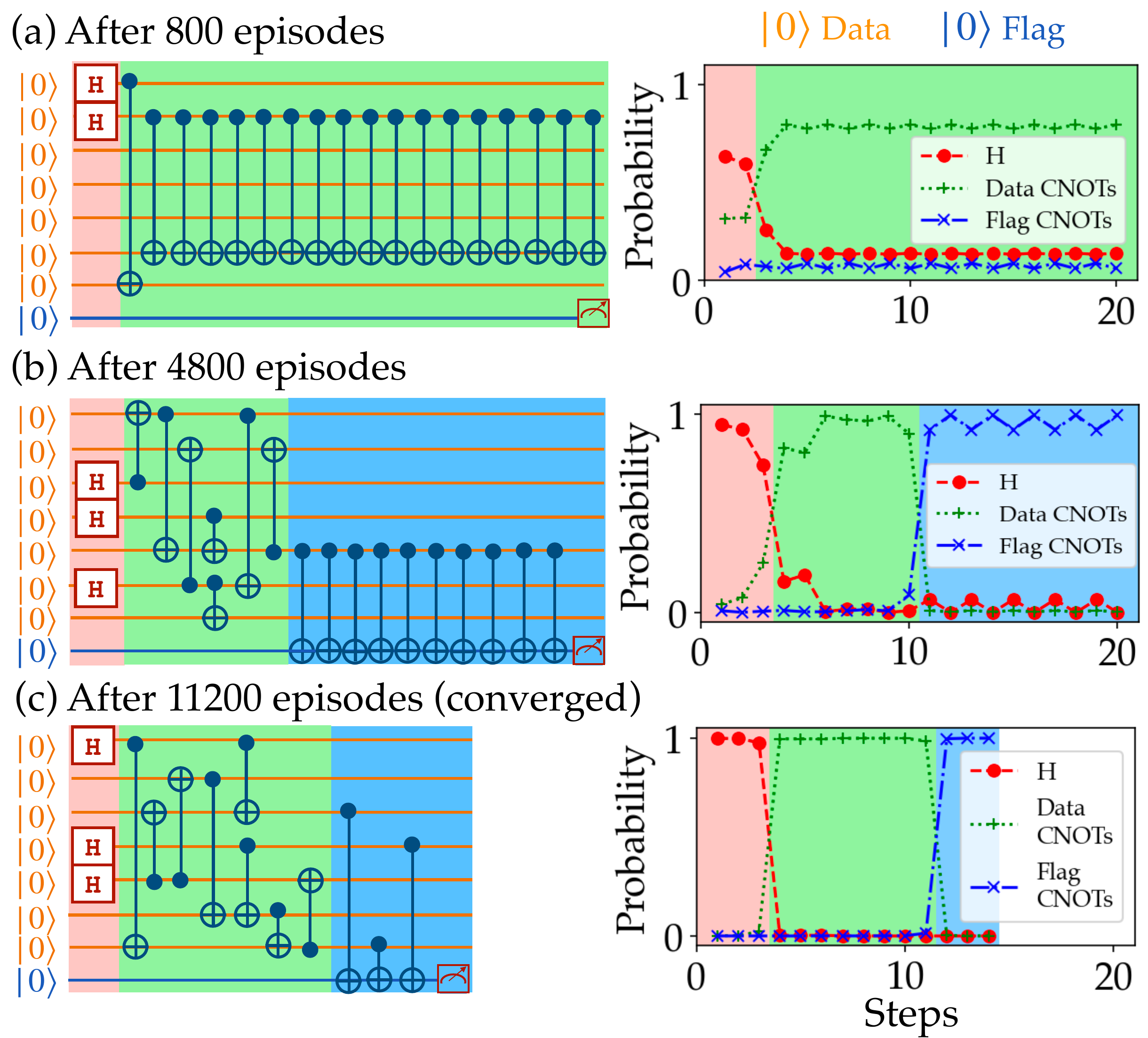}
	\caption{\label{fig:ft-logical-strategy} Evolution of the learned strategy during training. We train an RL agent for the integrated fault-tolerant $|0\rangle_L$ state preparation of the $[[7,1,3]]$ Steane code, assuming all-to-all qubit connectivity. The left part of the figure shows the circuit prepared by the agent. The right part shows the probability of actions for each step. We group the actions into 3 main groups: applying Hadamard gates on some qubits (red), applying CNOTs between data qubits (green), and applying CNOTs between data qubits and flag qubits (blue). The background color indicates the most probable group of actions at that step. We can see the progression of the RL agent's learning process, starting from applying mostly Hadamard gates in the first few steps in (a), followed by learning how to prepare the logical state in (b), and finally learning how to prepare the state and flag the harmful errors after convergence in (c). We hypothesize that the agent applies long sequences of self-cancelling CNOTs in (a) and (b) because it has not yet learned what to do in the later time steps. The agent then chose a ``safe'' strategy by applying multiple CNOTs several times, which does not change the reward.}
\end{figure}

The logical state of a surface code is commonly prepared fault-tolerantly by using repeated stabilizer measurements~\cite{fowler_surface_2012,ye_logical_2023}. Here, we for the first time show RL-discovered flag-based fault-tolerant logical state preparation of a surface code in the standard 2D grid connectivity: We illustrate this by preparing the $|0\rangle_L$ of the $[[9,1,3]]$ Surface-17 code with the connectivity and qubit placement from~\cite{krinner2022realizing} in Fig.~\ref{fig:result-ft-logical}(e). The repeated stabilizer measurements preparation needs $8$ syndrome qubits, while RL-discovered circuit, although given $8$ flag qubits to use, only needs $4$ flag qubits and uses $16$ two qubit gates. It is only $5$ two-qubit gates more than preparing it in an all-to-all qubit connectivity as shown in Table~\ref{tab:comparison} (the circuit is shown in Appendix~\ref{app:ftlsp-circuits}) \footnote{Currently, the RL method only looks at weight-1 errors of $d=3$ codes, thus missing the possible correctable weight-2 errors in the surface code. The latter can be easily implemented in a version of the method tailored to specific codes. However, our goal is to have a method that is as general as possible.\label{footnote:surface}}. The discovered circuit shows a novel approach to fault-tolerantly prepare a logical Pauli state of the surface code.

In Appendix~\ref{app:ler-ar-circuits-connectivity}, we compare the logical error and acceptance rates of the circuits shown in Fig.~\ref{fig:result-ft-logical}. In all cases the logical error rate scales as $p^2$, confirming that the circuits are fault-tolerant, as desired. In Appendix~\ref{app:ftlsp-varymiu}, we show how different values of the weight $\mu$ in the reward affect the acceptance and logical error rates.

We now show that the RL approach allows a straightforward exploration of different qubit connectivity and placements, i.e.,~assignments of data and flag qubits to physical qubits of the underlying device, by training different RL agents. We illustrate this by preparing the $|0\rangle_L$ state for the $[[7,1,3]]$ Steane code. On a $3\times3$ grid based on the Google Sycamore device, there are ${9 \choose 7} = 36$ possible data and flag qubit placements. We train on all possible configurations and in Fig.~\ref{fig:ft-logical-connectivity}(a), we show $5$ qubit placements where the circuit discovered by the RL agent has the lowest number of two-qubit gates. Next, we illustrate the same preparation for heavy-hex connectivity based on the IBMQ Guadalupe device. We show $3$ data and flag qubit placements where the circuit has the lowest number of two-qubit gates in Fig.~\ref{fig:ft-logical-connectivity}(b). We have also tried different flag qubit placements, but sometimes the agent does not find an encoding circuit, especially when the flag qubit is not located in the crossing (i.e.~the flag qubit is connected to $3$ data qubits). This may indicate that the best flag qubit placement in the heavy-hex connectivity is in the crossings. We provide the circuits in Appendix~\ref{app:ftlsp-circuits-connectivity}.

One might expect that when the RL agent directly prepares a fault-tolerant logical state, it would try to detect some harmful errors in the middle of the preparation. This is indeed the case, for example, in the circuit shown in Fig.~\ref{fig:result-ft-logical}(c). However, we observe that most of the discovered circuits can be decomposed into a state preparation circuit $U$, followed by a verification circuit $V$ (e.g., the circuits shown in Fig.~\ref{fig:result-ft-logical}(a), (b), and (d)). 

We can try to investigate the strategy that the RL agent learns by looking at the circuits and the action probabilities during the training. We illustrate this in Fig.~\ref{fig:ft-logical-strategy}. We see that in the initial training steps, the agent applies Hadamard gates to initialize some qubits in the $|+\rangle$, which is a known strategy for CSS codes~\cite{amaro2020scalable}. Next, the agent learns to prepare the logical state circuit without flagging harmful errors. Finally, the agent learns to flag the harmful errors until the training converges. We illustrate this strategy for the integrated fault-tolerant preparation of the $|0\rangle_L$ state of the $[[7,1,3]]$ Steane code on all-to-all qubit connectivity in Fig.~\ref{fig:ft-logical-strategy}.

 \begin{figure}[tb]
	\includegraphics[width=.45\textwidth]{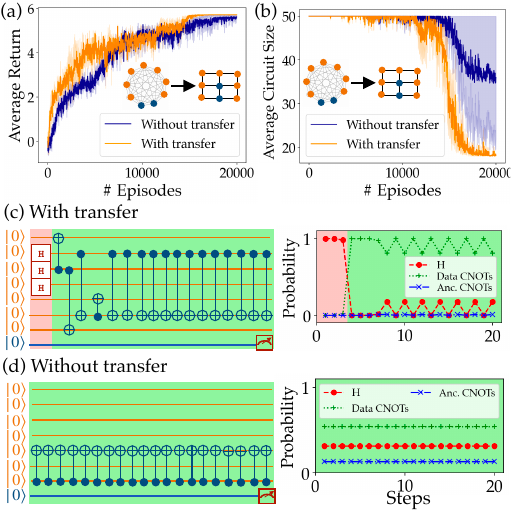}
	\caption{\label{fig:transfer-ftlsp} Transfer learning for integrated fault-tolerant logical state preparation from an all-to-all qubit connectivity to a 2D grid connectivity. Part (a) shows the average return and (b) shows the circuit size during training for the fault-tolerant $|0\rangle_L$ preparation of the $[[7,1,3]]$ Steane code with and without transfer learning. The training without transfer learning also converges, but requires more training. The central part (c) shows that when we directly use the transferred agent without training, the agent retains the knowledge of placing Hadamard gates as shown previously in Fig.~\ref{fig:ft-logical-strategy}. This is in contrast to the case without transfer learning shown in (d), where the agent applies only a long sequence of self-cancelling CNOTs. In this case, the agent has not learned anything, so the probability of each gate is still uniform. The CNOT gate between qubit 7 and 5 is just a random gate chosen by the agent.}
\end{figure}

Finally, we want to explore the possibility of transfer learning in this task. Transfer learning is a powerful technique in machine learning that leverages knowledge or a strategy gained from solving one problem to solve related but different problems. However, transfer learning does not always work effectively because it depends heavily on the similarity of the problems~\cite{taylor2009transfer}. 

We show a transfer learning technique where we reuse the agent that was trained to prepare a fault-tolerant logical state in an all-to-all qubit connectivity scenario to prepare the same state for the situation of restricted connectivity. The transfer learning process is explained in detail in Appendix~\ref{app:logical-state-transfer-connectivity}. We find that transfer learning helps to make the training converge faster. Additionally, we see that the transferred agent retains the strategy from the previous training. We illustrate this by comparing the integrated fault-tolerant preparation of the $|0\rangle_L$ state of the $[[7,1,3]]$ Steane code on a 2D grid without and with transfer learning from all-to-all qubit connectivity in Fig.~\ref{fig:transfer-ftlsp}. 

In summary, we have shown that the integrated fault-tolerant logical state preparation (IFT-LSP) approach always finds circuits with better or similar performance, both compared to circuits known in the literature as well as compared to circuits found by separating the task into the two subtasks (LSP+VCS). We have also demonstrated state-of-the-art RL-based fault-tolerant logical state preparation for restricted qubit connectivity scenarios with different connectivity and qubit placements. Furthermore, with transfer learning, we can reuse an RL agent trained for all-to-all qubit connectivity to accelerate the training for restricted qubit connectivity.

\section{Conclusions and Outlook}
\label{sec:conclusion}

We have presented reinforcement learning (RL) approaches to discover quantum circuits for fault-tolerant (FT) logical state preparation of QEC codes based on flag qubit protocols. We have started with the non-FT logical state preparation task and showed that RL prepares the logical state with a smaller circuit size than other methods for the all-to-all qubit connectivity scenarios. We have also highlighted that including the hardware constraint directly in the training yields quantum circuits with a smaller circuit size than transpiling a circuit for all-to-all qubit connectivity. We have then synthesized verification circuits to perform FT logical state preparation. We have demonstrated that RL can discover verification circuits that perform better than or equal to existing circuits in the literature. We have shown that the main approach that we advocate in this work, where we integrate the subtasks into the challenge of direct integrated fault-tolerant logical state preparation (IFT-LSP), performs even better than separating the tasks. Furthermore, we have demonstrated RL-based fault-tolerant logical state preparation under constrained connectivity for different qubit connectivity and placements. Finally, we have investigated and shown that transfer learning can help speed up the training process of the RL agent.

In this work, we have demonstrated the first steps in using an RL approach for the automatic discovery of quantum circuits for fault-tolerant protocols in quantum error correction. Our approach could naturally be extended and applied to different tasks, such as the discovery of quantum circuits for fault-tolerant magic state preparation~\cite{chamberland2019fault,shutty_decoding_2022}, syndrome measurement~\cite{delfosse2020short,bhatnagar_low-depth_2023}, logical gates, error correction cycles, and other quantum error correction subroutines. On the one hand, exploring these scenarios will not require a completely different approach, since verification-like circuits can be used to render the tasks fault-tolerant. However, such extensions will require a careful design of the appropriate reward function, set of actions, and observations to effectively train the RL agent. It will also be interesting to explore several avenues for scalability: on the one hand, this could include for instance devising modular formulations of the FT compilation tasks, e.g., by including more complex building blocks, such as sub-circuits for specific tasks such as stabilizer readout, in the set of actions available to the RL agent. On the other hand, it would be exciting to explore the potential of collaborative multi-agent RL scenarios, which may allow one to apply the techniques proposed in this work to larger-distance and concatenated error correction codes. 

\begin{acknowledgments}
We thank Sangkha Borah, Maximilian N\"{a}gele, Oleg Yevtushenko, Josias Old, Julio Carlos Magdalena de la Fuente, and Manuel Rispler for fruitful discussions. The research is part of the Munich Quantum Valley (K-4 and K-8), which is supported by the Bavarian state government with funds from the Hightech Agenda Bayern Plus. L.C. and M.M. furthermore acknowledge support by the US Army Research Office through Grant Number W911NF-21-1-0007. M.M. also acknowledges support by the European Union’s Horizon Europe research and innovation program under Grant Agreement Number 101114305 (``MILLENION-SGA1'' EU Project), the ERC Starting Grant QNets through Grant Number 804247, and by the Deutsche Forschungsgemeinschaft (DFG, German Research Foundation) under Germany's Excellence Strategy `Cluster of Excellence Matter and Light for Quantum Computing (ML4Q) EXC 2004/1' 390534769.
\end{acknowledgments}

\bibliography{main}
\pagebreak

\appendix

\section{Tableau representation}
\label{app:tableau}

Here, we give more details about the representation of quantum circuits as tableaus. Note that in this work, we omit the ``destabilizer'' generators in the tableau described in~\cite{aaronson2004improved}, since they are not useful for the task at hand. 

A tableau can be represented as a $n \times (2n + 1)$ matrix of binary variables $x_{ij}$, $z_{ij}, r_{i}$ for $i,j \in \{1,\dots, n\}$. Each row $i$ of the tableau $[x_{i1}, \dots x_{in}, z_{i1}, \dots, z_{in}, r_{i}]$ represents the Pauli matrix of the generators or the logical operators, where the $x_{ij}z_{ij}$ bits determine the $j$-th Pauli matrix, where $00$, $01$, $10$, and $11$ denote $I$, $Z$, $X$, and $Y$ Pauli, respectively, and $r_i$ denotes the phase ($1$ for negative phase and $0$ for positive phase). For instance, a binary vector $[10011 | 00110 | 1]$ represents the Pauli $-XIZYX$. 

For instance, the tableau for $|0\rangle_L$ of the $[[7,1,3]]$ Steane code~\cite{steane1996multiple} is a matrix of binary numbers of size $7 \times 15$ that contains $7-1=6$ stabilizer generators and $1$ logical operator $Z_L$.  Table~\ref{tab:stabilizer_generators} shows the generators of the $[[7,1,3]]$ Steane code. Eq.~(\ref{eq:steane-tableau}) shows an example of the tableau for $|0\rangle_L$   of the $[[7,1,3]]$ Steane code when we choose $Z_L = +ZZZZZZZ$. In this tableau, the first row represents the logical operator $Z_L = +ZZZZZZZ$, the second row represents the first stabilizer operator $+ZIZIZIZ$, and so on.

\begin{equation}
\label{eq:steane-tableau}
\left(\begin{array}{ccccccc|ccccccc|c}
0 & 0 & 0 & 0 & 0 & 0 & 0 & 1 & 1 & 1 & 1 & 1 & 1 & 1 & 0 \\
0 & 0 & 0 & 0 & 0 & 0 & 0 & 1 & 0 & 1 & 0 & 1 & 0 & 1 & 0 \\
1 & 0 & 1 & 0 & 1 & 0 & 1 & 0 & 0 & 0 & 0 & 0 & 0 & 0 & 0 \\
0 & 0 & 0 & 0 & 0 & 0 & 0 & 0 & 1 & 1 & 0 & 0 & 1 & 1 & 0 \\
0 & 1 & 1 & 0 & 0 & 1 & 1 & 0 & 0 & 0 & 0 & 0 & 0 & 0 & 0 \\
0 & 0 & 0 & 0 & 0 & 0 & 0 & 0 & 0 & 0 & 1 & 1 & 1 & 1 & 0 \\
0 & 0 & 0 & 1 & 1 & 1 & 1 & 0 & 0 & 0 & 0 & 0 & 0 & 0 & 0 \\
\end{array}
\right)
\end{equation}

As discussed in the main text, reordering the rows of the tableau represents the same state. However, this is not the case for the canonical tableau~\cite{aaronson2004improved}, which is a unique representation of a tableau that represents the same state. We can apply Gaussian elimination to the matrix in Eq.~(\ref{eq:steane-tableau}) and get the canonical tableau in Eq.~(\ref{eq:steane-tableau-canonical}). The Pauli string for this tableau is: $+XIXIXIX$, $+ZIIIIZZ$, $+IXXIIXX$, $+IZIIZIZ$, $+IIZIZZI$, $+IIIXXXX$, and $+IIIZZZZ$. We can reorder the rows or change a row by multiplying other rows in Eq.~(\ref{eq:steane-tableau}) and it will still give the same canonical tableau.

\begin{equation}
\label{eq:steane-tableau-canonical}
\left(\begin{array}{ccccccc|ccccccc|c}
1 & 0 & 1 & 0 & 1 & 0 & 1 & 0 & 0 & 0 & 0 & 0 & 0 & 0 & 0 \\
0 & 0 & 0 & 0 & 0 & 0 & 0 & 1 & 0 & 0 & 0 & 0 & 1 & 1 & 0 \\
0 & 1 & 1 & 0 & 0 & 1 & 1 & 0 & 0 & 0 & 0 & 0 & 0 & 0 & 0 \\
0 & 0 & 0 & 0 & 0 & 0 & 0 & 0 & 1 & 0 & 0 & 1 & 0 & 1 & 0 \\
0 & 0 & 0 & 0 & 0 & 0 & 0 & 0 & 0 & 1 & 0 & 1 & 1 & 0 & 0 \\
0 & 0 & 0 & 1 & 1 & 1 & 1 & 0 & 0 & 0 & 0 & 0 & 0 & 0 & 0 \\
0 & 0 & 0 & 0 & 0 & 0 & 0 & 0 & 0 & 0 & 1 & 1 & 1 & 1 & 0 \\
\end{array}
\right)
\end{equation}

We flatten this matrix into a vector and use it to compute the distance metric and as an input to the neural networks.

\section{Hyperparameters and details of the training}
\label{app:training-details}

We use the multilayer perceptron architecture for the critic and actor networks to train the PPO algorithm~\cite{sutton1999policy}. Both of the networks have two hidden layers with the $\text{ReLU}$ activation function. The number of hidden nodes is set to $128$. However, in cases where the number of physical qubits is more than $10$ with all-to-all qubit connectivity, we increase the number of hidden nodes to $256$. The weight matrices are initialized with a uniformly distributed orthogonal matrix with $0.01$ scale, while the biases are initialized to zero.

The hyperparameters of the PPO training are as follows~\cite{sutton1999policy}. We use the Adam optimizer with a learning rate of $0.001$ with an annealing learning rate. We train $10$ agents in parallel. Each agent sees batches of $16$ environments. We train the agent for a total of one million time steps with an entropy coefficient of $0.05$. The network is updated after every $4$ epochs and the number of minibatches is set to $4$. The discount factor $(\gamma)$ is set to $0.99$, the generalized advantage estimate (GAE) value ($\lambda$)  is set to $0.95$,  the clipping parameter ($\epsilon$) is set to $0.2$, the value function coefficient is set to $0.5$, and the maximum gradient norm clip value is set to $0.5$. For harder cases (e.g. larger physical qubits or restricted connectivity), we increase the total time steps to $10$ or $30$ million and change the learning rate to $0.0005$ and the entropy coefficient to $0.1$. All experimental results shown in this paper are done on NVIDIA Quadro RTX 6000 GPU. The training is done with the same seed value to ensure the same randomness.

For all experiments, we set the stopping threshold $\epsilon = 0.9999$ and the maximum steps in the trajectory $L = 50$ and in harder cases to $L = 100$.  For the reward of verification circuit synthesis in Eq.(~\ref{eq:vcs-reward}), we set $\mu_{f} = n$, $\mu_{d} = \lfloor n / 2 \rfloor$, and $\mu_{p} = 1$. For the reward of fault-tolerant logical state preparation also defined in Eq.(~\ref{eq:vcs-reward}), we set $\mu_{d} = n$, $\mu_{f} = \lfloor n / 2 \rfloor$, and $\mu_{p} = 1$. These values are determined from our numerical experiments by varying the weight, which we discuss in Appendix~\ref{app:vcs-varymiu} and Appendix~\ref{app:ftlsp-varymiu}.

\section{Comparison of distance functions}
\label{app:distance-comparison}

We propose to use the complementary distance $1-d_t$ between the target canonical tableau ($G_{\rm{target}}$) and the canonical tableau of the current circuit at time $t$ ($G_{\rm{t}}$) as a reward to the RL agent. We first convert the current and target canonical tableau matrices into binary vectors and compute the distance.  In principle, we can take any binary distance measure. A natural choice of metric is the Hamming distance, which fits the current application since it is mostly used in coding theory.  However, our empirical experiments showed that the Jaccard distance works better than the Hamming distance. 

Let $C_{11}$ be the number of elements in $G_{\rm{target}}$ and $G_{t}$ that have a value $1$ at the same position, $C_{00}$ the number of elements in  $G_{\rm{target}}$ and $G_{t}$ that have a value of $0$ at the same position, $C_{01}$ the number of elements that have a value of $0$  in  $G_{\rm{target}}$ and $1$ in $G_{t}$ at the same position, and $C_{10}$ the number of elements that have a value $1$  in  $G_{\rm{target}}$ and $0$ in $G_{t}$ at the same position.

The Hamming distance $d_H$ is defined as follows

\begin{equation}
	\label{eq:hamming-dist}	
	d_H  = \frac{C_{01} + C_{10}}{C_{00} + C_{01} + C_{10} + C_{11}},
\end{equation}

while the Jaccard distance~\cite{kosub_note_2016} $d_J$ is defined as

\begin{equation}
	\label{eq:jaccard-dist}	
	d_J  = \frac{C_{01} + C_{10}}{C_{01} + C_{10} + C_{11}}.
\end{equation}

\begin{figure}[!htb]
	\includegraphics[width=.45\textwidth]{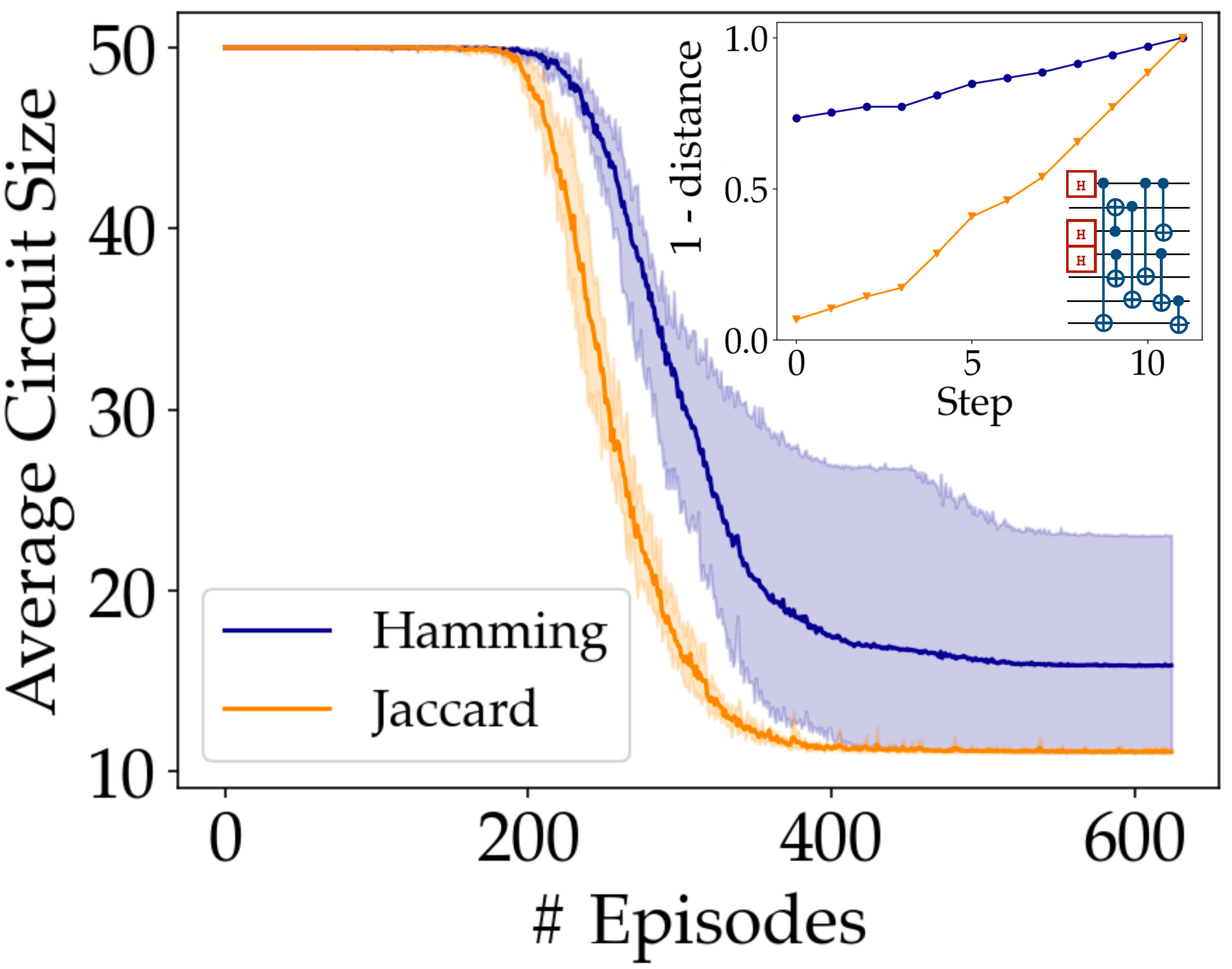}
	\caption{\label{fig:distance-comparison} Comparison of Hamming and Jaccard distance for the reward function. Average circuit size of training an RL agent to prepare the $|0\rangle_L$ state of the $[[7,1,3]]$ Steane code using the Hamming and Jaccard distance for the complementary tableau distance reward $1 -d_t$. The color shade shows the standard deviation over five different trainings. The inset shows how the inverse distance value evolves for each step of gate application for the circuit shown.    }
\end{figure}

We compare the RL training for preparing the $|0\rangle_L$ state of the $[[7,1,3]]$ Steane code using Hamming and Jaccard distance. In Fig.~\ref{fig:distance-comparison}, we see that results based on using the Jaccard distance converge faster and more stably than those obtained by using the Hamming distance. As seen in Eq.~(\ref{eq:jaccard-dist}), the computation of the Jaccard distance does not take into account the $C_{00}$. In stabilizer language, this means that we do not take into account when the Pauli identity $I$ matches in both target and current tableau. This also means that the Jaccard distance penalizes more dissimilarities compared to the Hamming distance.   We hypothesize that since we are using a reward shaping technique, the increase in the inverse distance is larger when using the Jaccard distance, which is better learned by the RL agent. We illustrate this in the inset of  Fig.~\ref{fig:distance-comparison}, where we test an RL-prepared circuit and compare how the value of inverse distance evolves for each step of applying a gate. We see that the Jaccard distance starts lower and increases more steeply than the Hamming distance. Therefore, we use the Jaccard distance $d_J$ as the distance metric $d_t$ for our experiments.

\section{Calculation of the complementary tableau distance and product state}
\label{app:reward-computation}

\begin{figure}[!htb]
	\includegraphics[width=.4\textwidth]{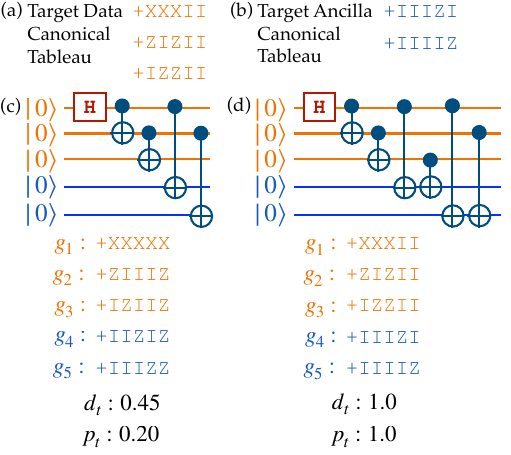}
	\caption{\label{fig:reward-calculation} Illustration of the complementary tableau distance ($d_t$) and product state ($p_t$)  calculation. We want to prepare the state $|000\rangle + |111\rangle$  with two ancillas $(n_A = 2)$.  The target canonical tableau can be separated according to the data (a) and ancilla (b) qubits. (c) shows an example of a circuit that almost prepares the target state, but the data qubits are entangled with the ancilla. (d) shows an example of a circuit that prepares the target state and is a product state of the data qubits and the ancillas. }
\end{figure}

Here we show an example of distance and product state calculations, as explained in Sec.~\ref{sec:verification-circuit}, which are part of the reward function for verification circuit synthesis and integrated fault-tolerant logical state preparation tasks. 

As an example, consider the preparation of the state $|000\rangle + |111\rangle$  with two ancillas $(n_A = 2)$. The state has the following target canonical tableau: $+XXX$, $+ZIZ$, and $+IZZ$. We need to append the last column of the target canonical tableau with $I^{\otimes n_A}$, so the target canonical tableau for the data qubits is now:  $+XXXII$, $+ZIZII$, $+IZZII$, as shown in Fig.~\ref{fig:reward-calculation}(a). As mentioned in Sec.~\ref{sec:verification-circuit}, the stabilizer generators of the ancilla must be of $Z$-type in the location of the ancilla and $I$ in the others for it to be a product state. Therefore, the target canonical tableau of the ancilla qubits must be: $+IIIZI$ and $+IIIIZ$, as shown in Fig.~\ref{fig:reward-calculation}(b). 

Fig.~\ref{fig:reward-calculation}(c) and (d) show two circuit examples. To compute the complementary tableau distance $d_t$, we take $g_1, g_2,$ and $g_3$ of the canonical tableau of the circuit and compute the complementary tableau distance with the target data canonical tableau. Since the ancilla qubits are always placed last, we can extract the canonical tableau of the data qubits by taking the submatrix of the canonical tableau from rows $1$ to $n$. To compute the product state $p_t$, we first take $g_4$ and $g_5$ from the canonical tableau of the circuit. We then compute the $d_t$ between this subtableau and the target ancilla canonical tableau.  We can see that the circuit in Fig.~\ref{fig:reward-calculation}(c) is still far from the target state because the data qubits are entangled with the ancilla, since both the $d_t$ and $p_t$ are below $1$. While the circuit in Fig.~\ref{fig:reward-calculation}(d) correctly prepares the state ($d_t=1$) and is a product state of the data qubits and the ancillas ($p_t=1$).

\section{Minimizing the number of gates in the reward function}
\label{app:reward-explanation}

\begin{figure}[!htb]
	\includegraphics[width=.35\textwidth]{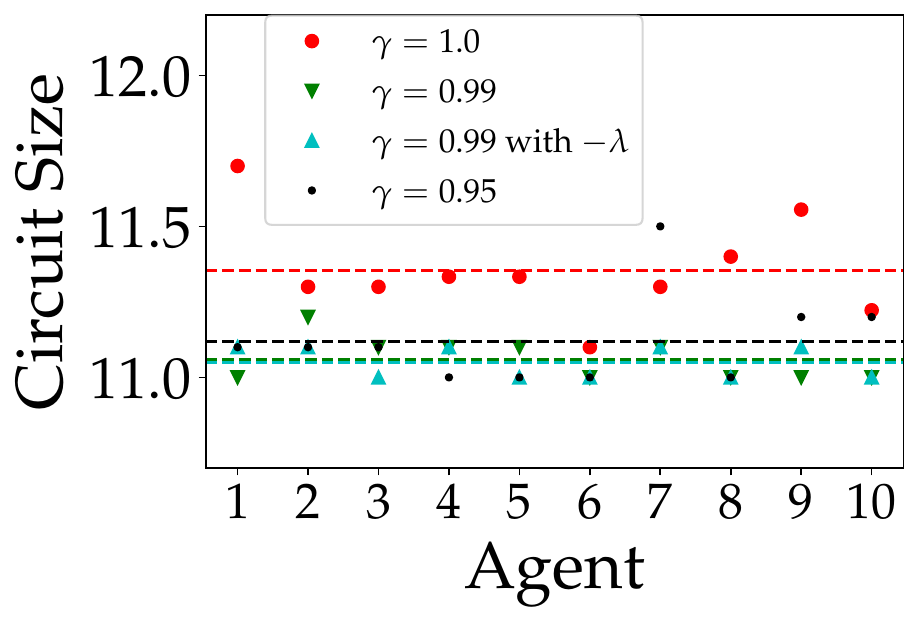}
	\caption{\label{fig:gamma} Illustration of how the discount factor $\gamma$ affects the circuit size. We train 10 different agents for preparing the $|0\rangle_L$ state of the $[[7,1,3]]$ Steane code with $\gamma = \{ 0.95 , 0.99, 1\}$. Additionally, we also tried to set $\gamma$ to $0.99$ and add the $\lambda = - 1 / 50$ (referred in the figure as $\gamma = 0.99$ with $-\lambda$) in the reward function to penalize longer trajectories. The dashed line shows the average circuit size. }
\end{figure}

One might notice that the reward function in the logical state preparation task does not include a term that minimizes the number of gates to make the preparation circuit compact. The most common technique is to add an extra term $-\lambda$ at each time step to penalize longer sequences, so that the reward function in Eq.~\ref{eq:logical-state-reward} is $r_t = d_{t-1} - d_t - \lambda$. The problem is that $\lambda$ is now a hyperparameter that must be tuned so that it does not become stronger than the complementary tableau distance reward.

However, without this term, one can also use the discount factor $\gamma$ in the cumulative reward~\cite{sutton2018reinforcement}. Instead of maximizing $\mathcal{J(\theta)} = \mathbb{E}_{\tau \sim \pi_{\theta}}[\sum_{t=0}^{T} r_t]$, we instead maximize $\mathcal{J(\theta)} = \mathbb{E}_{\tau \sim \pi_{\theta}}[\sum_{t=0}^{T} \gamma^t r_t]$. The $\gamma$ value ranges from $0$ to $1$. It determines how much future rewards are reduced in value compared to immediate rewards. When $\gamma = 1$, the agent values future rewards as much as present rewards, which can lead to longer trajectories. When $\gamma < 1$, the agent places more emphasis on the long-term rewards and may take more steps initially to reach a state that yields higher rewards in the long run, which can lead to shorter trajectories.

We show this empirically in Fig.~\ref{fig:gamma}. We see that $\gamma = 1$ leads to a longer circuit than $\gamma < 1$. Furthermore, adding the $-\lambda$ term to penalize longer sequences does not further reduce the circuit size. For all experiments, we do not include the $-\lambda$ term and set $\gamma = 0.99$.

\section{List of stabilizer generators}
\label{app:stabilizer-generators}

\begin{table}[H]
	\caption{\label{tab:stabilizer_generators} Here we list the stabilizer generators for the 
		$[[5,1,3]]$ perfect code~\cite{laflamme1996perfect}, $[[7,1,3]]$ Steane code~\cite{steane1996multiple}, $[[9,1,3]]$ Shor code~\cite{shor1995scheme}, $[[9,1,3]]$ Surface-17 code~\cite{kitaev_fault-tolerant_2003},  $[[15,1,3]]$ quantum Reed-Muller code or 3D color code~\cite{anderson2014fault}, and  the $[[17,1,5]]$ 2D color code~\cite{bombin2015gauge}. For the logical operators, we choose $Z_L$ to be $Z^{\otimes n}$ and $X_L$ as $X^{\otimes n}$, where $n$ is the number of physical qubits of the respective code.}

	\small 
 \centering
 \setlength{\tabcolsep}{1em}

		\begin{tabular}{cccc}

$[[5,1,3]]$ & $[[7,1,3]]$ & $[[9,1,3]]$ & $[[9,1,3]]$ \\
perfect & Steane & Shor & Surface-17\\\hline \hline
\verb+IXZZX+       & \verb+ZIZIZIZ+     & \verb+ZZIIIIIII+   & \verb+ZIIZIIIII+   \\
\verb+XZZXI+       & \verb+XIXIXIX+     & \verb+ZIZIIIIII+   & \verb+IIIZZIZZI+   \\
\verb+ZZXIX+       & \verb+IZZIIZZ+     & \verb+XXXXXXIII+   & \verb+IZZIZZIII+   \\
\verb+ZXIXZ+       & \verb+IXXIIXX+     & \verb+IIIZZIIII+   & \verb+IIIIIZIIZ+   \\
            & \verb+IIIZZZZ+     & \verb+IIIZIZIII+   & \verb+IXXIIIIII+   \\
            & \verb+IIIXXXX+     & \verb+XXXIIIXXX+   & \verb+XXIXXIIII+   \\
            &             & \verb+IIIIIIZZI+   & \verb+IIIIXXIXX+   \\
            &             & \verb+IIIIIIZIZ+   & \verb+IIIIIIXXI+   \\
            & & & \\
 		\end{tabular}

	\begin{tabular}{cc}
 $[[15,1,3]]$ Reed-Muller    & $[[17,1,5]]$  2D Color     \\\hline \hline
\verb+ZIZIZIZIZIZIZIZ+ & \verb+XXXXIIIIIIIIIIIII+ \\
\verb+XIXIXIXIXIXIXIX+ & \verb+ZZZZIIIIIIIIIIIII+ \\
\verb+IZZIIZZIIZZIIZZ+ & \verb+XIXIXXIIIIIIIIIII+ \\
\verb+IXXIIXXIIXXIIXX+ & \verb+ZIZIZZIIIIIIIIIII+ \\
\verb+IIIZZZZIIIIZZZZ+ & \verb+IIIIXXIIXXIIIIIII+ \\
\verb+IIIXXXXIIIIXXXX+ & \verb+IIIIZZIIZZIIIIIII+ \\
\verb+IIIIIIIZZZZZZZZ+ & \verb+IIIIIIXXIIXXIIIII+ \\
\verb+IIIIIIIXXXXXXXX+ & \verb+IIIIIIZZIIZZIIIII+ \\
\verb+IIZIIIZIIIZIIIZ+ & \verb+IIIIIIIIXXIIXXIII+ \\
\verb+IIIIZIZIIIIIZIZ+ & \verb+IIIIIIIIZZIIZZIII+ \\
\verb+IIIIIZZIIIIIIZZ+ & \verb+IIIIIIIIIIXXIIXXI+ \\
\verb+IIIIIIIIIZZIIZZ+ & \verb+IIIIIIIIIIZZIIZZI+ \\
\verb+IIIIIIIIIIIZZZZ+ & \verb+IIIIIIIXIIIXIIIXX+ \\
\verb+IIIIIIIIZIZIZIZ+ & \verb+IIIIIIIZIIIZIIIZZ+ \\
                & \verb+IIXXIXXIIXXIIXXII+ \\
                & \verb+IIZZIZZIIZZIIZZII+ 
                \end{tabular}

\end{table}

\section{Logical state preparation circuits for all-to-all qubit connectivity with standard gate sets}
\label{app:lsp-circuits}

We show some examples of learned circuits for all-to-all qubit connectivity with the standard gate sets. For all circuits shown below, we choose $Z_L = Z^{\otimes n}$. These circuits are also available online~\footnotemark[1]. 

\textbf{The $\mathbf{|0\rangle_L}$ state of the $\mathbf{[[5,1,3]]}$ perfect code}. We see that the prepared circuit has a pattern of two $S$ gates at the end of the circuit. This is because the RL agent would first aim to get the correct stabilizer generators without worrying about the sign and then apply $Z$ gates (which can be decomposed into two $S$ gates) to fix the sign. 

\noindent \scalebox{0.8}{
	\Qcircuit @C=0.1em @R=0.1em @!R { \\
		\nghost{{q}_{0} :  } & \lstick{{q}_{0} :  } & \qw & \targ & \qw & \ctrl{2} & \qw & \qw & \qw & \qw & \qw & \qw & \qw\\
		\nghost{{q}_{1} :  } & \lstick{{q}_{1} :  } & \gate{\mathrm{H}} & \qw & \ctrl{2} & \qw & \qw & \ctrl{2} & \qw & \qw & \qw & \qw & \qw\\
		\nghost{{q}_{2} :  } & \lstick{{q}_{2} :  } & \qw & \qw & \qw & \targ & \gate{\mathrm{H}} & \qw & \ctrl{1} & \gate{\mathrm{S}} & \gate{\mathrm{S}} & \qw & \qw\\
		\nghost{{q}_{3} :  } & \lstick{{q}_{3} :  } & \qw & \qw & \targ & \gate{\mathrm{H}} & \ctrl{1} & \targ & \targ & \qw & \qw & \qw & \qw\\
		\nghost{{q}_{4} :  } & \lstick{{q}_{4} :  } & \gate{\mathrm{H}} & \ctrl{-4} & \qw & \qw & \targ & \gate{\mathrm{S}} & \gate{\mathrm{S}} & \qw & \qw & \qw & \qw\\
		\\ }}\scalebox{0.8}{
	\Qcircuit @C=0.1em @R=0.1em @!R { \\
		\nghost{{q}_{0} :  } & \lstick{{q}_{0} :  } & \gate{\mathrm{H}} & \qw & \qw & \ctrl{4} & \ctrl{2} & \gate{\mathrm{S}} & \gate{\mathrm{S}} & \qw & \qw & \qw & \qw\\
		\nghost{{q}_{1} :  } & \lstick{{q}_{1} :  } & \qw & \qw & \targ & \qw & \qw & \qw & \qw & \qw & \qw & \qw & \qw\\
		\nghost{{q}_{2} :  } & \lstick{{q}_{2} :  } & \qw & \qw & \qw & \qw & \targ & \gate{\mathrm{H}} & \ctrl{1} & \qw & \qw & \qw & \qw\\
		\nghost{{q}_{3} :  } & \lstick{{q}_{3} :  } & \gate{\mathrm{H}} & \ctrl{1} & \ctrl{-2} & \qw & \gate{\mathrm{H}} & \ctrl{1} & \targ & \gate{\mathrm{S}} & \gate{\mathrm{S}} & \qw & \qw\\
		\nghost{{q}_{4} :  } & \lstick{{q}_{4} :  } & \qw & \targ & \qw & \targ & \qw & \targ & \qw & \qw & \qw & \qw & \qw\\
		\\ }}

\textbf{The $\mathbf{|0\rangle_L}$ state of the $\mathbf{[[7,1,3]]}$ Steane code}. We see that the RL agent learns from scratch to prepare part of the initial state of the physical qubits to $|+\rangle$ by applying an $H$ gate, similar to the strategy in~\cite{amaro2020scalable}. We find that we can exploit this observation by using the following alternative strategy to speed up the training process. We first select a random subset of physical qubits to $|+\rangle$ and then allow the agent to apply gates other than $H$.

\noindent \scalebox{1.0}{
	\Qcircuit @C=0.1em @R=0.1em @!R { \\
		\nghost{{q}_{0} :  } & \lstick{{q}_{0} :  } & \gate{\mathrm{H}} & \ctrl{6} & \qw & \ctrl{4} & \qw & \ctrl{2} & \qw & \qw & \qw\\
		\nghost{{q}_{1} :  } & \lstick{{q}_{1} :  } & \gate{\mathrm{H}} & \qw & \ctrl{4} & \qw & \ctrl{1} & \qw & \qw & \qw & \qw\\
		\nghost{{q}_{2} :  } & \lstick{{q}_{2} :  } & \qw & \qw & \qw & \qw & \targ & \targ & \qw & \qw & \qw\\
		\nghost{{q}_{3} :  } & \lstick{{q}_{3} :  } & \gate{\mathrm{H}} & \qw & \qw & \qw & \ctrl{1} & \ctrl{2} & \qw & \qw & \qw\\
		\nghost{{q}_{4} :  } & \lstick{{q}_{4} :  } & \qw & \qw & \qw & \targ & \targ & \qw & \qw & \qw & \qw\\
		\nghost{{q}_{5} :  } & \lstick{{q}_{5} :  } & \qw & \qw & \targ & \qw & \qw & \targ & \ctrl{1} & \qw & \qw\\
		\nghost{{q}_{6} :  } & \lstick{{q}_{6} :  } & \qw & \targ & \qw & \qw & \qw & \qw & \targ & \qw & \qw\\
		\\ }} \scalebox{1.0}{
	\Qcircuit @C=0.1em @R=0.1em @!R { \\
		\nghost{{q}_{0} :  } & \lstick{{q}_{0} :  } & \gate{\mathrm{H}} & \ctrl{6} & \qw & \qw & \qw & \qw & \qw & \ctrl{2} & \qw & \qw\\
		\nghost{{q}_{1} :  } & \lstick{{q}_{1} :  } & \gate{\mathrm{H}} & \qw & \ctrl{4} & \qw & \qw & \targ & \ctrl{1} & \qw & \qw & \qw\\
		\nghost{{q}_{2} :  } & \lstick{{q}_{2} :  } & \qw & \qw & \qw & \qw & \qw & \qw & \targ & \targ & \qw & \qw\\
		\nghost{{q}_{3} :  } & \lstick{{q}_{3} :  } & \gate{\mathrm{H}} & \qw & \qw & \qw & \ctrl{1} & \ctrl{-2} & \qw & \qw & \qw & \qw\\
		\nghost{{q}_{4} :  } & \lstick{{q}_{4} :  } & \qw & \qw & \qw & \targ & \targ & \qw & \qw & \qw & \qw & \qw\\
		\nghost{{q}_{5} :  } & \lstick{{q}_{5} :  } & \qw & \qw & \targ & \qw & \ctrl{1} & \qw & \qw & \qw & \qw & \qw\\
		\nghost{{q}_{6} :  } & \lstick{{q}_{6} :  } & \qw & \targ & \qw & \ctrl{-2} & \targ & \qw & \qw & \qw & \qw & \qw\\
		\\ }}

\textbf{The $\mathbf{|+\rangle_L}$ state of the $\mathbf{[[9,1,3]]}$ Shor code.}

\noindent \scalebox{1.0}{
	\Qcircuit @C=0.1em @R=0.1em @!R { \\
		\nghost{{q}_{0} :  } & \lstick{{q}_{0} :  } & \qw & \targ & \targ & \ctrl{1} & \ctrl{2} & \qw & \qw\\
		\nghost{{q}_{1} :  } & \lstick{{q}_{1} :  } & \qw & \qw & \qw & \targ & \qw & \qw & \qw\\
		\nghost{{q}_{2} :  } & \lstick{{q}_{2} :  } & \qw & \qw & \qw & \qw & \targ & \qw & \qw\\
		\nghost{{q}_{3} :  } & \lstick{{q}_{3} :  } & \gate{\mathrm{H}} & \ctrl{-3} & \qw & \ctrl{2} & \ctrl{1} & \qw & \qw\\
		\nghost{{q}_{4} :  } & \lstick{{q}_{4} :  } & \qw & \qw & \qw & \qw & \targ & \qw & \qw\\
		\nghost{{q}_{5} :  } & \lstick{{q}_{5} :  } & \qw & \qw & \qw & \targ & \qw & \qw & \qw\\
		\nghost{{q}_{6} :  } & \lstick{{q}_{6} :  } & \gate{\mathrm{H}} & \qw & \ctrl{-6} & \ctrl{2} & \ctrl{1} & \qw & \qw\\
		\nghost{{q}_{7} :  } & \lstick{{q}_{7} :  } & \qw & \qw & \qw & \qw & \targ & \qw & \qw\\
		\nghost{{q}_{8} :  } & \lstick{{q}_{8} :  } & \qw & \qw & \qw & \targ & \qw & \qw & \qw\\
		\\ }} \scalebox{1.0}{
	\Qcircuit @C=0.1em @R=0.1em @!R { \\
		\nghost{{q}_{0} :  } & \lstick{{q}_{0} :  } & \gate{\mathrm{H}} & \ctrl{8} & \ctrl{1} & \ctrl{2} & \qw & \qw & \qw & \qw\\
		\nghost{{q}_{1} :  } & \lstick{{q}_{1} :  } & \qw & \qw & \targ & \qw & \qw & \qw & \qw & \qw\\
		\nghost{{q}_{2} :  } & \lstick{{q}_{2} :  } & \qw & \qw & \qw & \targ & \qw & \qw & \qw & \qw\\
		\nghost{{q}_{3} :  } & \lstick{{q}_{3} :  } & \gate{\mathrm{H}} & \qw & \ctrl{2} & \ctrl{1} & \qw & \qw & \qw & \qw\\
		\nghost{{q}_{4} :  } & \lstick{{q}_{4} :  } & \qw & \qw & \qw & \targ & \qw & \qw & \qw & \qw\\
		\nghost{{q}_{5} :  } & \lstick{{q}_{5} :  } & \qw & \qw & \targ & \ctrl{3} & \qw & \qw & \qw & \qw\\
		\nghost{{q}_{6} :  } & \lstick{{q}_{6} :  } & \qw & \qw & \qw & \qw & \targ & \qw & \qw & \qw\\
		\nghost{{q}_{7} :  } & \lstick{{q}_{7} :  } & \qw & \qw & \qw & \qw & \qw & \targ & \qw & \qw\\
		\nghost{{q}_{8} :  } & \lstick{{q}_{8} :  } & \qw & \targ & \qw & \targ & \ctrl{-2} & \ctrl{-1} & \qw & \qw\\
		\\ }}

\textbf{The $\mathbf{|0\rangle_L}$ state of the $\mathbf{[[15,1,3]]}$ Reed-Muller or distance-3 3D color code}.

\noindent  \scalebox{1.0}{
	\Qcircuit @C=0.1em @R=0.1em @!R { \\
		\nghost{{q}_{0} :  } & \lstick{{q}_{0} :  } & \qw & \targ & \qw & \qw & \qw & \qw & \ctrl{14} & \ctrl{3} & \qw & \qw & \qw & \qw & \qw & \qw & \ctrl{5} & \qw & \qw\\
		\nghost{{q}_{1} :  } & \lstick{{q}_{1} :  } & \qw & \qw & \targ & \qw & \qw & \qw & \qw & \qw & \qw & \targ & \ctrl{5} & \qw & \ctrl{8} & \ctrl{7} & \qw & \qw & \qw\\
		\nghost{{q}_{2} :  } & \lstick{{q}_{2} :  } & \gate{\mathrm{H}} & \qw & \ctrl{-1} & \qw & \ctrl{10} & \qw & \qw & \qw & \qw & \qw & \qw & \qw & \qw & \qw & \qw & \qw & \qw\\
		\nghost{{q}_{3} :  } & \lstick{{q}_{3} :  } & \qw & \qw & \qw & \qw & \qw & \targ & \qw & \targ & \qw & \qw & \qw & \qw & \qw & \qw & \qw & \qw & \qw\\
		\nghost{{q}_{4} :  } & \lstick{{q}_{4} :  } & \gate{\mathrm{H}} & \qw & \ctrl{9} & \ctrl{2} & \qw & \qw & \qw & \qw & \qw & \qw & \qw & \qw & \qw & \qw & \qw & \qw & \qw\\
		\nghost{{q}_{5} :  } & \lstick{{q}_{5} :  } & \qw & \qw & \qw & \qw & \qw & \qw & \qw & \qw & \qw & \qw & \qw & \targ & \qw & \qw & \targ & \qw & \qw\\
		\nghost{{q}_{6} :  } & \lstick{{q}_{6} :  } & \qw & \qw & \qw & \targ & \qw & \qw & \qw & \qw & \qw & \qw & \targ & \ctrl{-1} & \qw & \qw & \qw & \qw & \qw\\
		\nghost{{q}_{7} :  } & \lstick{{q}_{7} :  } & \qw & \qw & \qw & \targ & \qw & \qw & \qw & \qw & \targ & \qw & \ctrl{2} & \qw & \qw & \qw & \qw & \qw & \qw\\
		\nghost{{q}_{8} :  } & \lstick{{q}_{8} :  } & \qw & \qw & \qw & \qw & \qw & \qw & \qw & \qw & \qw & \qw & \qw & \targ & \qw & \targ & \qw & \qw & \qw\\
		\nghost{{q}_{9} :  } & \lstick{{q}_{9} :  } & \qw & \qw & \qw & \qw & \qw & \qw & \qw & \qw & \qw & \qw & \targ & \qw & \targ & \qw & \qw & \qw & \qw\\
		\nghost{{q}_{10} :  } & \lstick{{q}_{10} :  } & \gate{\mathrm{H}} & \qw & \qw & \ctrl{-3} & \qw & \qw & \qw & \ctrl{3} & \qw & \qw & \qw & \ctrl{-2} & \qw & \qw & \qw & \qw & \qw\\
		\nghost{{q}_{11} :  } & \lstick{{q}_{11} :  } & \gate{\mathrm{H}} & \ctrl{-11} & \qw & \qw & \qw & \qw & \qw & \qw & \qw & \qw & \qw & \qw & \targ & \qw & \qw & \qw & \qw\\
		\nghost{{q}_{12} :  } & \lstick{{q}_{12} :  } & \qw & \qw & \qw & \qw & \targ & \qw & \qw & \qw & \ctrl{-5} & \qw & \qw & \targ & \ctrl{-1} & \qw & \qw & \qw & \qw\\
		\nghost{{q}_{13} :  } & \lstick{{q}_{13} :  } & \qw & \qw & \targ & \qw & \qw & \ctrl{-10} & \qw & \targ & \qw & \qw & \ctrl{1} & \ctrl{-1} & \qw & \qw & \qw & \qw & \qw\\
		\nghost{{q}_{14} :  } & \lstick{{q}_{14} :  } & \qw & \qw & \qw & \qw & \qw & \qw & \targ & \qw & \qw & \ctrl{-13} & \targ & \qw & \qw & \qw & \qw & \qw & \qw\\
		\\ }} 

\noindent \scalebox{1.0}{
	\Qcircuit @C=0.1em @R=0.1em @!R { \\
		\nghost{{q}_{0} :  } & \lstick{{q}_{0} :  } & \qw & \targ & \qw & \ctrl{14} & \ctrl{1} & \qw & \qw & \qw & \qw & \ctrl{11} & \qw & \ctrl{4} & \ctrl{8} & \qw & \qw & \qw\\
		\nghost{{q}_{1} :  } & \lstick{{q}_{1} :  } & \qw & \qw & \targ & \qw & \targ & \qw & \qw & \qw & \qw & \qw & \ctrl{4} & \qw & \qw & \ctrl{8} & \qw & \qw\\
		\nghost{{q}_{2} :  } & \lstick{{q}_{2} :  } & \gate{\mathrm{H}} & \qw & \ctrl{-1} & \qw & \ctrl{11} & \qw & \ctrl{4} & \qw & \qw & \qw & \qw & \qw & \qw & \qw & \qw & \qw\\
		\nghost{{q}_{3} :  } & \lstick{{q}_{3} :  } & \qw & \qw & \targ & \qw & \qw & \qw & \qw & \qw & \ctrl{3} & \qw & \qw & \qw & \qw & \qw & \qw & \qw\\
		\nghost{{q}_{4} :  } & \lstick{{q}_{4} :  } & \gate{\mathrm{H}} & \qw & \ctrl{-1} & \qw & \qw & \ctrl{6} & \qw & \ctrl{1} & \qw & \qw & \qw & \targ & \qw & \qw & \qw & \qw\\
		\nghost{{q}_{5} :  } & \lstick{{q}_{5} :  } & \qw & \qw & \qw & \qw & \qw & \qw & \qw & \targ & \qw & \qw & \targ & \qw & \qw & \qw & \qw & \qw\\
		\nghost{{q}_{6} :  } & \lstick{{q}_{6} :  } & \qw & \qw & \qw & \qw & \qw & \qw & \targ & \qw & \targ & \qw & \qw & \qw & \qw & \qw & \qw & \qw\\
		\nghost{{q}_{7} :  } & \lstick{{q}_{7} :  } & \qw & \qw & \targ & \qw & \qw & \qw & \targ & \ctrl{1} & \ctrl{2} & \qw & \qw & \qw & \qw & \qw & \qw & \qw\\
		\nghost{{q}_{8} :  } & \lstick{{q}_{8} :  } & \qw & \qw & \qw & \qw & \qw & \qw & \qw & \targ & \qw & \qw & \qw & \qw & \targ & \qw & \qw & \qw\\
		\nghost{{q}_{9} :  } & \lstick{{q}_{9} :  } & \qw & \qw & \qw & \qw & \qw & \qw & \qw & \qw & \targ & \qw & \qw & \qw & \qw & \targ & \qw & \qw\\
		\nghost{{q}_{10} :  } & \lstick{{q}_{10} :  } & \gate{\mathrm{H}} & \ctrl{-10} & \qw & \qw & \qw & \targ & \ctrl{-3} & \qw & \targ & \qw & \qw & \qw & \qw & \qw & \qw & \qw\\
		\nghost{{q}_{11} :  } & \lstick{{q}_{11} :  } & \gate{\mathrm{H}} & \qw & \ctrl{-4} & \qw & \qw & \ctrl{1} & \ctrl{2} & \qw & \qw & \targ & \qw & \qw & \qw & \qw & \qw & \qw\\
		\nghost{{q}_{12} :  } & \lstick{{q}_{12} :  } & \qw & \qw & \qw & \qw & \qw & \targ & \qw & \qw & \qw & \qw & \qw & \qw & \qw & \qw & \qw & \qw\\
		\nghost{{q}_{13} :  } & \lstick{{q}_{13} :  } & \qw & \qw & \qw & \qw & \targ & \qw & \targ & \ctrl{1} & \ctrl{-3} & \qw & \qw & \qw & \qw & \qw & \qw & \qw\\
		\nghost{{q}_{14} :  } & \lstick{{q}_{14} :  } & \qw & \qw & \qw & \targ & \qw & \qw & \qw & \targ & \qw & \qw & \qw & \qw & \qw & \qw & \qw & \qw\\
		\\ }}

\textbf{The $\mathbf{|0\rangle_L}$ state of the $\mathbf{[[17,1,5]]}$ 2D color code}

\noindent  \scalebox{1.0}{
	\Qcircuit @C=0.1em @R=0.1em @!R { \\
		\nghost{{q}_{0} :  } & \lstick{{q}_{0} :  } & \qw & \qw & \targ & \ctrl{3} & \qw & \qw & \qw & \qw & \ctrl{15} & \qw & \ctrl{5} & \qw & \qw & \targ & \qw & \qw & \qw\\
		\nghost{{q}_{1} :  } & \lstick{{q}_{1} :  } & \gate{\mathrm{H}} & \ctrl{12} & \qw & \qw & \targ & \qw & \qw & \qw & \qw & \qw & \qw & \targ & \ctrl{2} & \qw & \qw & \qw & \qw\\
		\nghost{{q}_{2} :  } & \lstick{{q}_{2} :  } & \gate{\mathrm{H}} & \qw & \qw & \qw & \qw & \qw & \qw & \ctrl{11} & \qw & \qw & \qw & \qw & \qw & \ctrl{-2} & \qw & \qw & \qw\\
		\nghost{{q}_{3} :  } & \lstick{{q}_{3} :  } & \qw & \qw & \qw & \targ & \qw & \qw & \qw & \qw & \qw & \ctrl{5} & \qw & \qw & \targ & \qw & \qw & \qw & \qw\\
		\nghost{{q}_{4} :  } & \lstick{{q}_{4} :  } & \gate{\mathrm{H}} & \qw & \qw & \ctrl{1} & \ctrl{-3} & \qw & \qw & \qw & \qw & \qw & \qw & \qw & \qw & \qw & \qw & \qw & \qw\\
		\nghost{{q}_{5} :  } & \lstick{{q}_{5} :  } & \qw & \qw & \qw & \targ & \qw & \qw & \qw & \qw & \qw & \qw & \targ & \qw & \qw & \qw & \qw & \qw & \qw\\
		\nghost{{q}_{6} :  } & \lstick{{q}_{6} :  } & \gate{\mathrm{H}} & \qw & \qw & \ctrl{3} & \qw & \targ & \qw & \qw & \qw & \qw & \qw & \qw & \targ & \targ & \qw & \qw & \qw\\
		\nghost{{q}_{7} :  } & \lstick{{q}_{7} :  } & \qw & \qw & \qw & \qw & \targ & \qw & \qw & \qw & \qw & \qw & \qw & \qw & \qw & \qw & \targ & \qw & \qw\\
		\nghost{{q}_{8} :  } & \lstick{{q}_{8} :  } & \qw & \qw & \qw & \qw & \qw & \qw & \targ & \qw & \qw & \targ & \qw & \ctrl{-7} & \ctrl{-2} & \qw & \qw & \qw & \qw\\
		\nghost{{q}_{9} :  } & \lstick{{q}_{9} :  } & \qw & \qw & \qw & \targ & \qw & \qw & \ctrl{-1} & \qw & \qw & \qw & \qw & \qw & \qw & \qw & \qw & \qw & \qw\\
		\nghost{{q}_{10} :  } & \lstick{{q}_{10} :  } & \qw & \qw & \qw & \targ & \qw & \qw & \qw & \qw & \qw & \qw & \qw & \qw & \qw & \ctrl{-4} & \ctrl{-3} & \qw & \qw\\
		\nghost{{q}_{11} :  } & \lstick{{q}_{11} :  } & \gate{\mathrm{H}} & \qw & \qw & \ctrl{-1} & \qw & \qw & \qw & \qw & \qw & \qw & \qw & \targ & \targ & \qw & \qw & \qw & \qw\\
		\nghost{{q}_{12} :  } & \lstick{{q}_{12} :  } & \gate{\mathrm{H}} & \qw & \ctrl{-12} & \qw & \qw & \qw & \qw & \qw & \qw & \targ & \qw & \qw & \qw & \qw & \qw & \qw & \qw\\
		\nghost{{q}_{13} :  } & \lstick{{q}_{13} :  } & \qw & \targ & \qw & \qw & \qw & \qw & \qw & \targ & \qw & \ctrl{-1} & \qw & \qw & \qw & \qw & \qw & \qw & \qw\\
		\nghost{{q}_{14} :  } & \lstick{{q}_{14} :  } & \gate{\mathrm{H}} & \qw & \qw & \qw & \qw & \ctrl{-8} & \qw & \qw & \qw & \qw & \ctrl{2} & \qw & \ctrl{-3} & \qw & \qw & \qw & \qw\\
		\nghost{{q}_{15} :  } & \lstick{{q}_{15} :  } & \qw & \qw & \qw & \qw & \qw & \qw & \qw & \qw & \targ & \targ & \qw & \ctrl{-4} & \qw & \qw & \qw & \qw & \qw\\
		\nghost{{q}_{16} :  } & \lstick{{q}_{16} :  } & \gate{\mathrm{H}} & \qw & \qw & \qw & \ctrl{-9} & \qw & \qw & \qw & \qw & \ctrl{-1} & \targ & \qw & \qw & \qw & \qw & \qw & \qw\\
		\\ }}

\section{Transfer learning for different logical states}
\label{app:logical-state-transfer}

\begin{figure}[htb]
	\includegraphics[width=.5\textwidth]{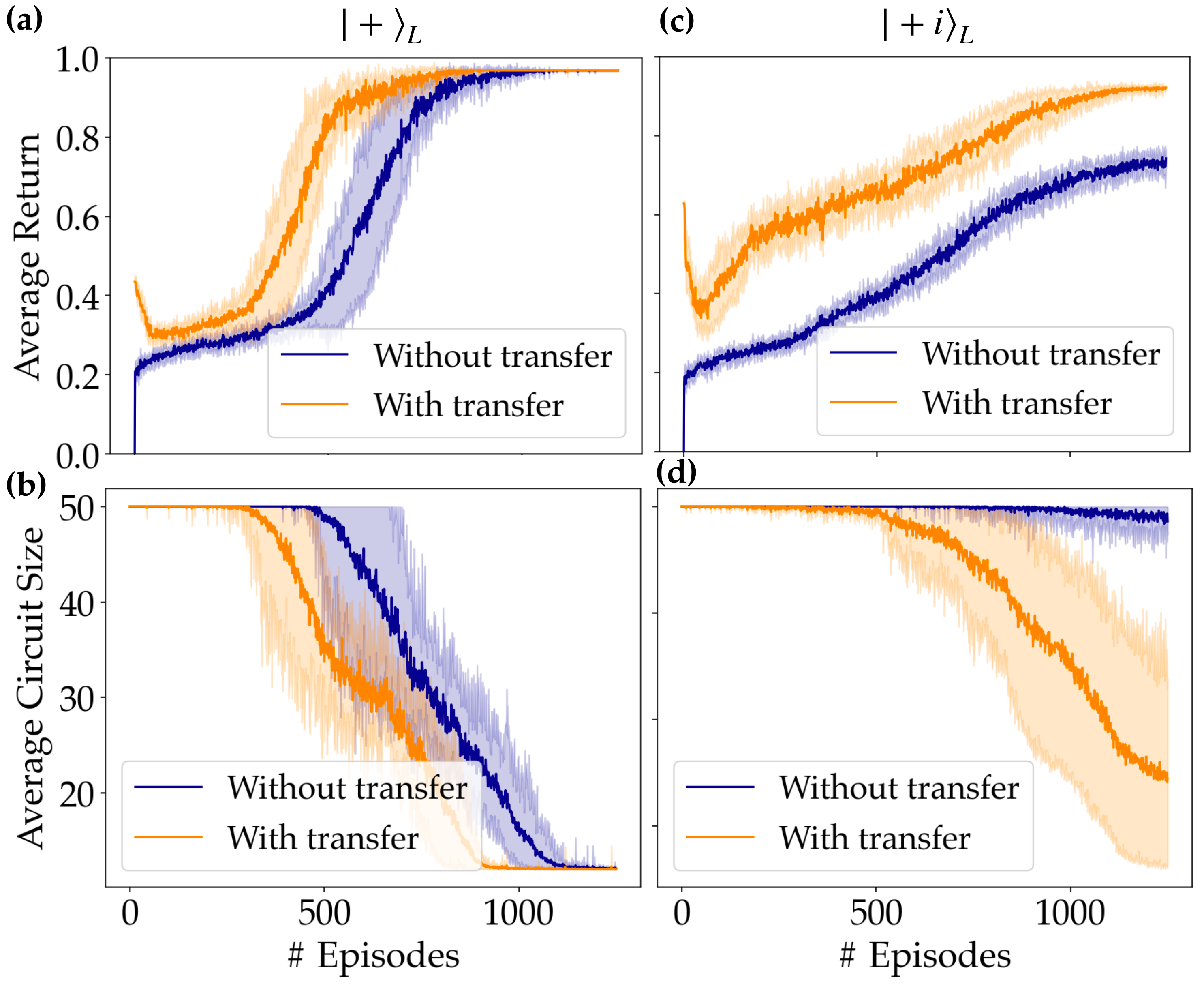}
	\caption{\label{fig:transfer-logical} Transfer learning results for different logical state preparation tasks. We first train the agent to prepare the $|0\rangle_L$ for the $[[7,1,3]]$ Steane code with $Z_L = Z^{\bigotimes 7}$ and then reuse and retrain the agent to prepare $|+\rangle_L$  with $X_L = X^{\bigotimes 7}$  and then successively $|+i\rangle_L$  with $Y_L = -Y^{\bigotimes 7}$. (a) and (b) show the average return and circuit size of training without and with transfer learning to prepare $|+\rangle_L$.  (c) and (d) show the same for the preparation of $|+i\rangle_L$. The shaded area shows the standard deviation of training $10$ different agents.  }
\end{figure}

Here, we show an application of transfer learning where we can reuse an RL agent trained on one logical state to prepare another logical state with the same code. One might argue that we can apply the logical $H$ gate to the $|0\rangle_L$ state to prepare the $|+\rangle_L$ state, and then apply the logical $S$ gate to prepare  $|+i\rangle_L$ state. However, in some codes, the logical $H$ and $S$ gates themselves are not transversal.

We illustrate this by reusing the agent trained to prepare the $|0\rangle_L$ state of the $[[7,1,3]]$ Steane code to prepare the states $|+\rangle_L$  and $|+i\rangle_L$ of the same code. In particular, this transfer learning is called a one-to-one policy transfer via policy reuse~\cite{zhu2023transfer}. In this case, we can directly reuse the networks of the RL agent since the number of input and output nodes does not change.

We compare the preparation of the $|+\rangle_L$ and the $|+i\rangle_L$ state of the $[[7,1,3]]$ Steane code without and with transfer learning in Fig.~\ref{fig:transfer-logical}. Without transfer learning means that the RL agent is trained from scratch to prepare the states. On the other hand, with transfer learning means that we first train the RL agent to prepare $|0\rangle_L$, and then retrain it to prepare $|+\rangle_L$, and consecutively retrain it to prepare $|+i\rangle_L$. 

Fig.~\ref{fig:transfer-logical}(a) and \ref{fig:transfer-logical}(b) show the evolution of the average return and the average circuit size evolution during training for the preparation of $|+\rangle_L$ with and without transfer learning. Fig.~\ref{fig:transfer-logical}(c) and (d) show the same but for the preparation of $|+i\rangle_L$. Here, we qualitatively compare the performance based on the metrics used in~\cite{zhu2023transfer} to evaluate transfer learning for deep reinforcement learning. The first metric is the jumpstart performance, which is the initial return value of the agent. We can see that the average return of the RL agent with transfer learning is higher than without.  The second metric is the time to threshold, which is the learning time required for the agent to reach a certain performance. We also see qualitatively that the value of the average return with transfer learning is almost always above without transfer learning. Therefore, we have shown that the proposed transfer learning technique is useful to efficiently prepare different logical states.

\section{Logical state preparation circuits for IBM Quantum devices}
\label{app:lsp-circuits-ibm}

We show here more circuit examples for logical state preparation based on the IBM Quantum device connectivity and gate set. The qubit placement is also given according to the notation in the \texttt{Qiskit} library~\cite{qiskit}. If the qubit placement is given as $a,b,c,\dots$, then it means that qubit $0$ ($q_0$) in the circuit is placed in qubit $a$ on the device, qubit $1$ ($q_1$)  is placed in qubit $b$ on the device, and so on. These circuits are also available online~\footnotemark[1]. 

\textbf{The $\mathbf{|0\rangle_L}$ state of the $\mathbf{[[5,1,3]]}$ perfect code} on the IBMQ Manila~\cite{manila} connectivity. The qubit placement is 4,3,0,2,1.

\noindent \scalebox{0.9}{
\Qcircuit @C=0.1em @R=0.1em @!R { \\
	 	\nghost{{q}_{0} :  } & \lstick{{q}_{0} :  } & \qw & \qw & \qw & \targ & \qw & \qw & \ctrl{1} & \qw & \qw & \qw & \qw\\
	 	\nghost{{q}_{1} :  } & \lstick{{q}_{1} :  } & \gate{\mathrm{\sqrt{X}}} & \qw & \targ & \ctrl{-1} & \gate{\mathrm{\sqrt{X}}} & \ctrl{2} & \targ & \gate{\mathrm{\sqrt{X}}} & \qw & \qw & \qw\\
	 	\nghost{{q}_{2} :  } & \lstick{{q}_{2} :  } & \gate{\mathrm{\sqrt{X}}} & \qw & \qw & \targ & \qw & \qw & \qw & \qw & \qw & \qw & \qw\\
	 	\nghost{{q}_{3} :  } & \lstick{{q}_{3} :  } & \gate{\mathrm{\sqrt{X}}} & \ctrl{1} & \ctrl{-2} & \qw & \qw & \targ & \ctrl{1} & \qw & \qw & \qw & \qw\\
	 	\nghost{{q}_{4} :  } & \lstick{{q}_{4} :  } & \qw & \targ & \qw & \ctrl{-2} & \qw & \qw & \targ & \gate{\mathrm{\sqrt{X}}} & \gate{\mathrm{X}} & \qw & \qw\\
\\ }}

\noindent \scalebox{0.9}{
\Qcircuit @C=0.1em @R=0.1em @!R { \\
	 	\nghost{{q}_{0} :  } & \lstick{{q}_{0} :  } & \gate{\mathrm{\sqrt{X}}} & \ctrl{1} & \gate{\mathrm{\sqrt{X}}} & \qw & \qw & \qw & \qw & \qw & \qw & \qw & \qw\\
	 	\nghost{{q}_{1} :  } & \lstick{{q}_{1} :  } & \qw & \targ & \targ & \qw & \ctrl{2} & \qw & \qw & \targ & \qw & \qw & \qw\\
	 	\nghost{{q}_{2} :  } & \lstick{{q}_{2} :  } & \gate{\mathrm{\sqrt{X}}} & \qw & \qw & \qw & \qw & \qw & \targ & \qw & \gate{\mathrm{X}} & \qw & \qw\\
	 	\nghost{{q}_{3} :  } & \lstick{{q}_{3} :  } & \gate{\mathrm{\sqrt{X}}} & \ctrl{1} & \ctrl{-2} & \gate{\mathrm{\sqrt{X}}} & \targ & \ctrl{1} & \qw & \ctrl{-2} & \ctrl{1} & \qw & \qw\\
	 	\nghost{{q}_{4} :  } & \lstick{{q}_{4} :  } & \qw & \targ & \qw & \qw & \qw & \targ & \ctrl{-2} & \gate{\mathrm{\sqrt{X}}} & \targ & \qw & \qw\\
\\ }}

\textbf{The $\mathbf{|0\rangle_L}$ state of the $\mathbf{[[7,1,3]]}$ Steane code} on the IBMQ Jakarta~\cite{jakarta} connectivity. The qubit placement is 5,6,4,3,0,1,2.

\noindent \scalebox{0.9}{
\Qcircuit @C=0.1em @R=0.1em @!R { \\
	 	\nghost{{q}_{0} :  } & \lstick{{q}_{0} :  } & \gate{\mathrm{\sqrt{X}}} & \ctrl{1} & \qw & \targ & \ctrl{3} & \ctrl{2} & \targ & \gate{\mathrm{S}} & \qw & \qw\\
	 	\nghost{{q}_{1} :  } & \lstick{{q}_{1} :  } & \qw & \targ & \qw & \qw & \qw & \qw & \ctrl{-1} & \gate{\mathrm{S}} & \qw & \qw\\
	 	\nghost{{q}_{2} :  } & \lstick{{q}_{2} :  } & \qw & \qw & \qw & \qw & \qw & \targ & \qw & \qw & \qw & \qw\\
	 	\nghost{{q}_{3} :  } & \lstick{{q}_{3} :  } & \qw & \qw & \targ & \ctrl{-3} & \targ & \ctrl{2} & \targ & \qw & \qw & \qw\\
	 	\nghost{{q}_{4} :  } & \lstick{{q}_{4} :  } & \gate{\mathrm{\sqrt{X}}} & \qw & \qw & \ctrl{1} & \gate{\mathrm{S}} & \qw & \qw & \qw & \qw & \qw\\
	 	\nghost{{q}_{5} :  } & \lstick{{q}_{5} :  } & \gate{\mathrm{\sqrt{X}}} & \ctrl{1} & \ctrl{-2} & \targ & \qw & \targ & \ctrl{-2} & \ctrl{1} & \qw & \qw\\
	 	\nghost{{q}_{6} :  } & \lstick{{q}_{6} :  } & \qw & \targ & \qw & \qw & \qw & \qw & \qw & \targ & \qw & \qw\\
\\ }}\scalebox{0.9}{
\Qcircuit @C=0.1em @R=0.1em @!R { \\
	 	\nghost{{q}_{0} :  } & \lstick{{q}_{0} :  } & \qw & \targ & \ctrl{1} & \qw & \targ & \ctrl{2} & \targ & \qw & \qw & \qw & \qw\\
	 	\nghost{{q}_{1} :  } & \lstick{{q}_{1} :  } & \qw & \qw & \targ & \qw & \qw & \qw & \ctrl{-1} & \gate{\mathrm{S}} & \qw & \qw & \qw\\
	 	\nghost{{q}_{2} :  } & \lstick{{q}_{2} :  } & \qw & \qw & \qw & \qw & \qw & \targ & \gate{\mathrm{S}} & \qw & \qw & \qw & \qw\\
	 	\nghost{{q}_{3} :  } & \lstick{{q}_{3} :  } & \gate{\mathrm{\sqrt{X}}} & \ctrl{-3} & \ctrl{2} & \gate{\mathrm{\sqrt{X}}} & \ctrl{-3} & \targ & \ctrl{2} & \qw & \qw & \qw & \qw\\
	 	\nghost{{q}_{4} :  } & \lstick{{q}_{4} :  } & \qw & \qw & \qw & \targ & \qw & \qw & \qw & \qw & \qw & \qw & \qw\\
	 	\nghost{{q}_{5} :  } & \lstick{{q}_{5} :  } & \qw & \targ & \targ & \ctrl{-1} & \qw & \ctrl{-2} & \targ & \targ & \qw & \qw & \qw\\
	 	\nghost{{q}_{6} :  } & \lstick{{q}_{6} :  } & \gate{\mathrm{\sqrt{X}}} & \ctrl{-1} & \qw & \qw & \qw & \qw & \qw & \ctrl{-1} & \gate{\mathrm{S}} & \qw & \qw\\
\\ }}

\textbf{The $\mathbf{|+\rangle_L}$ state of the $\mathbf{[[9,1,3]]}$ Shor code} on the  IBMQ Guadalupe~\cite{guadalupe} connectivity. The qubit placement is 5,3,8,7,6,4,0,1,2. 

\noindent \scalebox{0.9}{
\Qcircuit @C=0.1em @R=0.1em @!R { \\
	 	\nghost{{q}_{0} :  } & \lstick{{q}_{0} :  } & \qw & \qw & \targ & \ctrl{2} & \qw & \qw & \qw & \qw & \qw\\
	 	\nghost{{q}_{1} :  } & \lstick{{q}_{1} :  } & \gate{\mathrm{\sqrt{X}}} & \ctrl{7} & \ctrl{-1} & \qw & \ctrl{7} & \gate{\mathrm{S}} & \qw & \qw & \qw\\
	 	\nghost{{q}_{2} :  } & \lstick{{q}_{2} :  } & \qw & \qw & \qw & \targ & \qw & \qw & \qw & \qw & \qw\\
	 	\nghost{{q}_{3} :  } & \lstick{{q}_{3} :  } & \gate{\mathrm{\sqrt{X}}} & \qw & \ctrl{2} & \ctrl{1} & \qw & \qw & \qw & \qw & \qw\\
	 	\nghost{{q}_{4} :  } & \lstick{{q}_{4} :  } & \qw & \qw & \qw & \targ & \qw & \qw & \qw & \qw & \qw\\
	 	\nghost{{q}_{5} :  } & \lstick{{q}_{5} :  } & \qw & \qw & \targ & \ctrl{2} & \qw & \gate{\mathrm{S}} & \qw & \qw & \qw\\
	 	\nghost{{q}_{6} :  } & \lstick{{q}_{6} :  } & \qw & \qw & \qw & \qw & \qw & \qw & \targ & \qw & \qw\\
	 	\nghost{{q}_{7} :  } & \lstick{{q}_{7} :  } & \qw & \qw & \targ & \targ & \qw & \ctrl{1} & \ctrl{-1} & \qw & \qw\\
	 	\nghost{{q}_{8} :  } & \lstick{{q}_{8} :  } & \qw & \targ & \ctrl{-1} & \qw & \targ & \targ & \qw & \qw & \qw\\
\\ }} \scalebox{0.9}{
\Qcircuit @C=0.1em @R=0.1em @!R { \\
	 	\nghost{{q}_{0} :  } & \lstick{{q}_{0} :  } & \qw & \qw & \targ & \ctrl{2} & \qw & \qw & \qw & \qw & \qw\\
	 	\nghost{{q}_{1} :  } & \lstick{{q}_{1} :  } & \gate{\mathrm{\sqrt{X}}} & \ctrl{7} & \ctrl{-1} & \qw & \qw & \qw & \qw & \qw & \qw\\
	 	\nghost{{q}_{2} :  } & \lstick{{q}_{2} :  } & \qw & \qw & \qw & \targ & \gate{\mathrm{S}} & \qw & \qw & \qw & \qw\\
	 	\nghost{{q}_{3} :  } & \lstick{{q}_{3} :  } & \gate{\mathrm{\sqrt{X}}} & \qw & \ctrl{2} & \ctrl{1} & \gate{\mathrm{S}} & \qw & \qw & \qw & \qw\\
	 	\nghost{{q}_{4} :  } & \lstick{{q}_{4} :  } & \qw & \qw & \qw & \targ & \qw & \qw & \qw & \qw & \qw\\
	 	\nghost{{q}_{5} :  } & \lstick{{q}_{5} :  } & \qw & \qw & \targ & \qw & \ctrl{2} & \qw & \qw & \qw & \qw\\
	 	\nghost{{q}_{6} :  } & \lstick{{q}_{6} :  } & \qw & \qw & \qw & \qw & \qw & \qw & \targ & \qw & \qw\\
	 	\nghost{{q}_{7} :  } & \lstick{{q}_{7} :  } & \qw & \qw & \targ & \ctrl{1} & \targ & \ctrl{1} & \ctrl{-1} & \qw & \qw\\
	 	\nghost{{q}_{8} :  } & \lstick{{q}_{8} :  } & \qw & \targ & \ctrl{-1} & \targ & \qw & \targ & \qw & \qw & \qw\\
\\ }}

\textbf{The $\mathbf{|0\rangle_L}$ state of the $\mathbf{[[15,1,3]]}$ Reed-Muller code} on the IBMQ Tokyo~\cite{tokyo} connectivity. The qubit placement is 5,17,2,9,10,13,1,11,7,16,4,3,8,6,12.

\noindent \scalebox{0.85}{
\Qcircuit @C=0.1em @R=0.1em @!R { \\
	 	\nghost{{q}_{0} :  } & \lstick{{q}_{0} :  } & \gate{\mathrm{S}} & \gate{\mathrm{\sqrt{X}}} & \gate{\mathrm{S}} & \qw & \ctrl{13} & \qw & \ctrl{4} & \qw & \qw & \qw & \qw & \qw & \qw & \qw & \qw & \qw & \qw & \qw & \qw & \qw & \qw & \qw & \qw & \qw & \qw & \qw\\
	 	\nghost{{q}_{1} :  } & \lstick{{q}_{1} :  } & \qw & \qw & \qw & \qw & \qw & \qw & \qw & \qw & \targ & \qw & \qw & \qw & \qw & \qw & \qw & \qw & \qw & \qw & \qw & \qw & \qw & \qw & \qw & \qw & \qw & \qw\\
	 	\nghost{{q}_{2} :  } & \lstick{{q}_{2} :  } & \gate{\mathrm{S}} & \gate{\mathrm{\sqrt{X}}} & \gate{\mathrm{S}} & \ctrl{6} & \qw & \qw & \qw & \qw & \qw & \qw & \qw & \qw & \targ & \ctrl{4} & \qw & \qw & \qw & \qw & \ctrl{9} & \qw & \qw & \qw & \qw & \qw & \qw & \qw\\
	 	\nghost{{q}_{3} :  } & \lstick{{q}_{3} :  } & \qw & \qw & \qw & \qw & \qw & \targ & \qw & \qw & \qw & \qw & \qw & \qw & \qw & \qw & \qw & \qw & \qw & \qw & \qw & \qw & \qw & \qw & \qw & \qw & \qw & \qw\\
	 	\nghost{{q}_{4} :  } & \lstick{{q}_{4} :  } & \qw & \qw & \qw & \qw & \qw & \qw & \targ & \qw & \qw & \qw & \qw & \qw & \qw & \qw & \qw & \qw & \targ & \qw & \qw & \qw & \targ & \qw & \qw & \qw & \qw & \qw\\
	 	\nghost{{q}_{5} :  } & \lstick{{q}_{5} :  } & \gate{\mathrm{S}} & \gate{\mathrm{\sqrt{X}}} & \gate{\mathrm{S}} & \qw & \qw & \qw & \ctrl{9} & \ctrl{3} & \qw & \qw & \qw & \qw & \qw & \qw & \qw & \qw & \qw & \qw & \qw & \targ & \qw & \qw & \qw & \qw & \qw & \qw\\
	 	\nghost{{q}_{6} :  } & \lstick{{q}_{6} :  } & \qw & \qw & \qw & \qw & \qw & \qw & \qw & \qw & \qw & \qw & \qw & \qw & \qw & \targ & \qw & \qw & \qw & \qw & \qw & \qw & \qw & \qw & \targ & \qw & \qw & \qw\\
	 	\nghost{{q}_{7} :  } & \lstick{{q}_{7} :  } & \qw & \qw & \qw & \qw & \qw & \qw & \qw & \qw & \qw & \qw & \qw & \qw & \qw & \qw & \targ & \ctrl{2} & \ctrl{-3} & \qw & \qw & \qw & \qw & \targ & \qw & \qw & \qw & \qw\\
	 	\nghost{{q}_{8} :  } & \lstick{{q}_{8} :  } & \qw & \qw & \qw & \targ & \qw & \qw & \qw & \targ & \qw & \ctrl{4} & \ctrl{5} & \qw & \qw & \qw & \qw & \qw & \targ & \ctrl{5} & \qw & \ctrl{-3} & \qw & \qw & \ctrl{-2} & \targ & \qw & \qw\\
	 	\nghost{{q}_{9} :  } & \lstick{{q}_{9} :  } & \qw & \qw & \qw & \qw & \qw & \qw & \qw & \qw & \qw & \qw & \qw & \qw & \qw & \qw & \qw & \targ & \qw & \qw & \qw & \qw & \qw & \qw & \qw & \qw & \qw & \qw\\
	 	\nghost{{q}_{10} :  } & \lstick{{q}_{10} :  } & \gate{\mathrm{S}} & \gate{\mathrm{\sqrt{X}}} & \gate{\mathrm{S}} & \qw & \qw & \ctrl{-7} & \qw & \qw & \qw & \qw & \qw & \ctrl{2} & \qw & \qw & \qw & \targ & \qw & \qw & \qw & \qw & \qw & \qw & \qw & \qw & \qw & \qw\\
	 	\nghost{{q}_{11} :  } & \lstick{{q}_{11} :  } & \qw & \qw & \qw & \qw & \qw & \qw & \qw & \qw & \qw & \qw & \qw & \qw & \qw & \targ & \qw & \ctrl{-1} & \qw & \qw & \targ & \qw & \qw & \qw & \qw & \qw & \qw & \qw\\
	 	\nghost{{q}_{12} :  } & \lstick{{q}_{12} :  } & \qw & \qw & \qw & \qw & \qw & \qw & \qw & \qw & \qw & \targ & \qw & \targ & \qw & \ctrl{-1} & \qw & \qw & \ctrl{-4} & \qw & \targ & \qw & \qw & \qw & \ctrl{2} & \ctrl{-4} & \qw & \qw\\
	 	\nghost{{q}_{13} :  } & \lstick{{q}_{13} :  } & \qw & \qw & \qw & \qw & \targ & \qw & \qw & \qw & \qw & \qw & \targ & \qw & \ctrl{-11} & \qw & \ctrl{-6} & \qw & \qw & \targ & \qw & \qw & \ctrl{-9} & \qw & \qw & \qw & \qw & \qw\\
	 	\nghost{{q}_{14} :  } & \lstick{{q}_{14} :  } & \qw & \qw & \qw & \qw & \qw & \qw & \targ & \qw & \ctrl{-13} & \qw & \qw & \qw & \qw & \qw & \qw & \qw & \qw & \qw & \ctrl{-2} & \qw & \qw & \ctrl{-7} & \targ & \qw & \qw & \qw\\
\\ }}

\section{Transfer learning to different connectivity}
\label{app:logical-state-transfer-connectivity}

\begin{figure}[t]
	\includegraphics[width=.5\textwidth]{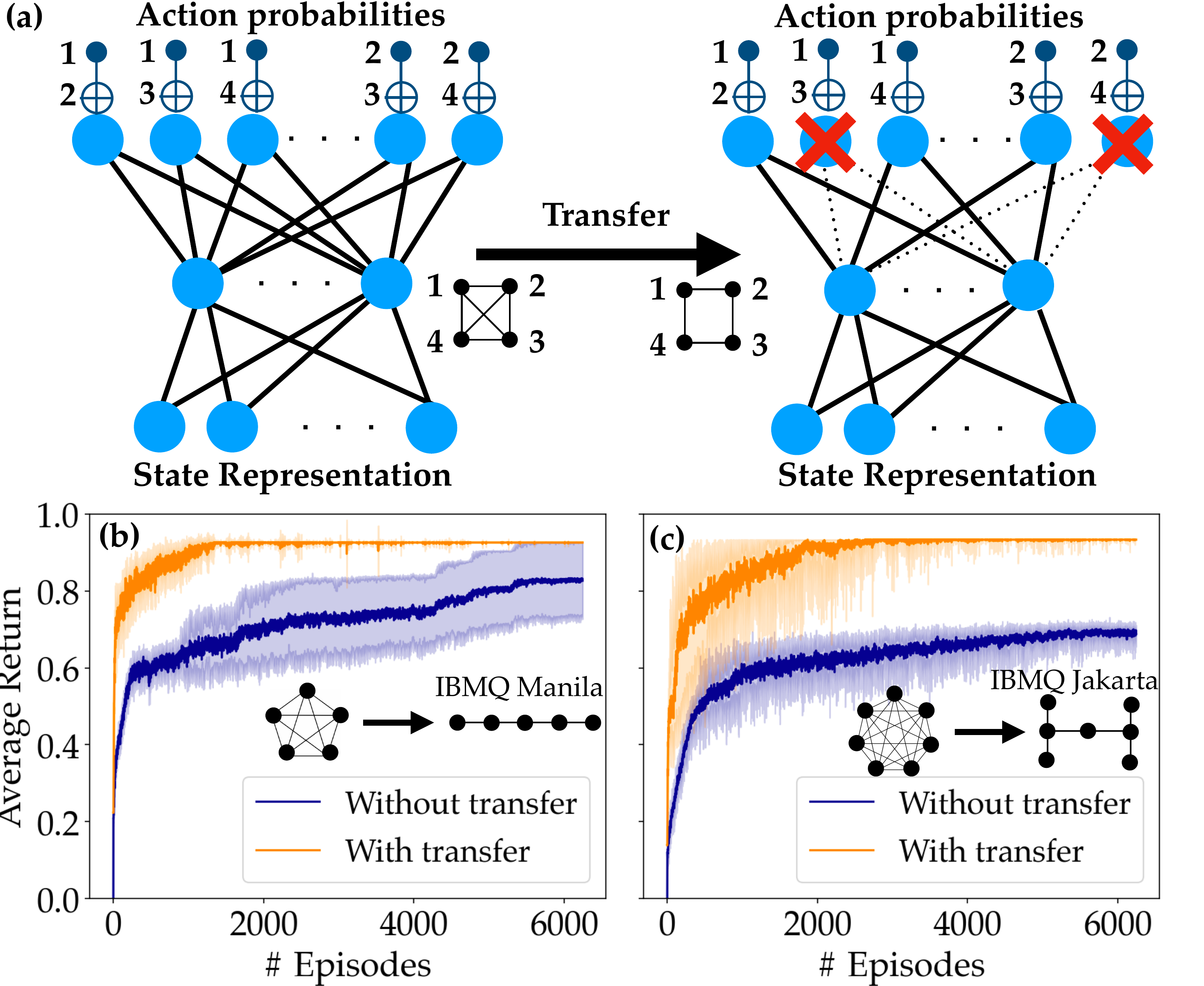}
	\caption{\label{fig:transfer-connectivity} Results of transfer learning for logical state preparation on IBM quantum devices. We first train the agent to prepare $|0\rangle_L$ in a setting of all-to-all qubit connectivity, and then reuse and retrain it to prepare the same $|0\rangle_L$ under a different, restricted qubit connectivity. (a) The sketch of the transfer learning on the network trained on four qubits with all-to-all qubit connectivity to square connectivity by removing the output nodes corresponding to invalid actions. (b) The training evolution of preparing $|0\rangle_L$ of the $[[5,1,3]]$ perfect code with $Z_L = Z^{\bigotimes 5}$ on all-to-all qubit connectivity and transferring it to IBMQ Lima~\cite{lima} connectivity and without transfer. (c) The training evolution of preparing $|0\rangle_L$ of $[[7,1,3]]$ Steane code with $Z_L = Z^{\bigotimes 7}$ on all-to-all qubit connectivity and transferring it to IBMQ Jakarta~\cite{jakarta} connectivity and without transfer.}
\end{figure}

We show a transfer learning technique that can reuse and retrain the agent trained to prepare a logical state for qubit connectivity $G$ to prepare the same state with different qubit connectivity $G'$. We assume that $G'$ is a spanning subgraph (having the same number of qubits but only some of the edges) of $G$. 

Since we transfer to a connectivity that is a spanning subgraph of the original connectivity, the possible action space in the new connectivity is a subset of the possible action space in the original connectivity. This is equivalent to removing some of the output nodes in the actor network that correspond to invalid actions in the new connectivity. The input nodes, hidden nodes, and the value network remain the same and can be transferred directly. After the transfer, we use this network as the initial network and fine tune it for the new connectivity.

We illustrate this by choosing a case where RL agents trained for all-to-all qubit connectivity are transferred to a more restricted connectivity. For example, the sketch of the transfer learning technique from four qubits with all-to-all qubit connectivity to a new connectivity where the connections between qubit $1$ and $3$ and the connections between qubit $2$ and $4$ are removed is shown in Fig.~\ref{fig:transfer-connectivity}(a). Assuming that we are using $\text{CNOT}$ gates, we need to remove the output nodes of the actor network corresponding to the following actions: $\text{CNOT}(1,3), \text{CNOT}(3,1), \text{CNOT}(2,4)$, and $CNOT(4,2)$, and keep the others. We then fine-tune this network to prepare the state for the new connectivity.

We compare the preparation of $|0\rangle_L$ of the $[[5,1,3]]$ code on IBMQ Lima~\cite{lima} connectivity and $[[7,1,3]]$ code on IBMQ Jakarta~\cite{jakarta} connectivity without and with transfer learning. Without transfer learning means that the networks are randomly initialized at the start, while with transfer learning means that we reuse networks that were trained for all-to-all qubit connectivity.
Fig.~\ref{fig:transfer-connectivity}(b) and (c) shows the average return during the training of the $[[5,1,3]]$ perfect code and the $[[7,1,3]]$ Steane code with and without transfer learning. We see that with transfer learning, the average return initially starts a little bit higher, and more importantly, converges faster than without transfer learning. The agents without transfer learning need much more training time for convergence.

\section{Varying the verification circuit synthesis task reward weights}
\label{app:vcs-varymiu}

Here, we vary the weights for the flag reward $\mu_f$, the complementary distance reward $\mu_d$, and the product state reward $\mu_p$ of the reward function that is defined in Eq.~(\ref{eq:vcs-reward}) for the verification circuit synthesis task. We then evaluate how this affects the acceptance and logical error rates. 

Effectively, only the weight ratios matter, since scaling the reward function generally does not affect the performance of the reinforcement learning training. We vary $\mu_f / \mu_d$ and $\mu_p / \mu_d$ and synthesize the verification circuit at each point. We then compute the acceptance and logical error rates and fit them with the exponential and quadratic functions, respectively. We then compare the average coefficients over $10$ different circuits.

\begin{figure}[htb]
	\includegraphics[width=.5\textwidth]{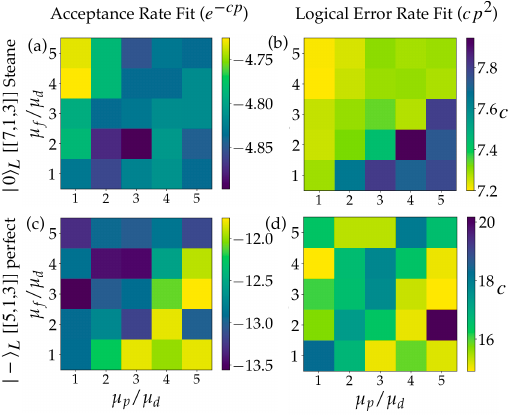}
	\caption{\label{fig:vary-miu-vcs} Varying the weights of the reward function for verification circuit synthesis. We vary the $\mu_f / \mu_d$ and $\mu_p / \mu_d$ ranging from $1$ to $5$ with an interval of $1$. The heatmap shows the average fitting coefficients for the acceptance ((a) and (c), the higher the better) and the logical error rate ((b) and (d), the lower the better). We evaluate for the verification circuit synthesis from the non-FT $|0\rangle_L$ of the $[[7,1,3]]$ Steane code taken from~\cite{goto_minimizing_2016}  ((a) and (b)) and $|-\rangle_L$ of the $[[5,1,3]]$ perfect code taken from~\cite{chao_quantum_2018}  ((c) and (d)).  }
\end{figure}

In Fig.~\ref{fig:vary-miu-vcs}(a) and (b), we see that the best strategy for the $|0\rangle_L$ state of the $[[7,1,3]]$ Steane code is to prioritize the weight of the flag reward $\mu_f$.  In contrast, one needs to prioritize the weight of the product state reward $\mu_p$ for $|-\rangle_L$ of the $[[5,1,3]]$ perfect code, as can be seen in Fig.~\ref{fig:vary-miu-vcs}(c) and (d).

\section{Examples of RL-discovered circuits for the verification circuit synthesis task}
\label{app:vcs-circuits}

Here we show more examples of RL-discovered circuits for the verification circuit synthesis task. We are particularly interested in showing the ability of RL to explore and present variants of circuits that minimize the fit coefficients of the acceptance or logical error rate. These circuits are also available online~\footnotemark[1]. 

We first show the synthesis of the verification circuit for the $|0\rangle_L$ state of the $[[7,1,3]]$ Steane code by taking the non-FT logical state preparation circuit from Ref.~\cite{goto_minimizing_2016}. First, we show that the RL method also rediscovers the same circuit used in Ref.~\cite{goto_minimizing_2016} shown below, which measures the stabilizer $Z$ logical operator $ZIIIIZZ$:

\noindent \scalebox{1.0}{
\Qcircuit @C=0.2em @R=0.1em @!R { \\
	 	\nghost{{q}_{0} :  } & \lstick{{q}_{0} :  } & \qw & \targ & \qw & \qw & \targ & \qw & \qw \barrier[0em]{7} & \qw & \ctrl{7} & \qw & \qw & \qw & \qw\\
	 	\nghost{{q}_{1} :  } & \lstick{{q}_{1} :  } & \gate{\mathrm{H}} & \ctrl{-1} & \qw & \ctrl{3} & \qw & \ctrl{4} & \qw & \qw & \qw & \qw & \qw & \qw & \qw\\
	 	\nghost{{q}_{2} :  } & \lstick{{q}_{2} :  } & \gate{\mathrm{H}} & \qw & \ctrl{4} & \qw & \ctrl{-2} & \qw & \qw & \qw & \qw & \qw & \qw & \qw & \qw\\
	 	\nghost{{q}_{3} :  } & \lstick{{q}_{3} :  } & \gate{\mathrm{H}} & \ctrl{2} & \qw & \qw & \ctrl{3} & \qw & \qw & \qw & \qw & \qw & \qw & \qw & \qw\\
	 	\nghost{{q}_{4} :  } & \lstick{{q}_{4} :  } & \qw & \qw & \qw & \targ & \qw & \qw & \targ & \qw & \qw & \qw & \qw & \qw & \qw\\
	 	\nghost{{q}_{5} :  } & \lstick{{q}_{5} :  } & \qw & \targ & \qw & \qw & \qw & \targ & \qw & \qw & \qw & \qw & \ctrl{2} & \qw & \qw\\
	 	\nghost{{q}_{6} :  } & \lstick{{q}_{6} :  } & \qw & \qw & \targ & \qw & \targ & \qw & \ctrl{-2} & \qw & \qw & \ctrl{1} & \qw & \qw & \qw\\
	 	\nghost{{q}_{7} :  } & \lstick{{q}_{7} :  } & \qw & \qw & \qw & \qw & \qw & \qw & \qw & \qw & \targ & \targ & \targ & \qw & \qw\\
\\ }}

The circuit with the lowest fitting coefficients of the acceptance and logical error rate is the circuit shown in Fig.~\ref{fig:result-flag-fully-connected}(a), which measures the stabilizer $Z$ logical operator $IIZIZZI$. The other circuits are just permutations of the CNOT gates in the verification circuit. 

We now show the synthesis of the verification circuit for the $|-\rangle_L$ state of the $[[5,1,3]]$ perfect code, using the non-FT logical state preparation circuit from Ref.~\cite{chao_quantum_2018}. 

The circuit with the lowest acceptance rate fitting coefficients is the circuit shown in Fig.~\ref{fig:result-flag-fully-connected}(b). Below we show another circuit with two flag qubits and similar fitting coefficients.

\noindent \scalebox{1.0}{
\Qcircuit @C=0.2em @R=0.1em @!R { \\
	 	\nghost{{q}_{0} :  } & \lstick{{q}_{0} :  } & \gate{\mathrm{H}} & \ctrl{1} & \qw & \ctrl{4} \barrier[0em]{6} & \qw & \qw & \qw & \qw & \qw & \qw & \control\qw & \qw & \control\qw & \qw & \qw & \qw\\
	 	\nghost{{q}_{1} :  } & \lstick{{q}_{1} :  } & \gate{\mathrm{H}} & \control\qw & \ctrl{1} & \qw & \qw & \qw & \qw & \qw & \qw & \control\qw & \qw & \targ & \qw & \qw & \qw & \qw\\
	 	\nghost{{q}_{2} :  } & \lstick{{q}_{2} :  } & \gate{\mathrm{H}} & \ctrl{1} & \control\qw & \qw & \qw & \qw & \targ & \qw & \control\qw & \qw & \qw & \qw & \qw & \qw & \qw & \qw\\
	 	\nghost{{q}_{3} :  } & \lstick{{q}_{3} :  } & \gate{\mathrm{H}} & \control\qw & \ctrl{1} & \qw & \qw & \qw & \qw & \qw & \qw & \qw & \qw & \qw & \qw & \qw & \qw & \qw\\
	 	\nghost{{q}_{4} :  } & \lstick{{q}_{4} :  } & \gate{\mathrm{H}} & \qw & \control\qw & \control\qw & \qw & \qw & \qw & \targ & \qw & \qw & \qw & \qw & \qw & \qw & \qw & \qw\\
	 	\nghost{{q}_{5} :  } & \lstick{{q}_{5} :  } & \qw & \qw & \qw & \qw & \qw & \gate{\mathrm{H}} & \qw & \qw & \ctrl{-3} & \qw & \ctrl{-5} & \ctrl{-4} & \qw & \gate{\mathrm{H}} & \qw & \qw\\
	 	\nghost{{q}_{6} :  } & \lstick{{q}_{6} :  } & \qw & \qw & \qw & \qw & \qw & \gate{\mathrm{H}} & \ctrl{-4} & \ctrl{-2} & \qw & \ctrl{-5} & \qw & \qw & \ctrl{-6} & \gate{\mathrm{H}} & \qw & \qw\\
\\ }}

The circuit with the lowest fitting coefficients of the logical error rate is the circuit shown below. The circuit needs $3$ flag qubits.  The acceptance and logical error rate fitting coefficients are $-14.1$ and $12.1$, respectively. One of the interesting properties that the RL agent learned is that it uses an extra flag qubit (second flag qubit) to fault-tolerantly measure stabilizer logical $X$ operators $IIZXZ$ following the protocol in Ref.~\cite{chao_quantum_2018} in the first flag qubit and $XIXZZ$ in the third flag qubit.

\noindent \scalebox{1.0}{
\Qcircuit @C=0.2em @R=0.1em @!R { \\
	 	\nghost{{q}_{0} :  } & \lstick{{q}_{0} :  } & \gate{\mathrm{H}} & \ctrl{1} & \qw & \ctrl{4} \barrier[0em]{7} & \qw & \qw & \qw & \qw & \targ & \qw & \qw & \qw & \qw & \qw & \qw & \qw & \qw & \qw\\
	 	\nghost{{q}_{1} :  } & \lstick{{q}_{1} :  } & \gate{\mathrm{H}} & \control\qw & \ctrl{1} & \qw & \qw & \qw & \qw & \qw & \qw & \qw & \qw & \qw & \qw & \qw & \qw & \qw & \qw & \qw\\
	 	\nghost{{q}_{2} :  } & \lstick{{q}_{2} :  } & \gate{\mathrm{H}} & \ctrl{1} & \control\qw & \qw & \qw & \qw & \targ & \qw & \qw & \qw & \qw & \qw & \control\qw & \qw & \qw & \qw & \qw & \qw\\
	 	\nghost{{q}_{3} :  } & \lstick{{q}_{3} :  } & \gate{\mathrm{H}} & \control\qw & \ctrl{1} & \qw & \qw & \qw & \qw & \qw & \qw & \control\qw & \qw & \targ & \qw & \qw & \qw & \qw & \qw & \qw\\
	 	\nghost{{q}_{4} :  } & \lstick{{q}_{4} :  } & \gate{\mathrm{H}} & \qw & \control\qw & \control\qw & \qw & \qw & \qw & \qw & \qw & \qw & \control\qw & \qw & \qw & \control\qw & \qw & \qw & \qw & \qw\\
	 	\nghost{{q}_{5} :  } & \lstick{{q}_{5} :  } & \qw & \qw & \qw & \qw & \qw & \gate{\mathrm{H}} & \qw & \ctrl{1} & \qw & \qw & \qw & \ctrl{-2} & \ctrl{-3} & \ctrl{-1} & \ctrl{1} & \gate{\mathrm{H}} & \qw & \qw\\
	 	\nghost{{q}_{6} :  } & \lstick{{q}_{6} :  } & \qw & \qw & \qw & \qw & \qw & \qw & \qw & \targ & \qw & \qw & \qw & \qw & \qw & \qw & \targ & \qw & \qw & \qw\\
	 	\nghost{{q}_{7} :  } & \lstick{{q}_{7} :  } & \qw & \qw & \qw & \qw & \qw & \gate{\mathrm{H}} & \ctrl{-5} & \qw & \ctrl{-7} & \ctrl{-4} & \ctrl{-3} & \gate{\mathrm{H}} & \qw & \qw & \qw & \qw & \qw & \qw\\
\\ }}

We show another circuit with three flag qubits and similar fitting coefficients. The RL agent learned to measure the stabilizer operators $IIZXZ$ in the first flag qubit and $IXZZX$ in the third flag qubit. 

\noindent \scalebox{1.0}{
\Qcircuit @C=0.2em @R=0.1em @!R { \\
	 	\nghost{{q}_{0} :  } & \lstick{{q}_{0} :  } & \gate{\mathrm{H}} & \ctrl{1} & \qw & \ctrl{4} \barrier[0em]{7} & \qw & \qw & \qw & \qw & \qw & \qw & \qw & \qw & \qw & \qw & \qw & \qw & \qw & \qw\\
	 	\nghost{{q}_{1} :  } & \lstick{{q}_{1} :  } & \gate{\mathrm{H}} & \control\qw & \ctrl{1} & \qw & \qw & \qw & \qw & \targ & \qw & \qw & \qw & \qw & \qw & \qw & \qw & \qw & \qw & \qw\\
	 	\nghost{{q}_{2} :  } & \lstick{{q}_{2} :  } & \gate{\mathrm{H}} & \ctrl{1} & \control\qw & \qw & \qw & \qw & \qw & \qw & \qw & \control\qw & \qw & \qw & \control\qw & \qw & \qw & \qw & \qw & \qw\\
	 	\nghost{{q}_{3} :  } & \lstick{{q}_{3} :  } & \gate{\mathrm{H}} & \control\qw & \ctrl{1} & \qw & \qw & \qw & \qw & \qw & \control\qw & \qw & \targ & \qw & \qw & \qw & \qw & \qw & \qw & \qw\\
	 	\nghost{{q}_{4} :  } & \lstick{{q}_{4} :  } & \gate{\mathrm{H}} & \qw & \control\qw & \control\qw & \qw & \qw & \targ & \qw & \qw & \qw & \qw & \qw & \qw & \control\qw & \qw & \qw & \qw & \qw\\
	 	\nghost{{q}_{5} :  } & \lstick{{q}_{5} :  } & \qw & \qw & \qw & \qw & \qw & \gate{\mathrm{H}} & \qw & \qw & \qw & \qw & \ctrl{-2} & \ctrl{1} & \ctrl{-3} & \ctrl{-1} & \ctrl{1} & \gate{\mathrm{H}} & \qw & \qw\\
	 	\nghost{{q}_{6} :  } & \lstick{{q}_{6} :  } & \qw & \qw & \qw & \qw & \qw & \qw & \qw & \qw & \qw & \qw & \qw & \targ & \qw & \qw & \targ & \qw & \qw & \qw\\
	 	\nghost{{q}_{7} :  } & \lstick{{q}_{7} :  } & \qw & \qw & \qw & \qw & \qw & \gate{\mathrm{H}} & \ctrl{-3} & \ctrl{-6} & \ctrl{-4} & \ctrl{-5} & \gate{\mathrm{H}} & \qw & \qw & \qw & \qw & \qw & \qw & \qw\\
\\ }}

\section{Examples of RL-discovered fault-tolerant logical state preparation circuits shown in Table~\ref{tab:comparison}}
\label{app:ftlsp-circuits}

Here, we show circuit examples from Table~\ref{tab:comparison} with two RL approaches: logical state preparation followed by verification circuit synthesis (LSP + VCS) and integrated fault-tolerant logical state preparation (IFT-LSP). These circuits are also available online~\footnotemark[1].  

\textbf{The $\mathbf{|1\rangle_L}$ state of the $\mathbf{[[5,1,3]]}$ perfect code}. The circuit obtained with the LSP + VCS approach has 14 two-qubit gates and 2 flag qubits:

\noindent \scalebox{1.0}{
\Qcircuit @C=0.2em @R=0.2em @!R { \\
\nghost{{q}_{0} : }	 & \lstick{{q}_{0} : }	 & \gate{\mathrm{H}}	 & \ctrl{2}	 & \qw	 & \qw	 & \qw	 & \targ	 & \targ \barrier[0em]{6} 	 & \qw 	 & \qw	 & \control\qw	 & \qw	 & \targ	 & \qw	 & \qw	 & \qw	 & \qw	 & \qw	 & \qw	 & \qw	 & \qw\\
\nghost{{q}_{1} : }	 & \lstick{{q}_{1} : }	 & \qw	 & \qw	 & \targ	 & \gate{\mathrm{H}}	 & \qw	 & \qw	 & \ctrl{-1}	 & \qw	 & \qw	 & \qw	 & \qw	 & \qw	 & \qw	 & \qw	 & \qw	 & \qw	 & \qw	 & \qw	 & \qw	 & \qw\\
\nghost{{q}_{2} : }	 &\lstick{{q}_{2} : }	 & \qw	 & \targ	 & \qw	 & \ctrl{2}	 & \gate{\mathrm{H}}	 & \ctrl{-2}	 & \gate{\mathrm{H}}	 & \qw	 & \qw	 & \qw	 & \qw	 & \qw	 & \control\qw	 & \targ	 & \control\qw	 & \qw	 & \control\qw	 & \qw	 & \qw	 & \qw\\
\nghost{{q}_{3} : }	 &\lstick{{q}_{3} : }	 & \gate{\mathrm{H}}	 & \qw	 & \ctrl{-2}	 & \qw	 & \qw	 & \ctrl{1}	 & \qw	 & \qw	 & \qw	 & \qw	 & \targ	 & \qw	 & \qw	 & \qw	 & \qw	 & \qw	 & \qw	 & \qw	 & \qw	 & \qw\\
\nghost{{q}_{4} : }	 & \lstick{{q}_{4} : }	 & \qw	 & \qw	 & \qw	 & \targ	 & \gate{\mathrm{H}}	 & \targ	 & \qw	 & \qw	 & \qw	 & \qw	 & \qw	 & \qw	 & \qw	 & \qw	 & \qw	 & \targ	 & \qw	 & \qw	 & \qw	 & \qw\\
\nghost{{q}_{5} : }	 &\lstick{{q}_{5} : }	 & \qw	 & \qw	 & \qw	 & \qw	 & \qw	 & \qw	 & \qw	 & \qw	 & \gate{\mathrm{H}}	 & \qw	 & \qw	 & \ctrl{-5}	 & \qw	 & \qw	 & \ctrl{-3}	 & \ctrl{-1}	 & \qw	 & \gate{\mathrm{H}}	 & \qw	 & \qw\\
\nghost{{q}_{6} : }	 & \lstick{{q}_{6} : }	 & \qw	 & \qw	 & \qw	 & \qw	 & \qw	 & \qw	 & \qw	 & \qw	 & \gate{\mathrm{H}}	 & \ctrl{-6}	 & \ctrl{-3}	 & \qw	 & \ctrl{-4}	 & \ctrl{-4}	 & \qw	 & \qw	 & \ctrl{-4}	 & \gate{\mathrm{H}}	 & \qw	 & \qw\\
\\ }}

The circuit obtained using the IFT-LSP approach has a smaller number of only 12 two-qubit gates and 2 flag qubits:

\noindent \scalebox{1.0}{
\Qcircuit @C=0.2em @R=0.2em @!R { \\
\nghost{{q}_{0} : }	 & \lstick{{q}_{0} : }	 & \gate{\mathrm{H}}	 & \ctrl{2}	 & \qw	 & \qw	 & \ctrl{4}	 & \qw	 & \targ	 & \qw	 & \control\qw	 & \qw	 & \qw	 & \qw	 & \qw	 & \targ	 & \qw	 & \qw	 & \qw	 & \qw\\
\nghost{{q}_{1} : }	 & \lstick{{q}_{1} : }	 & \qw	 & \qw	 & \targ	 & \qw	 & \qw	 & \ctrl{3}	 & \qw	 & \gate{\mathrm{H}}	 & \qw	 & \qw	 & \qw	 & \qw	 & \targ	 & \qw	 & \targ	 & \qw	 & \qw	 & \qw\\
\nghost{{q}_{2} : }	 & \lstick{{q}_{2} : }	 & \qw	 & \targ	 & \qw	 & \gate{\mathrm{H}}	 & \qw	 & \qw	 & \qw	 & \qw	 & \qw	 & \qw	 & \qw	 & \qw	 & \qw	 & \qw	 & \qw	 & \qw	 & \qw	 & \qw\\
\nghost{{q}_{3} : }	 & \lstick{{q}_{3} : }	 & \gate{\mathrm{H}}	 & \qw	 & \ctrl{-2}	 & \gate{\mathrm{H}}	 & \qw	 & \qw	 & \ctrl{-3}	 & \gate{\mathrm{H}}	 & \qw	 & \qw	 & \targ	 & \control\qw	 & \qw	 & \qw	 & \qw	 & \qw	 & \qw	 & \qw\\
\nghost{{q}_{4} : }	 & \lstick{{q}_{4} : }	 & \qw	 & \qw	 & \qw	 & \qw	 & \targ	 & \targ	 & \qw	 & \qw	 & \qw	 & \control\qw	 & \qw	 & \qw	 & \qw	 & \qw	 & \qw	 & \qw	 & \qw	 & \qw\\
\nghost{{q}_{5} : }	 & \lstick{{q}_{5} : }	 & \qw	 & \qw	 & \qw	 & \qw	 & \qw	 & \qw	 & \qw	 & \gate{\mathrm{H}}	 & \ctrl{-5}	 & \ctrl{-1}	 & \ctrl{-2}	 & \qw	 & \qw	 & \qw	 & \ctrl{-4}	 & \gate{\mathrm{H}}	 & \qw	 & \qw\\
\nghost{{q}_{6} : }	 & \lstick{{q}_{6} : }	 & \qw	 & \qw	 & \qw	 & \qw	 & \qw	 & \qw	 & \qw	 & \gate{\mathrm{H}}	 & \qw	 & \qw	 & \qw	 & \ctrl{-3}	 & \ctrl{-5}	 & \ctrl{-6}	 & \gate{\mathrm{H}}	 & \qw	 & \qw	 & \qw\\
\\ }}

\textbf{The $\mathbf{|-\rangle_L}$ state of the $\mathbf{[[5,1,3]]}$ perfect code}.  The circuit with the LSP + VCS approach has 12 two-qubit gates and 2 flag qubits:

\noindent \scalebox{1.0}{
\Qcircuit @C=0.2em @R=0.2em @!R { \\
	 	\nghost{{q}_{0} :  } & \lstick{{q}_{0} :  } & \gate{\mathrm{H}} & \ctrl{4} & \gate{\mathrm{H}} & \targ & \gate{\mathrm{H}} & \qw & \qw & \qw & \qw \barrier[0em]{6} & \qw & \qw & \qw & \targ & \qw & \qw & \qw & \qw & \qw & \qw & \qw & \qw\\
	 	\nghost{{q}_{1} :  } & \lstick{{q}_{1} :  } & \qw & \qw & \targ & \qw & \qw & \qw & \qw & \qw & \qw & \qw & \qw & \qw & \qw & \qw & \qw & \control\qw & \qw & \control\qw & \qw & \qw & \qw\\
	 	\nghost{{q}_{2} :  } & \lstick{{q}_{2} :  } & \qw & \qw & \qw & \qw & \qw & \targ & \qw & \qw & \qw & \qw & \qw & \qw & \qw & \qw & \control\qw & \qw & \targ & \qw & \qw & \qw & \qw\\
	 	\nghost{{q}_{3} :  } & \lstick{{q}_{3} :  } & \gate{\mathrm{H}} & \qw & \ctrl{-2} & \ctrl{-3} & \gate{\mathrm{H}} & \ctrl{-1} & \gate{\mathrm{H}} & \ctrl{1} & \qw & \qw & \qw & \targ & \qw & \control\qw & \qw & \qw & \qw & \qw & \qw & \qw & \qw\\
	 	\nghost{{q}_{4} :  } & \lstick{{q}_{4} :  } & \qw & \targ & \qw & \qw & \qw & \qw & \qw & \targ & \gate{\mathrm{H}} & \qw & \qw & \qw & \qw & \qw & \qw & \qw & \qw & \qw & \qw & \qw & \qw\\
	 	\nghost{{q}_{5} :  } & \lstick{{q}_{5} :  } & \qw & \qw & \qw & \qw & \qw & \qw & \qw & \qw & \qw & \qw & \gate{\mathrm{H}} & \qw & \qw & \ctrl{-2} & \qw & \ctrl{-4} & \ctrl{-3} & \qw & \gate{\mathrm{H}} & \qw & \qw\\
	 	\nghost{{q}_{6} :  } & \lstick{{q}_{6} :  } & \qw & \qw & \qw & \qw & \qw & \qw & \qw & \qw & \qw & \qw & \gate{\mathrm{H}} & \ctrl{-3} & \ctrl{-6} & \qw & \ctrl{-4} & \qw & \qw & \ctrl{-5} & \gate{\mathrm{H}} & \qw & \qw\\
\\ }}

The circuit with the IFT-LSP approach has 12 two-qubit gates and 2 flag qubits:

\noindent \scalebox{1.0}{
\Qcircuit @C=0.2em @R=0.2em @!R { \\
\nghost{{q}_{0} :  }	 & \lstick{{q}_{0} :  }	 & \gate{\mathrm{H}}	 & \ctrl{3}	 & \qw	 & \ctrl{4}	 & \targ	 & \gate{\mathrm{H}}	 & \qw	 & \targ	 & \qw	 & \qw	 & \qw	 & \qw	 & \qw	 & \qw	 & \qw	 & \qw\\
\nghost{{q}_{1} :  }	 & \lstick{{q}_{1} :  }	 & \qw	 & \qw	 & \targ	 & \qw	 & \ctrl{-1}	 & \qw	 & \qw	 & \qw	 & \qw	 & \qw	 & \control\qw	 & \qw	 & \control\qw	 & \qw	 & \qw	 & \qw\\
\nghost{{q}_{2} :  }	 & \lstick{{q}_{2} :  }	 & \gate{\mathrm{H}}	 & \qw	 & \ctrl{-1}	 & \qw	 & \targ	 & \gate{\mathrm{H}}	 & \control\qw	 & \qw	 & \targ	 & \qw	 & \qw	 & \qw	 & \qw	 & \qw	 & \qw	 & \qw\\
\nghost{{q}_{3} :  }	 & \lstick{{q}_{3} :  }	 & \qw	 & \targ	 & \gate{\mathrm{H}}	 & \qw	 & \ctrl{-1}	 & \qw	 & \qw	 & \qw	 & \qw	 & \targ	 & \qw	 & \control\qw	 & \qw	 & \qw	 & \qw	 & \qw\\
\nghost{{q}_{4} :  }	 & \lstick{{q}_{4} :  }	 & \qw	 & \qw	 & \qw	 & \targ	 & \qw	 & \qw	 & \qw	 & \qw	 & \qw	 & \qw	 & \qw	 & \qw	 & \qw	 & \qw	 & \qw	 & \qw\\
\nghost{{q}_{5} :  }	 & \lstick{{q}_{5} :  }	 & \qw	 & \qw	 & \qw	 & \qw	 & \qw	 & \gate{\mathrm{H}}	 & \qw	 & \qw	 & \ctrl{-3}	 & \qw	 & \ctrl{-4}	 & \ctrl{-2}	 & \qw	 & \gate{\mathrm{H}}	 & \qw	 & \qw\\
\nghost{{q}_{6} :  }	 & \lstick{{q}_{6} :  }	 & \qw	 & \qw	 & \qw	 & \qw	 & \qw	 & \gate{\mathrm{H}}	 & \ctrl{-4}	 & \ctrl{-6}	 & \qw	 & \ctrl{-3}	 & \qw	 & \qw	 & \ctrl{-5}	 & \gate{\mathrm{H}}	 & \qw	 & \qw\\
\\ }}

\textbf{The $\mathbf{|0\rangle_L}$ state of the $\mathbf{[[7,1,3]]}$ Steane code}. The circuit found with the LSP + VCS approach has 11 two-qubit gates and 1 flag qubit:

\noindent  \scalebox{1.0}{
\Qcircuit @C=0.2em @R=0.2em @!R { \\
	 	\nghost{{q}_{0} :  } & \lstick{{q}_{0} :  } & \gate{\mathrm{H}} & \ctrl{4} & \qw & \ctrl{6} & \qw & \qw & \qw & \ctrl{2} \barrier[0em]{7} & \qw & \qw & \qw & \qw & \qw & \qw\\
	 	\nghost{{q}_{1} :  } & \lstick{{q}_{1} :  } & \gate{\mathrm{H}} & \qw & \ctrl{4} & \qw & \qw & \targ & \ctrl{1} & \qw & \qw & \qw & \qw & \qw & \qw & \qw\\
	 	\nghost{{q}_{2} :  } & \lstick{{q}_{2} :  } & \qw & \qw & \qw & \qw & \qw & \qw & \targ & \targ & \qw & \qw & \ctrl{5} & \qw & \qw & \qw\\
	 	\nghost{{q}_{3} :  } & \lstick{{q}_{3} :  } & \gate{\mathrm{H}} & \qw & \qw & \qw & \ctrl{1} & \ctrl{-2} & \qw & \qw & \qw & \qw & \qw & \qw & \qw & \qw\\
	 	\nghost{{q}_{4} :  } & \lstick{{q}_{4} :  } & \qw & \targ & \qw & \qw & \targ & \qw & \qw & \qw & \qw & \qw & \qw & \ctrl{3} & \qw & \qw\\
	 	\nghost{{q}_{5} :  } & \lstick{{q}_{5} :  } & \qw & \qw & \targ & \qw & \ctrl{1} & \qw & \qw & \qw & \qw & \ctrl{2} & \qw & \qw & \qw & \qw\\
	 	\nghost{{q}_{6} :  } & \lstick{{q}_{6} :  } & \qw & \qw & \qw & \targ & \targ & \qw & \qw & \qw & \qw & \qw & \qw & \qw & \qw & \qw\\
	 	\nghost{{q}_{7} :  } & \lstick{{q}_{7} :  } & \qw & \qw & \qw & \qw & \qw & \qw & \qw & \qw & \qw & \targ & \targ & \targ & \qw & \qw\\
\\ }}

The circuit learned with the IFT-LSP approach has 11 two-qubit gates and 1 flag qubit, while using a smaller number of single-qubit gates:

\noindent \scalebox{1.0}{
\Qcircuit @C=0.2em @R=0.2em @!R { \\
	 	\nghost{{q}_{0} :  } & \lstick{{q}_{0} :  } & \gate{\mathrm{H}} & \ctrl{2} & \qw & \ctrl{4} & \qw & \qw & \qw & \qw & \qw & \qw & \qw & \qw\\
	 	\nghost{{q}_{1} :  } & \lstick{{q}_{1} :  } & \gate{\mathrm{H}} & \qw & \ctrl{5} & \qw & \ctrl{4} & \ctrl{1} & \qw & \qw & \qw & \qw & \qw & \qw\\
	 	\nghost{{q}_{2} :  } & \lstick{{q}_{2} :  } & \qw & \targ & \qw & \qw & \qw & \targ & \qw & \qw & \qw & \ctrl{5} & \qw & \qw\\
	 	\nghost{{q}_{3} :  } & \lstick{{q}_{3} :  } & \gate{\mathrm{H}} & \ctrl{2} & \qw & \qw & \qw & \ctrl{1} & \qw & \ctrl{4} & \qw & \qw & \qw & \qw\\
	 	\nghost{{q}_{4} :  } & \lstick{{q}_{4} :  } & \qw & \qw & \qw & \targ & \qw & \targ & \ctrl{2} & \qw & \qw & \qw & \qw & \qw\\
	 	\nghost{{q}_{5} :  } & \lstick{{q}_{5} :  } & \qw & \targ & \qw & \qw & \targ & \qw & \qw & \qw & \qw & \qw & \qw & \qw\\
	 	\nghost{{q}_{6} :  } & \lstick{{q}_{6} :  } & \qw & \qw & \targ & \qw & \qw & \qw & \targ & \qw & \ctrl{1} & \qw & \qw & \qw\\
	 	\nghost{{q}_{7} :  } & \lstick{{q}_{7} :  } & \qw & \qw & \qw & \qw & \qw & \qw & \qw & \targ & \targ & \targ & \qw & \qw\\
\\ }}

\textbf{The $\mathbf{|+\rangle_L}$ state of the  $\mathbf{[[7,1,3]]}$ Steane code}. The circuit learned with the LSP + VCS approach has 11 two-qubit gates and 1 flag qubit:

\noindent \scalebox{1.0}{
\Qcircuit @C=0.2em @R=0.2em @!R { \\
	 	\nghost{{q}_{0} :  } & \lstick{{q}_{0} :  } & \gate{\mathrm{H}} & \ctrl{6} & \ctrl{5} & \qw & \qw & \qw & \targ & \targ & \qw \barrier[0em]{7} & \qw & \qw & \qw & \qw & \qw & \qw & \qw & \qw\\
	 	\nghost{{q}_{1} :  } & \lstick{{q}_{1} :  } & \gate{\mathrm{H}} & \qw & \qw & \ctrl{5} & \ctrl{3} & \qw & \qw & \qw & \targ & \qw & \qw & \qw & \qw & \qw & \qw & \qw & \qw\\
	 	\nghost{{q}_{2} :  } & \lstick{{q}_{2} :  } & \gate{\mathrm{H}} & \qw & \qw & \qw & \qw & \qw & \qw & \ctrl{-2} & \ctrl{-1} & \qw & \qw & \targ & \qw & \qw & \qw & \qw & \qw\\
	 	\nghost{{q}_{3} :  } & \lstick{{q}_{3} :  } & \gate{\mathrm{H}} & \qw & \qw & \qw & \qw & \ctrl{1} & \ctrl{-3} & \qw & \qw & \qw & \qw & \qw & \qw & \qw & \qw & \qw & \qw\\
	 	\nghost{{q}_{4} :  } & \lstick{{q}_{4} :  } & \qw & \qw & \qw & \qw & \targ & \targ & \qw & \qw & \qw & \qw & \qw & \qw & \targ & \qw & \qw & \qw & \qw\\
	 	\nghost{{q}_{5} :  } & \lstick{{q}_{5} :  } & \qw & \qw & \targ & \qw & \qw & \qw & \qw & \qw & \qw & \qw & \qw & \qw & \qw & \targ & \qw & \qw & \qw\\
	 	\nghost{{q}_{6} :  } & \lstick{{q}_{6} :  } & \qw & \targ & \qw & \targ & \qw & \qw & \qw & \qw & \qw & \qw & \qw & \qw & \qw & \qw & \qw & \qw & \qw\\
	 	\nghost{{q}_{7} :  } & \lstick{{q}_{7} :  } & \qw & \qw & \qw & \qw & \qw & \qw & \qw & \qw & \qw & \qw & \gate{\mathrm{H}} & \ctrl{-5} & \ctrl{-3} & \ctrl{-2} & \gate{\mathrm{H}} & \qw & \qw\\
\\ }}

The circuit found with the IFT-LSP approach has 11 two-qubit gates and 1 flag qubit:

\noindent \scalebox{1.0}{
\Qcircuit @C=0.2em @R=0.2em @!R { \\
	 	\nghost{{q}_{0} :  } & \lstick{{q}_{0} :  } & \qw & \targ & \qw & \qw & \qw & \targ & \targ & \qw & \qw & \qw & \qw & \qw\\
	 	\nghost{{q}_{1} :  } & \lstick{{q}_{1} :  } & \qw & \qw & \targ & \qw & \qw & \qw & \qw & \targ & \qw & \qw & \qw & \qw\\
	 	\nghost{{q}_{2} :  } & \lstick{{q}_{2} :  } & \gate{\mathrm{H}} & \qw & \qw & \qw & \qw & \qw & \ctrl{-2} & \ctrl{-1} & \targ & \qw & \qw & \qw\\
	 	\nghost{{q}_{3} :  } & \lstick{{q}_{3} :  } & \gate{\mathrm{H}} & \qw & \qw & \qw & \ctrl{1} & \ctrl{-3} & \targ & \qw & \qw & \qw & \qw & \qw\\
	 	\nghost{{q}_{4} :  } & \lstick{{q}_{4} :  } & \gate{\mathrm{H}} & \qw & \ctrl{-3} & \ctrl{2} & \targ & \qw & \qw & \qw & \qw & \qw & \qw & \qw\\
	 	\nghost{{q}_{5} :  } & \lstick{{q}_{5} :  } & \gate{\mathrm{H}} & \ctrl{-5} & \ctrl{1} & \qw & \qw & \qw & \qw & \qw & \qw & \qw & \qw & \qw\\
	 	\nghost{{q}_{6} :  } & \lstick{{q}_{6} :  } & \qw & \qw & \targ & \targ & \qw & \qw & \qw & \targ & \qw & \qw & \qw & \qw\\
	 	\nghost{{q}_{7} :  } & \lstick{{q}_{7} :  } & \gate{\mathrm{H}} & \qw & \qw & \qw & \qw & \qw & \ctrl{-4} & \ctrl{-1} & \ctrl{-5} & \gate{\mathrm{H}} & \qw & \qw\\
\\ }}

\textbf{The $\mathbf{|0\rangle_L}$ state of the $\mathbf{[[9,1,3]]}$ Shor code}. Both approaches discover the known fault-tolerant state preparation circuit that does not require any flag qubits at all, shown below.

\noindent\scalebox{1.0}{
\Qcircuit @C=0.2em @R=0.2em @!R { \\
	 	\nghost{{q}_{0} :  } & \lstick{{q}_{0} :  } & \gate{\mathrm{H}} & \ctrl{2} & \ctrl{1} & \qw & \qw\\
	 	\nghost{{q}_{1} :  } & \lstick{{q}_{1} :  } & \qw & \qw & \targ & \qw & \qw\\
	 	\nghost{{q}_{2} :  } & \lstick{{q}_{2} :  } & \qw & \targ & \qw & \qw & \qw\\
	 	\nghost{{q}_{3} :  } & \lstick{{q}_{3} :  } & \gate{\mathrm{H}} & \ctrl{2} & \ctrl{1} & \qw & \qw\\
	 	\nghost{{q}_{4} :  } & \lstick{{q}_{4} :  } & \qw & \qw & \targ & \qw & \qw\\
	 	\nghost{{q}_{5} :  } & \lstick{{q}_{5} :  } & \qw & \targ & \qw & \qw & \qw\\
	 	\nghost{{q}_{6} :  } & \lstick{{q}_{6} :  } & \gate{\mathrm{H}} & \ctrl{2} & \ctrl{1} & \qw & \qw\\
	 	\nghost{{q}_{7} :  } & \lstick{{q}_{7} :  } & \qw & \qw & \targ & \qw & \qw\\
	 	\nghost{{q}_{8} :  } & \lstick{{q}_{8} :  } & \qw & \targ & \qw & \qw & \qw\\
	 	\nghost{{q}_{9} :  } & \lstick{{q}_{9} :  } & \qw & \qw & \qw & \qw & \qw\\
\\ }}

\textbf{The $\mathbf{|+\rangle_L}$ state of the $\mathbf{[[9,1,3]]}$ Shor code}. The circuit learned with the LSP + VCS approach has 11 two-qubit gates and 1 flag qubit:

\noindent \scalebox{1.0}{
\Qcircuit @C=0.2em @R=0.2em @!R { \\
	 	\nghost{{q}_{0} :  } & \lstick{{q}_{0} :  } & \gate{\mathrm{H}} & \ctrl{8} & \qw & \qw & \qw & \qw & \qw \barrier[0em]{9} & \qw & \qw & \qw & \qw & \qw & \qw\\
	 	\nghost{{q}_{1} :  } & \lstick{{q}_{1} :  } & \qw & \qw & \qw & \targ & \ctrl{1} & \qw & \qw & \qw & \qw & \qw & \qw & \qw & \qw\\
	 	\nghost{{q}_{2} :  } & \lstick{{q}_{2} :  } & \qw & \qw & \qw & \qw & \targ & \qw & \qw & \qw & \qw & \qw & \ctrl{7} & \qw & \qw\\
	 	\nghost{{q}_{3} :  } & \lstick{{q}_{3} :  } & \gate{\mathrm{H}} & \qw & \ctrl{2} & \qw & \qw & \qw & \qw & \qw & \qw & \qw & \qw & \qw & \qw\\
	 	\nghost{{q}_{4} :  } & \lstick{{q}_{4} :  } & \qw & \qw & \qw & \qw & \qw & \targ & \qw & \qw & \ctrl{5} & \qw & \qw & \qw & \qw\\
	 	\nghost{{q}_{5} :  } & \lstick{{q}_{5} :  } & \qw & \qw & \targ & \qw & \ctrl{3} & \ctrl{-1} & \qw & \qw & \qw & \qw & \qw & \qw & \qw\\
	 	\nghost{{q}_{6} :  } & \lstick{{q}_{6} :  } & \qw & \qw & \qw & \qw & \qw & \targ & \qw & \qw & \qw & \qw & \qw & \qw & \qw\\
	 	\nghost{{q}_{7} :  } & \lstick{{q}_{7} :  } & \qw & \qw & \qw & \qw & \qw & \qw & \targ & \qw & \qw & \ctrl{2} & \qw & \qw & \qw\\
	 	\nghost{{q}_{8} :  } & \lstick{{q}_{8} :  } & \qw & \targ & \qw & \ctrl{-7} & \targ & \ctrl{-2} & \ctrl{-1} & \qw & \qw & \qw & \qw & \qw & \qw\\
	 	\nghost{{q}_{9} :  } & \lstick{{q}_{9} :  } & \qw & \qw & \qw & \qw & \qw & \qw & \qw & \qw & \targ & \targ & \targ & \qw & \qw\\
\\ }}

The circuit learned using the IFT-LSP approach has again 11 two-qubit gates and 1 flag qubit:

\noindent \scalebox{1.0}{
\Qcircuit @C=0.2em @R=0.2em @!R { \\
	 	\nghost{{q}_{0} :  } & \lstick{{q}_{0} :  } & \gate{\mathrm{H}} & \ctrl{8} & \qw & \ctrl{6} & \targ & \ctrl{1} & \qw & \qw & \qw & \qw & \qw\\
	 	\nghost{{q}_{1} :  } & \lstick{{q}_{1} :  } & \qw & \qw & \qw & \qw & \qw & \targ & \ctrl{1} & \qw & \qw & \qw & \qw\\
	 	\nghost{{q}_{2} :  } & \lstick{{q}_{2} :  } & \qw & \qw & \qw & \qw & \qw & \qw & \targ & \ctrl{7} & \qw & \qw & \qw\\
	 	\nghost{{q}_{3} :  } & \lstick{{q}_{3} :  } & \gate{\mathrm{H}} & \qw & \ctrl{2} & \qw & \qw & \qw & \qw & \qw & \qw & \qw & \qw\\
	 	\nghost{{q}_{4} :  } & \lstick{{q}_{4} :  } & \qw & \qw & \qw & \qw & \qw & \targ & \qw & \qw & \ctrl{5} & \qw & \qw\\
	 	\nghost{{q}_{5} :  } & \lstick{{q}_{5} :  } & \qw & \qw & \targ & \qw & \ctrl{-5} & \ctrl{-1} & \qw & \qw & \qw & \qw & \qw\\
	 	\nghost{{q}_{6} :  } & \lstick{{q}_{6} :  } & \qw & \qw & \qw & \targ & \ctrl{1} & \ctrl{3} & \qw & \qw & \qw & \qw & \qw\\
	 	\nghost{{q}_{7} :  } & \lstick{{q}_{7} :  } & \qw & \qw & \qw & \qw & \targ & \qw & \qw & \qw & \qw & \qw & \qw\\
	 	\nghost{{q}_{8} :  } & \lstick{{q}_{8} :  } & \qw & \targ & \qw & \qw & \qw & \qw & \qw & \qw & \qw & \qw & \qw\\
	 	\nghost{{q}_{9} :  } & \lstick{{q}_{9} :  } & \qw & \qw & \qw & \qw & \qw & \targ & \qw & \targ & \targ & \qw & \qw\\
\\ }}

\textbf{The $\mathbf{|0\rangle_L}$ state of the $\mathbf{[[9,1,3]]}$ Surface-17 code}. The circuit learned with the LSP + VCS approach has 11 two-qubit gates and 1 flag qubit:

\noindent \scalebox{1.0}{
\Qcircuit @C=0.2em @R=0.2em @!R { \\
	 	\nghost{{q}_{0} :  } & \lstick{{q}_{0} :  } & \gate{\mathrm{H}} & \ctrl{4} & \qw & \ctrl{3} & \ctrl{2} \barrier[0em]{9} & \qw & \ctrl{9} & \qw & \qw & \qw & \qw\\
	 	\nghost{{q}_{1} :  } & \lstick{{q}_{1} :  } & \qw & \qw & \targ & \qw & \qw & \qw & \qw & \qw & \qw & \qw & \qw\\
	 	\nghost{{q}_{2} :  } & \lstick{{q}_{2} :  } & \gate{\mathrm{H}} & \qw & \ctrl{-1} & \qw & \targ & \qw & \qw & \qw & \qw & \qw & \qw\\
	 	\nghost{{q}_{3} :  } & \lstick{{q}_{3} :  } & \qw & \qw & \qw & \targ & \qw & \qw & \qw & \qw & \qw & \qw & \qw\\
	 	\nghost{{q}_{4} :  } & \lstick{{q}_{4} :  } & \qw & \targ & \qw & \targ & \qw & \qw & \qw & \qw & \ctrl{5} & \qw & \qw\\
	 	\nghost{{q}_{5} :  } & \lstick{{q}_{5} :  } & \gate{\mathrm{H}} & \ctrl{3} & \qw & \qw & \qw & \qw & \qw & \qw & \qw & \qw & \qw\\
	 	\nghost{{q}_{6} :  } & \lstick{{q}_{6} :  } & \qw & \qw & \targ & \qw & \targ & \qw & \qw & \qw & \qw & \qw & \qw\\
	 	\nghost{{q}_{7} :  } & \lstick{{q}_{7} :  } & \gate{\mathrm{H}} & \qw & \ctrl{-1} & \qw & \qw & \qw & \qw & \qw & \qw & \qw & \qw\\
	 	\nghost{{q}_{8} :  } & \lstick{{q}_{8} :  } & \qw & \targ & \qw & \ctrl{-4} & \ctrl{-2} & \qw & \qw & \ctrl{1} & \qw & \qw & \qw\\
	 	\nghost{{q}_{9} :  } & \lstick{{q}_{9} :  } & \qw & \qw & \qw & \qw & \qw & \qw & \targ & \targ & \targ & \qw & \qw\\
\\ }}

The circuit learned using the IFT-LSP approach has again 11 two-qubit gates and 1 flag qubit:

\scalebox{1.0}{
\Qcircuit @C=0.2em @R=0.2em @!R { \\
	 	\nghost{{q}_{0} :  } & \lstick{{q}_{0} :  } & \gate{\mathrm{H}} & \ctrl{3} & \qw & \ctrl{4} & \ctrl{2} & \ctrl{9} & \qw & \qw & \qw & \qw & \qw\\
	 	\nghost{{q}_{1} :  } & \lstick{{q}_{1} :  } & \qw & \qw & \targ & \qw & \qw & \qw & \qw & \qw & \qw & \qw & \qw\\
	 	\nghost{{q}_{2} :  } & \lstick{{q}_{2} :  } & \gate{\mathrm{H}} & \qw & \ctrl{-1} & \qw & \targ & \qw & \qw & \qw & \qw & \qw & \qw\\
	 	\nghost{{q}_{3} :  } & \lstick{{q}_{3} :  } & \qw & \targ & \qw & \qw & \qw & \qw & \qw & \qw & \qw & \qw & \qw\\
	 	\nghost{{q}_{4} :  } & \lstick{{q}_{4} :  } & \qw & \qw & \qw & \targ & \targ & \qw & \qw & \qw & \ctrl{5} & \qw & \qw\\
	 	\nghost{{q}_{5} :  } & \lstick{{q}_{5} :  } & \gate{\mathrm{H}} & \ctrl{3} & \qw & \qw & \ctrl{-1} & \qw & \ctrl{1} & \ctrl{4} & \qw & \qw & \qw\\
	 	\nghost{{q}_{6} :  } & \lstick{{q}_{6} :  } & \qw & \qw & \targ & \qw & \qw & \qw & \targ & \qw & \qw & \qw & \qw\\
	 	\nghost{{q}_{7} :  } & \lstick{{q}_{7} :  } & \gate{\mathrm{H}} & \qw & \ctrl{-1} & \qw & \qw & \qw & \qw & \qw & \qw & \qw & \qw\\
	 	\nghost{{q}_{8} :  } & \lstick{{q}_{8} :  } & \qw & \targ & \qw & \qw & \qw & \qw & \qw & \qw & \qw & \qw & \qw\\
	 	\nghost{{q}_{9} :  } & \lstick{{q}_{9} :  } & \qw & \qw & \qw & \qw & \qw & \targ & \qw & \targ & \targ & \qw & \qw\\
\\ }}

\textbf{The $\mathbf{|+\rangle_L}$ state of the $\mathbf{[[9,1,3]]}$ Surface-17 code}. The circuit learned with the LSP + VCS approach has 11 two-qubit gates and 1 flag qubit:

\noindent  \scalebox{1.0}{
\Qcircuit @C=0.2em @R=0.2em @!R { \\
	 	\nghost{{q}_{0} :  } & \lstick{{q}_{0} :  } & \gate{\mathrm{H}} & \ctrl{7} & \qw & \qw & \ctrl{3} & \qw \barrier[0em]{9} & \qw & \qw & \qw & \qw & \qw & \qw & \qw & \qw\\
	 	\nghost{{q}_{1} :  } & \lstick{{q}_{1} :  } & \qw & \qw & \targ & \targ & \qw & \targ & \qw & \qw & \qw & \targ & \qw & \qw & \qw & \qw\\
	 	\nghost{{q}_{2} :  } & \lstick{{q}_{2} :  } & \gate{\mathrm{H}} & \qw & \qw & \qw & \qw & \ctrl{-1} & \qw & \qw & \qw & \qw & \qw & \qw & \qw & \qw\\
	 	\nghost{{q}_{3} :  } & \lstick{{q}_{3} :  } & \qw & \qw & \qw & \qw & \targ & \qw & \qw & \qw & \qw & \qw & \qw & \qw & \qw & \qw\\
	 	\nghost{{q}_{4} :  } & \lstick{{q}_{4} :  } & \gate{\mathrm{H}} & \qw & \qw & \ctrl{-3} & \qw & \ctrl{2} & \qw & \qw & \targ & \qw & \qw & \qw & \qw & \qw\\
	 	\nghost{{q}_{5} :  } & \lstick{{q}_{5} :  } & \qw & \qw & \qw & \targ & \qw & \qw & \qw & \qw & \qw & \qw & \qw & \qw & \qw & \qw\\
	 	\nghost{{q}_{6} :  } & \lstick{{q}_{6} :  } & \gate{\mathrm{H}} & \qw & \qw & \qw & \ctrl{1} & \targ & \qw & \qw & \qw & \qw & \qw & \qw & \qw & \qw\\
	 	\nghost{{q}_{7} :  } & \lstick{{q}_{7} :  } & \qw & \targ & \qw & \qw & \targ & \qw & \qw & \qw & \qw & \qw & \targ & \qw & \qw & \qw\\
	 	\nghost{{q}_{8} :  } & \lstick{{q}_{8} :  } & \gate{\mathrm{H}} & \qw & \ctrl{-7} & \ctrl{-3} & \qw & \qw & \qw & \qw & \qw & \qw & \qw & \qw & \qw & \qw\\
	 	\nghost{{q}_{9} :  } & \lstick{{q}_{9} :  } & \qw & \qw & \qw & \qw & \qw & \qw & \qw & \gate{\mathrm{H}} & \ctrl{-5} & \ctrl{-8} & \ctrl{-2} & \gate{\mathrm{H}} & \qw & \qw\\
\\ }}

The circuit learned using the IFT-LSP approach has 11 two-qubit gates and 1 flag qubit:

\noindent  \scalebox{1.0}{
\Qcircuit @C=1.0em @R=0.2em @!R { \\
	 	\nghost{{q}_{0} :  } & \lstick{{q}_{0} :  } & \gate{\mathrm{H}} & \ctrl{7} & \qw & \ctrl{3} & \qw & \qw & \qw & \qw & \qw & \qw & \qw & \qw\\
	 	\nghost{{q}_{1} :  } & \lstick{{q}_{1} :  } & \qw & \qw & \targ & \qw & \targ & \targ & \qw & \qw & \qw & \qw & \qw & \qw\\
	 	\nghost{{q}_{2} :  } & \lstick{{q}_{2} :  } & \gate{\mathrm{H}} & \qw & \qw & \qw & \qw & \ctrl{-1} & \qw & \targ & \qw & \qw & \qw & \qw\\
	 	\nghost{{q}_{3} :  } & \lstick{{q}_{3} :  } & \qw & \qw & \qw & \targ & \qw & \qw & \qw & \qw & \qw & \qw & \qw & \qw\\
	 	\nghost{{q}_{4} :  } & \lstick{{q}_{4} :  } & \gate{\mathrm{H}} & \qw & \qw & \ctrl{3} & \ctrl{-3} & \qw & \qw & \qw & \targ & \qw & \qw & \qw\\
	 	\nghost{{q}_{5} :  } & \lstick{{q}_{5} :  } & \qw & \qw & \qw & \qw & \targ & \qw & \qw & \qw & \qw & \qw & \qw & \qw\\
	 	\nghost{{q}_{6} :  } & \lstick{{q}_{6} :  } & \gate{\mathrm{H}} & \qw & \qw & \qw & \qw & \ctrl{1} & \targ & \qw & \qw & \qw & \qw & \qw\\
	 	\nghost{{q}_{7} :  } & \lstick{{q}_{7} :  } & \qw & \targ & \qw & \targ & \qw & \targ & \qw & \qw & \qw & \qw & \qw & \qw\\
	 	\nghost{{q}_{8} :  } & \lstick{{q}_{8} :  } & \gate{\mathrm{H}} & \qw & \ctrl{-7} & \qw & \ctrl{-3} & \qw & \qw & \qw & \qw & \qw & \qw & \qw\\
	 	\nghost{{q}_{9} :  } & \lstick{{q}_{9} :  } & \gate{\mathrm{H}} & \qw & \qw & \qw & \qw & \qw & \ctrl{-3} & \ctrl{-7} & \ctrl{-5} & \gate{\mathrm{H}} & \qw & \qw\\
\\ }}

\textbf{The $\mathbf{|0\rangle_L}$ state of the $\mathbf{[[15,1,3]]}$ Reed-Muller code}. The circuit found based on the LSP + VCS approach has 29 two-qubit gates and 2 flag qubits:

\noindent \scalebox{1.0}{
\Qcircuit @C=0.1em @R=0.1em @!R { \\
	 	\nghost{{q}_{0} :  } & \lstick{{q}_{0} :  } & \qw & \targ & \qw & \qw & \qw & \qw & \qw & \qw & \qw & \ctrl{12} & \ctrl{4} & \qw & \ctrl{2} & \ctrl{8} & \ctrl{6} \barrier[0em]{16} & \qw & \qw & \qw & \qw & \qw & \qw & \qw & \qw & \qw\\
	 	\nghost{{q}_{1} :  } & \lstick{{q}_{1} :  } & \qw & \qw & \targ & \qw & \qw & \qw & \qw & \qw & \qw & \qw & \qw & \ctrl{8} & \qw & \qw & \qw & \qw & \qw & \qw & \qw & \qw & \qw & \qw & \qw & \qw\\
	 	\nghost{{q}_{2} :  } & \lstick{{q}_{2} :  } & \gate{\mathrm{H}} & \qw & \ctrl{-1} & \ctrl{11} & \qw & \qw & \qw & \qw & \ctrl{4} & \qw & \qw & \qw & \targ & \qw & \qw & \qw & \qw & \qw & \qw & \qw & \qw & \qw & \qw & \qw\\
	 	\nghost{{q}_{3} :  } & \lstick{{q}_{3} :  } & \qw & \qw & \targ & \qw & \qw & \qw & \ctrl{7} & \ctrl{3} & \qw & \qw & \qw & \qw & \qw & \qw & \qw & \qw & \qw & \qw & \qw & \qw & \qw & \ctrl{12} & \qw & \qw\\
	 	\nghost{{q}_{4} :  } & \lstick{{q}_{4} :  } & \gate{\mathrm{H}} & \qw & \ctrl{-1} & \qw & \ctrl{3} & \qw & \qw & \qw & \qw & \qw & \targ & \qw & \qw & \qw & \qw & \qw & \qw & \qw & \qw & \qw & \qw & \qw & \qw & \qw\\
	 	\nghost{{q}_{5} :  } & \lstick{{q}_{5} :  } & \qw & \qw & \qw & \qw & \qw & \qw & \qw & \qw & \qw & \qw & \targ & \qw & \qw & \qw & \qw & \qw & \qw & \qw & \qw & \qw & \qw & \qw & \qw & \qw\\
	 	\nghost{{q}_{6} :  } & \lstick{{q}_{6} :  } & \qw & \qw & \qw & \qw & \qw & \qw & \qw & \targ & \targ & \qw & \ctrl{-1} & \qw & \qw & \qw & \targ & \qw & \ctrl{10} & \qw & \qw & \qw & \qw & \qw & \qw & \qw\\
	 	\nghost{{q}_{7} :  } & \lstick{{q}_{7} :  } & \qw & \qw & \qw & \qw & \targ & \targ & \qw & \ctrl{1} & \ctrl{2} & \qw & \qw & \qw & \qw & \qw & \qw & \qw & \qw & \qw & \qw & \qw & \ctrl{9} & \qw & \qw & \qw\\
	 	\nghost{{q}_{8} :  } & \lstick{{q}_{8} :  } & \qw & \qw & \qw & \qw & \qw & \qw & \qw & \targ & \qw & \qw & \qw & \qw & \qw & \targ & \qw & \qw & \qw & \qw & \qw & \qw & \qw & \qw & \qw & \qw\\
	 	\nghost{{q}_{9} :  } & \lstick{{q}_{9} :  } & \qw & \qw & \qw & \qw & \qw & \qw & \qw & \qw & \targ & \qw & \qw & \targ & \qw & \qw & \qw & \qw & \qw & \qw & \qw & \ctrl{6} & \qw & \qw & \qw & \qw\\
	 	\nghost{{q}_{10} :  } & \lstick{{q}_{10} :  } & \gate{\mathrm{H}} & \ctrl{-10} & \ctrl{4} & \qw & \qw & \qw & \targ & \qw & \qw & \qw & \targ & \qw & \qw & \qw & \qw & \qw & \qw & \qw & \qw & \qw & \qw & \qw & \qw & \qw\\
	 	\nghost{{q}_{11} :  } & \lstick{{q}_{11} :  } & \gate{\mathrm{H}} & \qw & \qw & \qw & \qw & \ctrl{-4} & \ctrl{1} & \ctrl{2} & \qw & \qw & \qw & \qw & \qw & \qw & \qw & \qw & \qw & \qw & \qw & \qw & \qw & \qw & \qw & \qw\\
	 	\nghost{{q}_{12} :  } & \lstick{{q}_{12} :  } & \qw & \qw & \qw & \qw & \qw & \qw & \targ & \qw & \qw & \targ & \qw & \qw & \qw & \qw & \qw & \qw & \qw & \qw & \qw & \qw & \qw & \qw & \qw & \qw\\
	 	\nghost{{q}_{13} :  } & \lstick{{q}_{13} :  } & \qw & \qw & \qw & \targ & \qw & \qw & \qw & \targ & \ctrl{1} & \qw & \ctrl{-3} & \qw & \qw & \qw & \qw & \qw & \qw & \ctrl{2} & \qw & \qw & \qw & \qw & \qw & \qw\\
	 	\nghost{{q}_{14} :  } & \lstick{{q}_{14} :  } & \qw & \qw & \targ & \qw & \qw & \qw & \qw & \qw & \targ & \qw & \qw & \qw & \qw & \qw & \qw & \qw & \qw & \qw & \ctrl{2} & \qw & \qw & \qw & \qw & \qw\\
	 	\nghost{{q}_{15} :  } & \lstick{{q}_{15} :  } & \qw & \qw & \qw & \qw & \qw & \qw & \qw & \qw & \qw & \qw & \qw & \qw & \qw & \qw & \qw & \qw & \qw & \targ & \qw & \targ & \qw & \targ & \qw & \qw\\
	 	\nghost{{q}_{16} :  } & \lstick{{q}_{16} :  } & \qw & \qw & \qw & \qw & \qw & \qw & \qw & \qw & \qw & \qw & \qw & \qw & \qw & \qw & \qw & \qw & \targ & \qw & \targ & \qw & \targ & \qw & \qw & \qw\\
\\ }}

In contrast, the circuit obtained with the IFT-LSP approach has 25 two-qubit gates (instead of 29) and only requires 1 flag qubit (instead of 2), illustrating the superior performance of the IFT-LSP over the LSP + VCS approach:

\noindent \scalebox{1.0}{
\Qcircuit @C=0.1em @R=0.1em @!R { \\
	 	\nghost{{q}_{0} :  } & \lstick{{q}_{0} :  } & \gate{\mathrm{H}} & \qw & \ctrl{14} & \qw & \ctrl{1} & \qw & \ctrl{3} & \qw & \qw & \qw & \qw & \ctrl{9} & \qw & \qw & \qw & \qw & \qw & \qw\\
	 	\nghost{{q}_{1} :  } & \lstick{{q}_{1} :  } & \qw & \qw & \qw & \targ & \targ & \qw & \qw & \qw & \ctrl{5} & \qw & \ctrl{7} & \qw & \ctrl{10} & \qw & \qw & \qw & \qw & \qw\\
	 	\nghost{{q}_{2} :  } & \lstick{{q}_{2} :  } & \gate{\mathrm{H}} & \ctrl{10} & \qw & \ctrl{-1} & \qw & \ctrl{5} & \qw & \qw & \qw & \ctrl{3} & \qw & \qw & \qw & \qw & \qw & \qw & \qw & \qw\\
	 	\nghost{{q}_{3} :  } & \lstick{{q}_{3} :  } & \qw & \qw & \qw & \targ & \qw & \qw & \targ & \qw & \qw & \qw & \qw & \qw & \qw & \qw & \qw & \qw & \qw & \qw\\
	 	\nghost{{q}_{4} :  } & \lstick{{q}_{4} :  } & \qw & \qw & \qw & \qw & \qw & \qw & \qw & \targ & \qw & \qw & \qw & \qw & \qw & \qw & \qw & \qw & \qw & \qw\\
	 	\nghost{{q}_{5} :  } & \lstick{{q}_{5} :  } & \gate{\mathrm{H}} & \qw & \qw & \ctrl{-2} & \ctrl{8} & \qw & \ctrl{1} & \ctrl{-1} & \qw & \targ & \qw & \qw & \qw & \qw & \ctrl{10} & \qw & \qw & \qw\\
	 	\nghost{{q}_{6} :  } & \lstick{{q}_{6} :  } & \qw & \qw & \qw & \qw & \qw & \qw & \targ & \qw & \targ & \qw & \qw & \qw & \qw & \qw & \qw & \qw & \qw & \qw\\
	 	\nghost{{q}_{7} :  } & \lstick{{q}_{7} :  } & \qw & \qw & \qw & \targ & \qw & \targ & \qw & \qw & \qw & \qw & \qw & \qw & \qw & \qw & \qw & \qw & \qw & \qw\\
	 	\nghost{{q}_{8} :  } & \lstick{{q}_{8} :  } & \qw & \qw & \qw & \qw & \qw & \qw & \targ & \qw & \qw & \qw & \targ & \qw & \qw & \qw & \qw & \qw & \qw & \qw\\
	 	\nghost{{q}_{9} :  } & \lstick{{q}_{9} :  } & \qw & \qw & \qw & \qw & \qw & \qw & \qw & \targ & \qw & \qw & \qw & \targ & \qw & \qw & \qw & \ctrl{6} & \qw & \qw\\
	 	\nghost{{q}_{10} :  } & \lstick{{q}_{10} :  } & \gate{\mathrm{H}} & \qw & \qw & \ctrl{-3} & \qw & \ctrl{3} & \ctrl{-2} & \ctrl{-1} & \qw & \qw & \qw & \qw & \qw & \qw & \qw & \qw & \qw & \qw\\
	 	\nghost{{q}_{11} :  } & \lstick{{q}_{11} :  } & \qw & \qw & \qw & \qw & \qw & \qw & \qw & \qw & \targ & \qw & \qw & \qw & \targ & \ctrl{4} & \qw & \qw & \qw & \qw\\
	 	\nghost{{q}_{12} :  } & \lstick{{q}_{12} :  } & \qw & \targ & \qw & \qw & \qw & \qw & \targ & \qw & \qw & \qw & \qw & \qw & \qw & \qw & \qw & \qw & \qw & \qw\\
	 	\nghost{{q}_{13} :  } & \lstick{{q}_{13} :  } & \qw & \qw & \qw & \qw & \targ & \targ & \ctrl{-1} & \ctrl{1} & \ctrl{-2} & \qw & \qw & \qw & \qw & \qw & \qw & \qw & \qw & \qw\\
	 	\nghost{{q}_{14} :  } & \lstick{{q}_{14} :  } & \qw & \qw & \targ & \qw & \qw & \qw & \qw & \targ & \qw & \qw & \qw & \qw & \qw & \qw & \qw & \qw & \qw & \qw\\
	 	\nghost{{q}_{15} :  } & \lstick{{q}_{15} :  } & \qw & \qw & \qw & \qw & \qw & \qw & \qw & \qw & \qw & \qw & \qw & \qw & \qw & \targ & \targ & \targ & \qw & \qw\\
\\ }}

\textbf{The $\mathbf{|+\rangle_L}$ state of the $\mathbf{[[15,1,3]]}$ Reed-Muller code}. The circuit found with the LSP + VCS approach has 31 two-qubit gates and 1 flag qubit:

\noindent \scalebox{0.9}{
\Qcircuit @C=0.1em @R=0.1em @!R { \\
	 	\nghost{{q}_{0} :  } & \lstick{{q}_{0} :  } & \qw & \targ & \qw & \qw & \qw & \qw & \qw & \ctrl{1} & \ctrl{2} & \qw & \qw & \qw & \qw & \qw & \qw \barrier[0em]{15} & \qw & \qw & \qw & \qw & \qw & \targ & \qw & \qw & \qw & \qw & \qw & \qw\\
	 	\nghost{{q}_{1} :  } & \lstick{{q}_{1} :  } & \qw & \qw & \qw & \qw & \targ & \qw & \ctrl{12} & \targ & \qw & \qw & \qw & \ctrl{3} & \qw & \qw & \qw & \qw & \qw & \qw & \qw & \qw & \qw & \qw & \qw & \qw & \qw & \qw & \qw\\
	 	\nghost{{q}_{2} :  } & \lstick{{q}_{2} :  } & \gate{\mathrm{H}} & \qw & \qw & \ctrl{12} & \qw & \qw & \qw & \qw & \targ & \ctrl{2} & \qw & \qw & \qw & \qw & \qw & \qw & \qw & \qw & \qw & \qw & \qw & \qw & \qw & \qw & \qw & \qw & \qw\\
	 	\nghost{{q}_{3} :  } & \lstick{{q}_{3} :  } & \qw & \qw & \qw & \qw & \qw & \targ & \qw & \qw & \qw & \qw & \qw & \qw & \qw & \qw & \qw & \qw & \qw & \qw & \qw & \qw & \qw & \qw & \qw & \targ & \qw & \qw & \qw\\
	 	\nghost{{q}_{4} :  } & \lstick{{q}_{4} :  } & \gate{\mathrm{H}} & \qw & \ctrl{8} & \qw & \qw & \ctrl{-1} & \qw & \qw & \qw & \targ & \ctrl{1} & \targ & \qw & \qw & \qw & \qw & \qw & \qw & \qw & \qw & \qw & \targ & \qw & \qw & \qw & \qw & \qw\\
	 	\nghost{{q}_{5} :  } & \lstick{{q}_{5} :  } & \gate{\mathrm{H}} & \ctrl{-5} & \qw & \qw & \qw & \ctrl{5} & \qw & \qw & \qw & \qw & \targ & \qw & \qw & \qw & \qw & \qw & \qw & \qw & \qw & \qw & \qw & \qw & \qw & \qw & \qw & \qw & \qw\\
	 	\nghost{{q}_{6} :  } & \lstick{{q}_{6} :  } & \gate{\mathrm{H}} & \qw & \qw & \qw & \qw & \qw & \qw & \qw & \ctrl{4} & \qw & \ctrl{6} & \qw & \qw & \targ & \qw & \qw & \qw & \qw & \qw & \qw & \qw & \qw & \qw & \qw & \qw & \qw & \qw\\
	 	\nghost{{q}_{7} :  } & \lstick{{q}_{7} :  } & \qw & \qw & \qw & \qw & \qw & \qw & \qw & \qw & \qw & \qw & \qw & \targ & \ctrl{2} & \qw & \qw & \qw & \qw & \qw & \qw & \qw & \qw & \qw & \qw & \qw & \qw & \qw & \qw\\
	 	\nghost{{q}_{8} :  } & \lstick{{q}_{8} :  } & \qw & \qw & \qw & \qw & \qw & \qw & \qw & \qw & \qw & \targ & \qw & \qw & \qw & \qw & \targ & \qw & \qw & \qw & \qw & \qw & \qw & \qw & \qw & \qw & \qw & \qw & \qw\\
	 	\nghost{{q}_{9} :  } & \lstick{{q}_{9} :  } & \qw & \qw & \qw & \qw & \qw & \qw & \qw & \targ & \qw & \ctrl{-1} & \qw & \qw & \targ & \qw & \qw & \qw & \qw & \qw & \qw & \qw & \qw & \qw & \targ & \qw & \qw & \qw & \qw\\
	 	\nghost{{q}_{10} :  } & \lstick{{q}_{10} :  } & \qw & \qw & \qw & \qw & \qw & \targ & \qw & \qw & \targ & \qw & \qw & \ctrl{-3} & \targ & \qw & \ctrl{-2} & \qw & \qw & \qw & \qw & \targ & \qw & \qw & \qw & \qw & \qw & \qw & \qw\\
	 	\nghost{{q}_{11} :  } & \lstick{{q}_{11} :  } & \gate{\mathrm{H}} & \qw & \qw & \qw & \ctrl{-10} & \qw & \qw & \qw & \qw & \qw & \qw & \qw & \ctrl{-1} & \qw & \targ & \qw & \qw & \qw & \qw & \qw & \qw & \qw & \qw & \qw & \qw & \qw & \qw\\
	 	\nghost{{q}_{12} :  } & \lstick{{q}_{12} :  } & \qw & \qw & \targ & \qw & \qw & \qw & \qw & \qw & \qw & \qw & \targ & \ctrl{1} & \ctrl{2} & \qw & \qw & \qw & \qw & \qw & \qw & \qw & \qw & \qw & \qw & \qw & \qw & \qw & \qw\\
	 	\nghost{{q}_{13} :  } & \lstick{{q}_{13} :  } & \qw & \qw & \qw & \qw & \qw & \qw & \targ & \qw & \qw & \qw & \qw & \targ & \qw & \ctrl{-7} & \qw & \qw & \qw & \targ & \qw & \qw & \qw & \qw & \qw & \qw & \qw & \qw & \qw\\
	 	\nghost{{q}_{14} :  } & \lstick{{q}_{14} :  } & \qw & \qw & \qw & \targ & \qw & \qw & \qw & \ctrl{-5} & \qw & \qw & \qw & \qw & \targ & \qw & \ctrl{-3} & \qw & \qw & \qw & \targ & \qw & \qw & \qw & \qw & \qw & \qw & \qw & \qw\\
	 	\nghost{{q}_{15} :  } & \lstick{{q}_{15} :  } & \qw & \qw & \qw & \qw & \qw & \qw & \qw & \qw & \qw & \qw & \qw & \qw & \qw & \qw & \qw & \qw & \gate{\mathrm{H}} & \ctrl{-2} & \ctrl{-1} & \ctrl{-5} & \ctrl{-15} & \ctrl{-11} & \ctrl{-6} & \ctrl{-12} & \gate{\mathrm{H}} & \qw & \qw\\
\\ }}

The circuit found with the IFT-LSP approach requires also 31 two-qubit gates and 1 flag qubit:

\noindent \scalebox{0.9}{
\Qcircuit @C=0.1em @R=0.1em @!R { \\
\nghost{{q}_{0} : }	 & \lstick{{q}_{0} : }	 & \qw	 & \targ	 & \qw	 & \qw	 & \qw	 & \qw	 & \qw	 & \qw	 & \qw	 & \targ	 & \qw	 & \qw	 & \ctrl{1}	 & \qw	 & \qw	 & \qw	 & \qw	 & \qw	 & \qw	 & \qw	 & \qw	 & \qw	 & \qw	 & \qw	 & \qw\\
\nghost{{q}_{1} : }	 & \lstick{{q}_{1} : }	 & \qw	 & \qw	 & \targ	 & \qw	 & \qw	 & \qw	 & \ctrl{12}	 & \qw	 & \qw	 & \qw	 & \qw	 & \ctrl{7}	 & \targ	 & \ctrl{3}	 & \qw	 & \qw	 & \qw	 & \targ	 & \qw	 & \qw	 & \qw	 & \qw	 & \qw	 & \qw	 & \qw\\
\nghost{{q}_{2} : }	 & \lstick{{q}_{2} : }	 & \gate{\mathrm{H}}	 & \qw	 & \qw	 & \ctrl{12}	 & \qw	 & \qw	 & \qw	 & \qw	 & \targ	 & \qw	 & \ctrl{3}	 & \qw	 & \ctrl{2}	 & \qw	 & \qw	 & \qw	 & \qw	 & \qw	 & \qw	 & \qw	 & \qw	 & \qw	 & \qw	 & \qw	 & \qw\\
\nghost{{q}_{3} : }	 & \lstick{{q}_{3} : }	 & \qw	 & \qw	 & \qw	 & \qw	 & \qw	 & \qw	 & \qw	 & \targ	 & \qw	 & \ctrl{-3}	 & \qw	 & \qw	 & \qw	 & \qw	 & \qw	 & \qw	 & \qw	 & \qw	 & \qw	 & \targ	 & \qw	 & \qw	 & \qw	 & \qw	 & \qw\\
\nghost{{q}_{4} : }	 & \lstick{{q}_{4} : }	 & \gate{\mathrm{H}}	 & \qw	 & \qw	 & \qw	 & \qw	 & \ctrl{2}	 & \qw	 & \ctrl{-1}	 & \qw	 & \qw	 & \qw	 & \qw	 & \targ	 & \targ	 & \qw	 & \qw	 & \qw	 & \qw	 & \qw	 & \qw	 & \qw	 & \qw	 & \qw	 & \qw	 & \qw\\
\nghost{{q}_{5} : }	 & \lstick{{q}_{5} : }	 & \gate{\mathrm{H}}	 & \ctrl{-5}	 & \qw	 & \qw	 & \ctrl{5}	 & \qw	 & \qw	 & \qw	 & \qw	 & \qw	 & \targ	 & \qw	 & \qw	 & \qw	 & \qw	 & \targ	 & \qw	 & \qw	 & \qw	 & \qw	 & \qw	 & \qw	 & \qw	 & \qw	 & \qw\\
\nghost{{q}_{6} : }	 & \lstick{{q}_{6} : }	 & \gate{\mathrm{H}}	 & \ctrl{6}	 & \qw	 & \qw	 & \qw	 & \targ	 & \qw	 & \qw	 & \ctrl{-4}	 & \qw	 & \qw	 & \qw	 & \qw	 & \qw	 & \targ	 & \qw	 & \qw	 & \qw	 & \qw	 & \qw	 & \qw	 & \qw	 & \qw	 & \qw	 & \qw\\
\nghost{{q}_{7} : }	 & \lstick{{q}_{7} : }	 & \qw	 & \qw	 & \qw	 & \qw	 & \qw	 & \qw	 & \qw	 & \qw	 & \qw	 & \qw	 & \targ	 & \qw	 & \targ	 & \qw	 & \qw	 & \qw	 & \qw	 & \qw	 & \qw	 & \qw	 & \targ	 & \qw	 & \qw	 & \qw	 & \qw\\
\nghost{{q}_{8} : }	 & \lstick{{q}_{8} : }	 & \qw	 & \qw	 & \qw	 & \qw	 & \qw	 & \qw	 & \qw	 & \targ	 & \qw	 & \qw	 & \qw	 & \targ	 & \qw	 & \qw	 & \qw	 & \qw	 & \qw	 & \qw	 & \qw	 & \qw	 & \qw	 & \qw	 & \qw	 & \qw	 & \qw\\
\nghost{{q}_{9} : }	 & \lstick{{q}_{9} : }	 & \qw	 & \qw	 & \qw	 & \qw	 & \qw	 & \targ	 & \qw	 & \ctrl{-1}	 & \qw	 & \qw	 & \qw	 & \qw	 & \qw	 & \qw	 & \qw	 & \qw	 & \qw	 & \qw	 & \qw	 & \qw	 & \qw	 & \targ	 & \qw	 & \qw	 & \qw\\
\nghost{{q}_{10} : }	 & \lstick{{q}_{10} : }	 & \qw	 & \qw	 & \qw	 & \qw	 & \targ	 & \qw	 & \qw	 & \ctrl{4}	 & \qw	 & \qw	 & \qw	 & \qw	 & \ctrl{-3}	 & \targ	 & \qw	 & \qw	 & \qw	 & \qw	 & \qw	 & \qw	 & \qw	 & \qw	 & \qw	 & \qw	 & \qw\\
\nghost{{q}_{11} : }	 & \lstick{{q}_{11} : }	 & \gate{\mathrm{H}}	 & \qw	 & \ctrl{-10}	 & \qw	 & \qw	 & \qw	 & \qw	 & \qw	 & \qw	 & \targ	 & \qw	 & \qw	 & \qw	 & \qw	 & \qw	 & \qw	 & \qw	 & \qw	 & \targ	 & \qw	 & \qw	 & \qw	 & \qw	 & \qw	 & \qw\\
\nghost{{q}_{12} : }	 & \lstick{{q}_{12} : }	 & \qw	 & \targ	 & \qw	 & \qw	 & \qw	 & \qw	 & \qw	 & \qw	 & \ctrl{1}	 & \qw	 & \ctrl{-5}	 & \qw	 & \qw	 & \qw	 & \qw	 & \qw	 & \qw	 & \qw	 & \qw	 & \qw	 & \qw	 & \qw	 & \qw	 & \qw	 & \qw\\
\nghost{{q}_{13} : }	 & \lstick{{q}_{13} : }	 & \qw	 & \qw	 & \qw	 & \qw	 & \qw	 & \qw	 & \targ	 & \qw	 & \targ	 & \qw	 & \qw	 & \qw	 & \qw	 & \ctrl{-3}	 & \ctrl{-7}	 & \qw	 & \targ	 & \qw	 & \qw	 & \qw	 & \qw	 & \qw	 & \qw	 & \qw	 & \qw\\
\nghost{{q}_{14} : }	 & \lstick{{q}_{14} : }	 & \qw	 & \qw	 & \qw	 & \targ	 & \qw	 & \ctrl{-5}	 & \qw	 & \targ	 & \qw	 & \ctrl{-3}	 & \qw	 & \qw	 & \qw	 & \qw	 & \qw	 & \qw	 & \qw	 & \qw	 & \qw	 & \qw	 & \qw	 & \qw	 & \qw	 & \qw	 & \qw\\
\nghost{{q}_{15} : }	 & \lstick{{q}_{15} : }	 & \qw	 & \qw	 & \qw	 & \qw	 & \qw	 & \qw	 & \qw	 & \qw	 & \qw	 & \qw	 & \qw	 & \qw	 & \qw	 & \qw	 & \gate{\mathrm{H}}	 & \ctrl{-10}	 & \ctrl{-2}	 & \ctrl{-14}	 & \ctrl{-4}	 & \ctrl{-12}	 & \ctrl{-8}	 & \ctrl{-6}	 & \gate{\mathrm{H}}	 & \qw	 & \qw\\
\\ }}

\section{Examples of when LSP+VCS fails with restricted connectivity}
\label{app:ftlsp-fail-cases}

\begin{figure}[htb]
	\includegraphics[width=.4\textwidth]{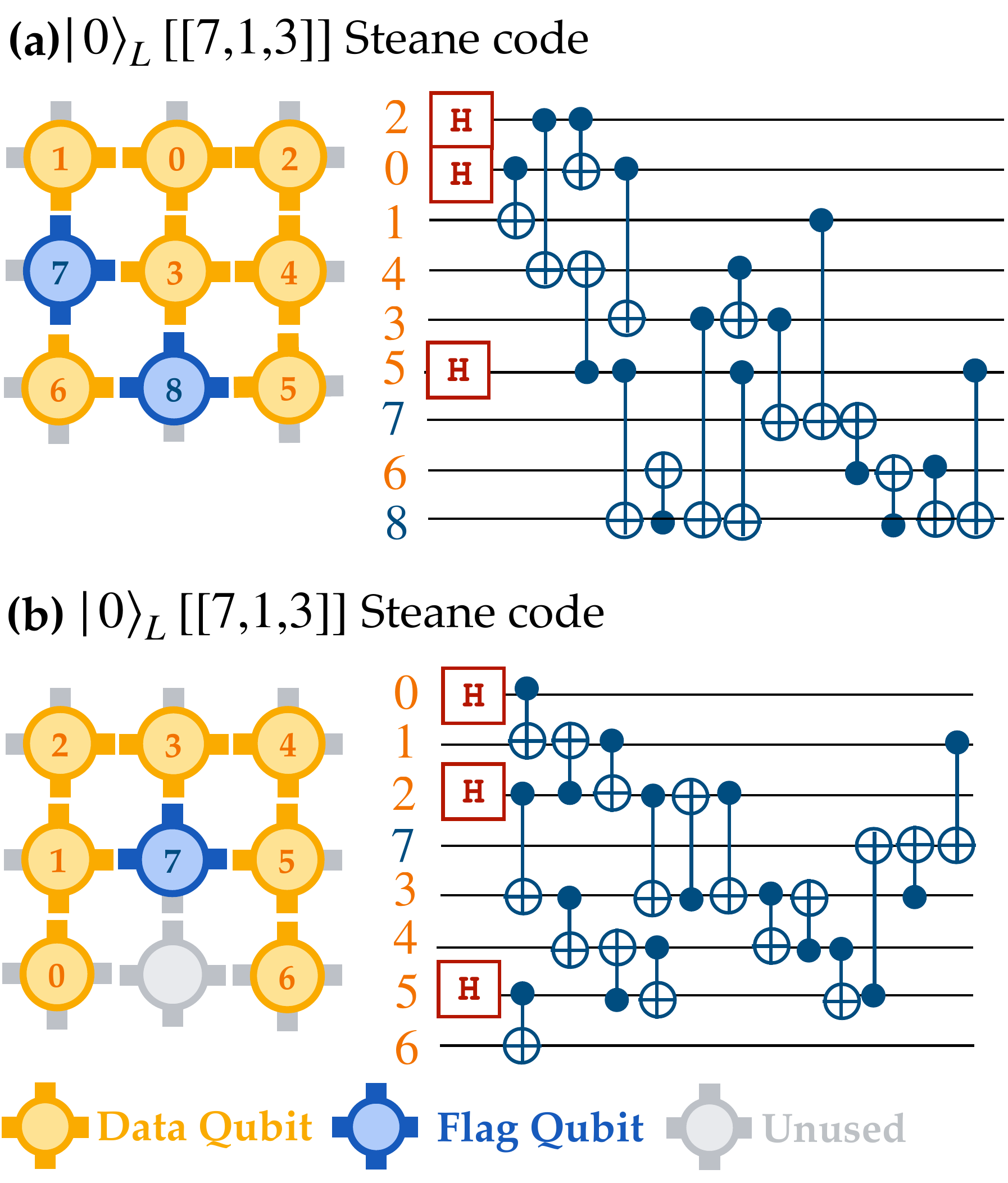}
	\caption{\label{fig:ftlsp-fail-cases}  Two cases where the LSP+VCS approach fails, but the IFT-LSP approach succeeds with restricted connectivity. We use the IFT-LSP approach to fault-tolerantly prepare the $|0\rangle_L$ state of the $[[7,1,3]]$ Steane code on a 2D grid. (a) A case where the data qubit is only connected to the flag qubits.  (b) A case where VCS does not find a verification circuit.   }
\end{figure}

Here, we discuss two cases where logical state preparation (LSP) followed by verification circuit synthesis (VCS) to prepare a fault-tolerant logical state would fail in restricted connectivity settings. Fig.~\ref{fig:ftlsp-fail-cases}(a) shows a case where a data qubit (qubit 6 in Fig.~\ref{fig:ftlsp-fail-cases}(a)) is only connected to the flag qubits, and the circuit on the right is the output of the integrated fault-tolerant logical state preparation (IFT-LSP) task. If we separate the task, the logical state preparation task fails to prepare the state in this case. Fig.~\ref{fig:ftlsp-fail-cases}(b) shows a case where if we separate the task, then the verification circuit synthesis fails, while the circuit on the right is the RL-prepared circuit with the fault-tolerant logical state preparation task. If we separate the task, then the state preparation does not know where the ancilla is. Therefore, the verification circuit synthesis fails to flag all of the harmful errors.

\section{Logical error and acceptance rate of circuits found by integrated fault-tolerant logical state preparation}
\label{app:ler-ar-circuits-connectivity}

\begin{figure}[htb]
	\includegraphics[width=.49\textwidth]{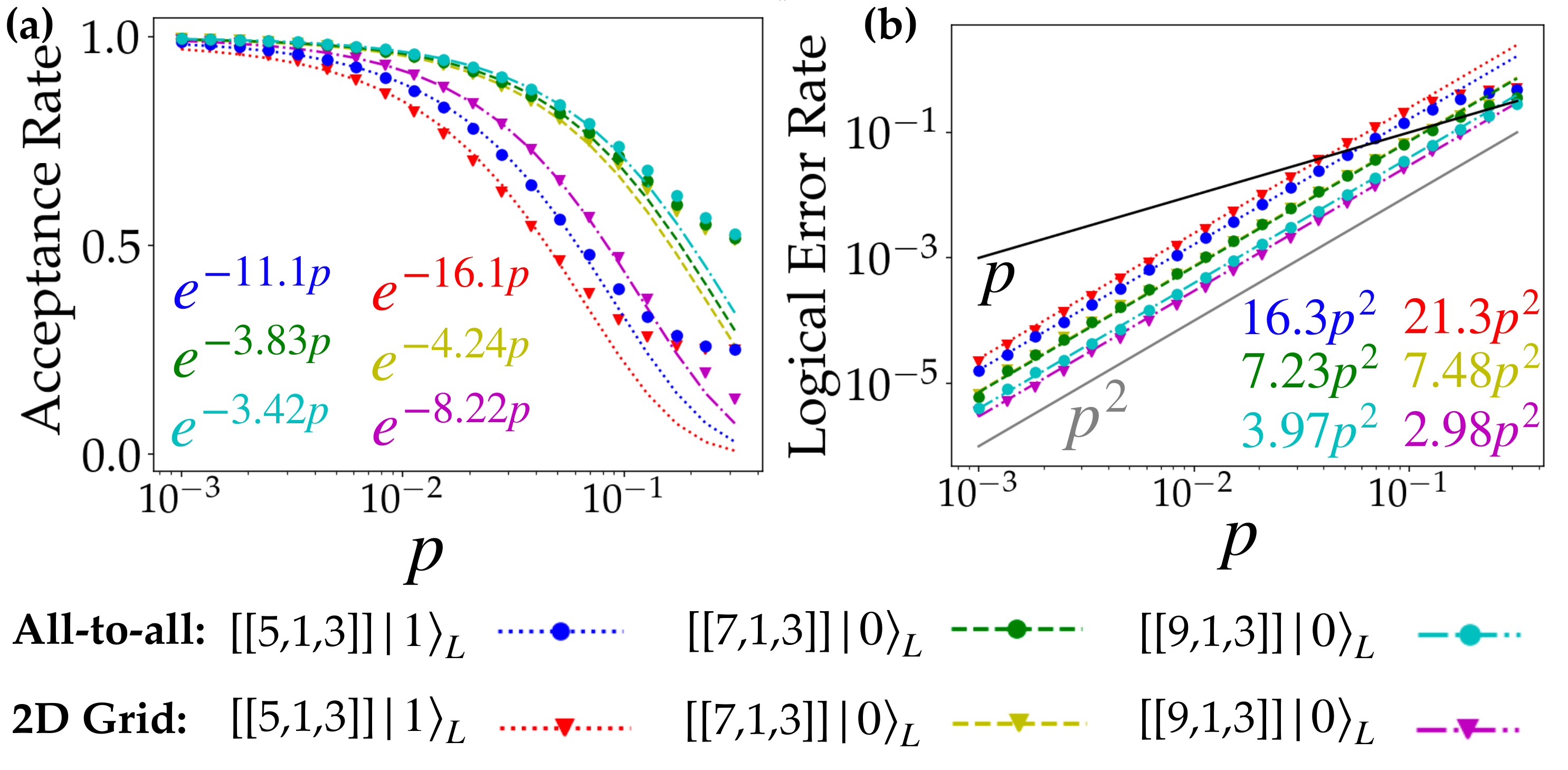}
	\caption{\label{fig:ler-ar-iftlsp} Logical error rate and acceptance rate of circuits found with integrated fault-tolerant logical state preparation. We take the circuits shown in Fig.~\ref{fig:result-ft-logical}. The circuits for $[[9,1,3]]$ Surface-17 in all-to-all qubit connectivity is shown in Appendix~\ref{app:ftlsp-circuits}.  }
\end{figure}

When using a restricted connectivity, we would expect a trade-off in the state acceptance and logical error rate compared to using all-to-all qubit connectivity.  In Fig.~\ref{fig:ler-ar-iftlsp}(a) and (b), we quantify and compare these two rates of circuits shown in Fig.~\ref{fig:result-ft-logical}. We see that the state acceptance rate of the circuit to prepare the $|0\rangle_L$ state of the $[[7,1,3]]$ Steane code on a 2D grid is only marginally lower than the one for the circuits with all-to-all qubit connectivity, but both are higher than the acceptance rate of the verification circuit synthesis task in Fig.~\ref{fig:result-flag-fully-connected}(c). Similarly, the logical error rate is only marginally higher. In the case of the $|1\rangle_L$ for the $[[5,1,3]]$ perfect code, the circuit with restricted connectivity requires 5 two-qubit gates more than the circuit shown with full connectivity. Nevertheless, the logical error rate is larger only by approximately 25\%. Interestingly, for the $|0\rangle_L$ of the $[[9,1,3]]$ Surface-17 code, the logical error rate is higher in the 2D grid connectivity than in the all-to-all qubit connectivity, but the acceptance rate is lower. We suggest that this has to do with the number of flag qubits. In the 2D grid, we need $4$ flag qubits, while we only need $1$ flag qubits in the all-to-all qubit connectivity case. We have also observed similar phenomenon in Appendix~\ref{app:vcs-circuits}, where using more flag qubits decrease the acceptance rate but increase the logical error rate. The relationship between the number of flag qubits, number of gates, acceptance, and logical error rate of different circuits is an interesting avenue for further study, but beyond the scope of this paper. Nevertheless, in all cases the logical error rate scales as $p^2$, confirming that the circuits are fault-tolerant, as desired.

\section{Circuits for fault-tolerant logical state preparation in restricted connectivity}
\label{app:ftlsp-circuits-connectivity}

Here, we show examples of fault-tolerant logical state preparation circuits with restricted connectivity as shown in Fig.~\ref{fig:ft-logical-connectivity}. If the qubit placement is given as $a,b,c,\dots$, then it means that qubit $0$ ($q_0$) in the circuit is placed in qubit $a$ on the device, qubit $1$ ($q_1$)  is placed in qubit $b$ on the device, and so on. These circuits are also available online~\footnotemark[1]. 

\subsection{The $|0\rangle_L$ state of the $[[7,1,3]]$ Steane code on 2D grid connectivity (Google Sycamore)}

For the qubit placement, we follow the row-major order starting with $0$ at the top left of the qubit and $9$ at the bottom right of the qubit. The last two qubits are always given as ancilla qubits. The RL agent can decide whether to use them or not.

The circuit for the first qubit placement shown in Fig.~\ref{fig:ft-logical-connectivity}(a) with $11$ two-qubit gates is already shown in Fig.~\ref{fig:result-ft-logical}(c).

The circuit for the second qubit placement shown in Fig.~\ref{fig:ft-logical-connectivity}(a) with $12$ two-qubit gates.  The qubit placement is: $2,5,6,1,0,4,7,3,8$.

\noindent \scalebox{1.0}{
\Qcircuit @C=0.1em @R=0.1em @!R { \\
	 	\nghost{{q}_{0} :  } & \lstick{{q}_{0} :  } & \gate{\mathrm{H}} & \ctrl{3} & \qw & \qw & \qw & \qw & \qw & \qw & \qw & \qw & \qw & \qw & \qw\\
	 	\nghost{{q}_{1} :  } & \lstick{{q}_{1} :  } & \qw & \qw & \qw & \qw & \targ & \qw & \qw & \qw & \qw & \qw & \qw & \qw & \qw\\
	 	\nghost{{q}_{2} :  } & \lstick{{q}_{2} :  } & \qw & \qw & \qw & \targ & \qw & \qw & \qw & \qw & \qw & \ctrl{5} & \qw & \qw & \qw\\
	 	\nghost{{q}_{3} :  } & \lstick{{q}_{3} :  } & \qw & \targ & \ctrl{2} & \qw & \qw & \targ & \ctrl{2} & \qw & \qw & \qw & \qw & \qw & \qw\\
	 	\nghost{{q}_{4} :  } & \lstick{{q}_{4} :  } & \gate{\mathrm{H}} & \qw & \qw & \qw & \qw & \ctrl{-1} & \qw & \qw & \qw & \qw & \ctrl{3} & \qw & \qw\\
	 	\nghost{{q}_{5} :  } & \lstick{{q}_{5} :  } & \gate{\mathrm{H}} & \ctrl{1} & \targ & \qw & \ctrl{-4} & \ctrl{1} & \targ & \ctrl{1} & \ctrl{2} & \qw & \qw & \qw & \qw\\
	 	\nghost{{q}_{6} :  } & \lstick{{q}_{6} :  } & \qw & \targ & \qw & \ctrl{-4} & \qw & \targ & \qw & \targ & \qw & \qw & \qw & \qw & \qw\\
	 	\nghost{{q}_{7} :  } & \lstick{{q}_{7} :  } & \qw & \qw & \qw & \qw & \qw & \qw & \qw & \qw & \targ & \targ & \targ & \qw & \qw\\
	 	\nghost{{q}_{8} :  } & \lstick{{q}_{8} :  } & \qw & \qw & \qw & \qw & \qw & \qw & \qw & \qw & \qw & \qw & \qw & \qw & \qw\\
\\ }}

The circuit for the third qubit placement shown in Fig.~\ref{fig:ft-logical-connectivity}(a) with $13$ two-qubit gates.  The qubit placement is: $2,5,8,1,0,4,3,7,6$.

\noindent \scalebox{1.0}{
\Qcircuit @C=0.1em @R=0.1em @!R { \\
	 	\nghost{{q}_{0} :  } & \lstick{{q}_{0} :  } & \gate{\mathrm{H}} & \ctrl{1} & \qw & \ctrl{3} & \qw & \qw & \qw & \qw & \qw & \qw & \qw & \qw & \qw & \qw\\
	 	\nghost{{q}_{1} :  } & \lstick{{q}_{1} :  } & \qw & \targ & \targ & \qw & \qw & \ctrl{4} & \qw & \qw & \qw & \qw & \qw & \qw & \qw & \qw\\
	 	\nghost{{q}_{2} :  } & \lstick{{q}_{2} :  } & \gate{\mathrm{H}} & \qw & \ctrl{-1} & \qw & \qw & \qw & \qw & \qw & \qw & \qw & \qw & \ctrl{5} & \qw & \qw\\
	 	\nghost{{q}_{3} :  } & \lstick{{q}_{3} :  } & \qw & \qw & \qw & \targ & \ctrl{1} & \qw & \targ & \ctrl{1} & \ctrl{2} & \qw & \qw & \qw & \qw & \qw\\
	 	\nghost{{q}_{4} :  } & \lstick{{q}_{4} :  } & \qw & \qw & \qw & \qw & \targ & \qw & \qw & \targ & \qw & \qw & \qw & \qw & \qw & \qw\\
	 	\nghost{{q}_{5} :  } & \lstick{{q}_{5} :  } & \gate{\mathrm{H}} & \ctrl{1} & \qw & \qw & \qw & \targ & \ctrl{-2} & \ctrl{2} & \targ & \targ & \ctrl{2} & \qw & \qw & \qw\\
	 	\nghost{{q}_{6} :  } & \lstick{{q}_{6} :  } & \qw & \targ & \qw & \qw & \qw & \qw & \qw & \qw & \qw & \ctrl{-1} & \qw & \qw & \qw & \qw\\
	 	\nghost{{q}_{7} :  } & \lstick{{q}_{7} :  } & \qw & \qw & \qw & \qw & \qw & \qw & \qw & \targ & \qw & \qw & \targ & \targ & \qw & \qw\\
	 	\nghost{{q}_{8} :  } & \lstick{{q}_{8} :  } & \qw & \qw & \qw & \qw & \qw & \qw & \qw & \qw & \qw & \qw & \qw & \qw & \qw & \qw\\
\\ }}

The circuit for the fourth qubit placement shown in Fig.~\ref{fig:ft-logical-connectivity}(a) with $14$ two-qubit gates. The qubit placement is: $0,2,8,6,4,1,7,5,3$.

\noindent  \scalebox{1.0}{
\Qcircuit @C=0.1em @R=0.1em @!R { \\
	 	\nghost{{q}_{0} :  } & \lstick{{q}_{0} :  } & \gate{\mathrm{H}} & \qw & \qw & \qw & \ctrl{5} & \qw & \qw & \ctrl{8} & \qw & \qw & \qw & \qw & \qw & \qw & \qw & \qw & \qw\\
	 	\nghost{{q}_{1} :  } & \lstick{{q}_{1} :  } & \qw & \qw & \qw & \qw & \qw & \qw & \targ & \qw & \qw & \qw & \qw & \qw & \qw & \qw & \qw & \qw & \qw\\
	 	\nghost{{q}_{2} :  } & \lstick{{q}_{2} :  } & \gate{\mathrm{H}} & \ctrl{4} & \qw & \qw & \qw & \qw & \qw & \qw & \targ & \qw & \qw & \qw & \qw & \qw & \qw & \qw & \qw\\
	 	\nghost{{q}_{3} :  } & \lstick{{q}_{3} :  } & \qw & \qw & \qw & \targ & \qw & \qw & \qw & \qw & \qw & \qw & \qw & \qw & \qw & \qw & \ctrl{5} & \qw & \qw\\
	 	\nghost{{q}_{4} :  } & \lstick{{q}_{4} :  } & \qw & \qw & \targ & \qw & \qw & \ctrl{2} & \qw & \qw & \qw & \targ & \ctrl{1} & \targ & \targ & \ctrl{4} & \qw & \qw & \qw\\
	 	\nghost{{q}_{5} :  } & \lstick{{q}_{5} :  } & \gate{\mathrm{H}} & \qw & \ctrl{-1} & \qw & \targ & \qw & \ctrl{-4} & \qw & \qw & \qw & \targ & \qw & \ctrl{-1} & \qw & \qw & \qw & \qw\\
	 	\nghost{{q}_{6} :  } & \lstick{{q}_{6} :  } & \qw & \targ & \qw & \ctrl{-3} & \qw & \targ & \qw & \qw & \ctrl{-4} & \ctrl{-2} & \qw & \ctrl{-2} & \qw & \qw & \qw & \qw & \qw\\
	 	\nghost{{q}_{7} :  } & \lstick{{q}_{7} :  } & \qw & \qw & \qw & \qw & \qw & \qw & \qw & \qw & \qw & \qw & \qw & \qw & \qw & \qw & \qw & \qw & \qw\\
	 	\nghost{{q}_{8} :  } & \lstick{{q}_{8} :  } & \qw & \qw & \qw & \qw & \qw & \qw & \qw & \targ & \qw & \qw & \qw & \qw & \qw & \targ & \targ & \qw & \qw\\
\\ }}

The circuit for the fifth qubit placement shown in Fig.~\ref{fig:ft-logical-connectivity}(a) with $14$ two-qubit gates. The qubit placement is: $2,5,0,4,3,6,7,1,8$.

\noindent \scalebox{1.0}{
\Qcircuit @C=0.1em @R=0.1em @!R { \\
	 	\nghost{{q}_{0} :  } & \lstick{{q}_{0} :  } & \qw & \targ & \qw & \qw & \qw & \qw & \qw & \qw & \qw & \qw & \qw & \qw & \qw\\
	 	\nghost{{q}_{1} :  } & \lstick{{q}_{1} :  } & \gate{\mathrm{H}} & \ctrl{-1} & \targ & \qw & \ctrl{2} & \qw & \qw & \qw & \qw & \qw & \ctrl{7} & \qw & \qw\\
	 	\nghost{{q}_{2} :  } & \lstick{{q}_{2} :  } & \qw & \qw & \qw & \targ & \qw & \qw & \qw & \qw & \qw & \qw & \qw & \qw & \qw\\
	 	\nghost{{q}_{3} :  } & \lstick{{q}_{3} :  } & \qw & \targ & \ctrl{-2} & \qw & \targ & \targ & \qw & \ctrl{3} & \qw & \qw & \qw & \qw & \qw\\
	 	\nghost{{q}_{4} :  } & \lstick{{q}_{4} :  } & \gate{\mathrm{H}} & \ctrl{-1} & \qw & \ctrl{-2} & \ctrl{1} & \qw & \targ & \qw & \qw & \qw & \qw & \qw & \qw\\
	 	\nghost{{q}_{5} :  } & \lstick{{q}_{5} :  } & \gate{\mathrm{H}} & \ctrl{1} & \qw & \qw & \targ & \qw & \ctrl{-1} & \qw & \ctrl{1} & \qw & \qw & \qw & \qw\\
	 	\nghost{{q}_{6} :  } & \lstick{{q}_{6} :  } & \qw & \targ & \qw & \qw & \qw & \ctrl{-3} & \ctrl{2} & \targ & \targ & \ctrl{2} & \qw & \qw & \qw\\
	 	\nghost{{q}_{7} :  } & \lstick{{q}_{7} :  } & \qw & \qw & \qw & \qw & \qw & \qw & \qw & \qw & \qw & \qw & \qw & \qw & \qw\\
	 	\nghost{{q}_{8} :  } & \lstick{{q}_{8} :  } & \qw & \qw & \qw & \qw & \qw & \qw & \targ & \qw & \qw & \targ & \targ & \qw & \qw\\
\\ }}

\subsection{The $|+\rangle_L$ state of the $[[7,1,3]]$ Steane code on 2D grid connectivity (Google Sycamore)} 

The qubit placement for the circuit below is: 4,3,8,2,7,0,5,1,6.

\scalebox{1.0}{
\Qcircuit @C=0.2em @R=0.2em @!R { \\
	 	\nghost{{q}_{0} :  } & \lstick{{q}_{0} :  } & \gate{\mathrm{H}} & \qw & \ctrl{1} & \ctrl{6} & \qw & \targ & \targ & \targ & \ctrl{6} & \ctrl{1} & \targ & \targ & \qw & \qw & \qw\\
	 	\nghost{{q}_{1} :  } & \lstick{{q}_{1} :  } & \qw & \targ & \targ & \qw & \qw & \qw & \qw & \ctrl{-1} & \qw & \targ & \ctrl{-1} & \qw & \qw & \qw & \qw\\
	 	\nghost{{q}_{2} :  } & \lstick{{q}_{2} :  } & \gate{\mathrm{H}} & \qw & \ctrl{4} & \qw & \qw & \qw & \qw & \qw & \qw & \qw & \qw & \qw & \qw & \qw & \qw\\
	 	\nghost{{q}_{3} :  } & \lstick{{q}_{3} :  } & \qw & \qw & \qw & \qw & \targ & \qw & \qw & \qw & \qw & \qw & \qw & \qw & \targ & \qw & \qw\\
	 	\nghost{{q}_{4} :  } & \lstick{{q}_{4} :  } & \gate{\mathrm{H}} & \qw & \qw & \qw & \qw & \ctrl{-4} & \qw & \qw & \qw & \qw & \qw & \qw & \qw & \qw & \qw\\
	 	\nghost{{q}_{5} :  } & \lstick{{q}_{5} :  } & \gate{\mathrm{H}} & \ctrl{-4} & \qw & \qw & \qw & \qw & \qw & \qw & \qw & \qw & \qw & \qw & \qw & \qw & \qw\\
	 	\nghost{{q}_{6} :  } & \lstick{{q}_{6} :  } & \qw & \qw & \targ & \targ & \ctrl{-3} & \qw & \qw & \qw & \targ & \qw & \qw & \qw & \qw & \qw & \qw\\
	 	\nghost{{q}_{7} :  } & \lstick{{q}_{7} :  } & \qw & \qw & \qw & \qw & \qw & \gate{\mathrm{H}} & \ctrl{-7} & \qw & \qw & \qw & \qw & \ctrl{-7} & \ctrl{-4} & \gate{\mathrm{H}} & \qw\\
	 	\nghost{{q}_{8} :  } & \lstick{{q}_{8} :  } & \qw & \qw & \qw & \qw & \qw & \qw & \qw & \qw & \qw & \qw & \qw & \qw & \qw & \qw & \qw\\
\\ }}

The qubit placement for the circuit below is: 5,4,7,6,1,0,8,2,3.

\scalebox{1.0}{
\Qcircuit @C=0.2em @R=0.2em @!R { \\
	 	\nghost{{q}_{0} :  } & \lstick{{q}_{0} :  } & \gate{\mathrm{H}} & \qw & \ctrl{1} & \qw & \ctrl{6} & \targ & \qw & \qw & \targ & \qw & \qw & \qw & \qw\\
	 	\nghost{{q}_{1} :  } & \lstick{{q}_{1} :  } & \qw & \targ & \targ & \ctrl{1} & \qw & \ctrl{-1} & \targ & \targ & \ctrl{-1} & \targ & \qw & \qw & \qw\\
	 	\nghost{{q}_{2} :  } & \lstick{{q}_{2} :  } & \gate{\mathrm{H}} & \qw & \ctrl{1} & \targ & \qw & \ctrl{1} & \qw & \ctrl{-1} & \qw & \qw & \qw & \qw & \qw\\
	 	\nghost{{q}_{3} :  } & \lstick{{q}_{3} :  } & \qw & \qw & \targ & \qw & \qw & \targ & \qw & \targ & \qw & \qw & \qw & \qw & \qw\\
	 	\nghost{{q}_{4} :  } & \lstick{{q}_{4} :  } & \gate{\mathrm{H}} & \ctrl{-3} & \targ & \qw & \qw & \qw & \ctrl{-3} & \qw & \qw & \qw & \qw & \qw & \qw\\
	 	\nghost{{q}_{5} :  } & \lstick{{q}_{5} :  } & \gate{\mathrm{H}} & \qw & \ctrl{-1} & \qw & \qw & \qw & \qw & \qw & \qw & \qw & \targ & \qw & \qw\\
	 	\nghost{{q}_{6} :  } & \lstick{{q}_{6} :  } & \qw & \qw & \qw & \qw & \targ & \qw & \qw & \qw & \qw & \qw & \qw & \qw & \qw\\
	 	\nghost{{q}_{7} :  } & \lstick{{q}_{7} :  } & \qw & \qw & \qw & \qw & \qw & \qw & \qw & \qw & \qw & \qw & \qw & \qw & \qw\\
	 	\nghost{{q}_{8} :  } & \lstick{{q}_{8} :  } & \qw & \qw & \qw & \qw & \qw & \qw & \gate{\mathrm{H}} & \ctrl{-5} & \qw & \ctrl{-7} & \ctrl{-3} & \gate{\mathrm{H}} & \qw\\
\\ }}

\subsection{The $|0\rangle_L$ state of the $[[7,1,3]]$ Steane code on heavy-hex connectivity (IBMQ Guadalupe)}

The circuit for the first qubit placements shown in Fig.~\ref{fig:ft-logical-connectivity}(b) has $22$ two-qubit gates. The qubit placement in the IBMQ Guadalupe~\cite{guadalupe} device is: $3,0,6,12,4,2,10,1,7$. The last two qubits are given as ancilla qubits.

\noindent \scalebox{1.0}{
\Qcircuit @C=0.05em @R=0.05em @!R { \\
	 	\nghost{{q}_{0} :  } & \lstick{{q}_{0} :  } & \qw & \targ & \qw & \qw & \qw & \qw & \qw & \qw & \qw & \qw & \qw & \qw & \qw & \qw & \qw & \qw & \qw & \qw & \qw & \qw & \qw & \qw & \qw & \qw & \qw\\
	 	\nghost{{q}_{1} :  } & \lstick{{q}_{1} :  } & \gate{\mathrm{H}} & \qw & \qw & \ctrl{6} & \qw & \ctrl{6} & \qw & \qw & \qw & \qw & \qw & \qw & \qw & \qw & \qw & \qw & \qw & \qw & \qw & \qw & \qw & \qw & \qw & \qw & \qw\\
	 	\nghost{{q}_{2} :  } & \lstick{{q}_{2} :  } & \qw & \qw & \qw & \qw & \qw & \qw & \qw & \qw & \targ & \qw & \qw & \qw & \qw & \qw & \qw & \qw & \qw & \qw & \qw & \qw & \qw & \qw & \qw & \qw & \qw\\
	 	\nghost{{q}_{3} :  } & \lstick{{q}_{3} :  } & \gate{\mathrm{H}} & \qw & \qw & \qw & \qw & \qw & \qw & \qw & \qw & \ctrl{3} & \qw & \qw & \qw & \qw & \qw & \qw & \qw & \qw & \qw & \qw & \qw & \qw & \qw & \qw & \qw\\
	 	\nghost{{q}_{4} :  } & \lstick{{q}_{4} :  } & \qw & \qw & \qw & \qw & \targ & \qw & \ctrl{4} & \qw & \qw & \qw & \ctrl{4} & \targ & \qw & \qw & \ctrl{4} & \targ & \qw & \ctrl{3} & \qw & \ctrl{3} & \targ & \ctrl{4} & \qw & \qw & \qw\\
	 	\nghost{{q}_{5} :  } & \lstick{{q}_{5} :  } & \gate{\mathrm{H}} & \ctrl{-5} & \ctrl{2} & \qw & \qw & \qw & \qw & \qw & \qw & \qw & \qw & \qw & \qw & \ctrl{2} & \qw & \qw & \qw & \qw & \targ & \qw & \qw & \qw & \qw & \qw & \qw\\
	 	\nghost{{q}_{6} :  } & \lstick{{q}_{6} :  } & \qw & \qw & \qw & \qw & \qw & \qw & \qw & \targ & \qw & \targ & \qw & \qw & \ctrl{2} & \qw & \qw & \qw & \ctrl{2} & \qw & \qw & \qw & \qw & \qw & \ctrl{2} & \qw & \qw\\
	 	\nghost{{q}_{7} :  } & \lstick{{q}_{7} :  } & \qw & \qw & \targ & \targ & \ctrl{-3} & \targ & \qw & \qw & \qw & \qw & \qw & \ctrl{-3} & \qw & \targ & \qw & \qw & \qw & \targ & \ctrl{-2} & \targ & \qw & \qw & \qw & \qw & \qw\\
	 	\nghost{{q}_{8} :  } & \lstick{{q}_{8} :  } & \qw & \qw & \qw & \qw & \qw & \qw & \targ & \ctrl{-2} & \ctrl{-6} & \qw & \targ & \qw & \targ & \qw & \targ & \ctrl{-4} & \targ & \qw & \qw & \qw & \ctrl{-4} & \targ & \targ & \qw & \qw\\
\\ }}

The circuit for the second qubit placement shown in Fig.~\ref{fig:ft-logical-connectivity}(b) with $27$ two-qubit gates. The qubit placement in the IBMQ Guadalupe~\cite{guadalupe} device is: $9,5,15,11,10,14,13,8,12$. The last two qubits are given as ancilla qubits.

\noindent \scalebox{0.9}{
\Qcircuit @C=0.05em @R=0.05em @!R { \\
	 	\nghost{{q}_{0} :  } & \lstick{{q}_{0} :  } & \gate{\mathrm{H}} & \qw & \ctrl{7} & \qw & \qw & \ctrl{7} & \qw & \qw & \qw & \qw & \qw & \qw & \qw & \qw & \qw & \qw & \qw & \qw & \qw & \qw & \qw & \qw & \targ & \qw & \qw & \qw & \qw & \qw & \qw\\
	 	\nghost{{q}_{1} :  } & \lstick{{q}_{1} :  } & \qw & \qw & \qw & \qw & \qw & \qw & \qw & \qw & \qw & \qw & \targ & \qw & \qw & \qw & \qw & \qw & \qw & \qw & \qw & \qw & \qw & \qw & \qw & \targ & \qw & \qw & \qw & \qw & \qw\\
	 	\nghost{{q}_{2} :  } & \lstick{{q}_{2} :  } & \gate{\mathrm{H}} & \qw & \qw & \qw & \qw & \qw & \qw & \qw & \ctrl{6} & \qw & \qw & \targ & \qw & \qw & \qw & \qw & \qw & \qw & \qw & \qw & \qw & \qw & \qw & \qw & \qw & \qw & \qw & \qw & \qw\\
	 	\nghost{{q}_{3} :  } & \lstick{{q}_{3} :  } & \qw & \qw & \qw & \qw & \targ & \qw & \targ & \qw & \qw & \ctrl{4} & \qw & \qw & \qw & \qw & \qw & \qw & \targ & \qw & \ctrl{2} & \qw & \ctrl{2} & \ctrl{4} & \qw & \qw & \ctrl{4} & \targ & \ctrl{4} & \qw & \qw\\
	 	\nghost{{q}_{4} :  } & \lstick{{q}_{4} :  } & \qw & \qw & \qw & \qw & \qw & \qw & \qw & \targ & \qw & \qw & \qw & \qw & \targ & \qw & \qw & \ctrl{4} & \qw & \qw & \qw & \qw & \qw & \qw & \qw & \qw & \qw & \qw & \qw & \qw & \qw\\
	 	\nghost{{q}_{5} :  } & \lstick{{q}_{5} :  } & \gate{\mathrm{H}} & \ctrl{1} & \qw & \qw & \qw & \qw & \ctrl{-2} & \qw & \qw & \qw & \qw & \qw & \qw & \qw & \targ & \qw & \ctrl{-2} & \ctrl{1} & \targ & \ctrl{1} & \targ & \qw & \qw & \qw & \qw & \ctrl{-2} & \qw & \qw & \qw\\
	 	\nghost{{q}_{6} :  } & \lstick{{q}_{6} :  } & \qw & \targ & \qw & \ctrl{2} & \qw & \qw & \qw & \qw & \qw & \qw & \qw & \qw & \qw & \targ & \ctrl{-1} & \qw & \qw & \targ & \ctrl{2} & \targ & \qw & \qw & \qw & \qw & \qw & \qw & \qw & \qw & \qw\\
	 	\nghost{{q}_{7} :  } & \lstick{{q}_{7} :  } & \qw & \qw & \targ & \qw & \ctrl{-4} & \targ & \qw & \qw & \qw & \targ & \ctrl{-6} & \qw & \qw & \qw & \qw & \qw & \qw & \qw & \qw & \qw & \qw & \targ & \ctrl{-7} & \ctrl{-6} & \targ & \qw & \targ & \qw & \qw\\
	 	\nghost{{q}_{8} :  } & \lstick{{q}_{8} :  } & \qw & \qw & \qw & \targ & \qw & \qw & \qw & \ctrl{-4} & \targ & \qw & \qw & \ctrl{-6} & \ctrl{-4} & \ctrl{-2} & \qw & \targ & \qw & \qw & \targ & \qw & \qw & \qw & \qw & \qw & \qw & \qw & \qw & \qw & \qw\\
\\ }}

The circuit for the third qubit placement shown in Fig.~\ref{fig:ft-logical-connectivity}(b) with $28$ two-qubit gates. The qubit placement in the IBMQ Guadalupe~\cite{guadalupe} device is: $6,15,0,2,13,10,4,1,7,12$. The last three qubits are given as ancilla qubits.

\noindent \scalebox{0.85}{
\Qcircuit @C=0.05em @R=0.05em @!R { \\
	 	\nghost{{q}_{0} :  } & \lstick{{q}_{0} :  } & \gate{\mathrm{H}} & \ctrl{8} & \qw & \qw & \qw & \qw & \qw & \qw & \qw & \ctrl{8} & \qw & \qw & \targ & \qw & \qw & \qw & \qw & \qw & \qw & \qw & \qw & \qw & \qw & \qw & \qw & \qw & \qw & \qw & \qw & \qw & \qw\\
	 	\nghost{{q}_{1} :  } & \lstick{{q}_{1} :  } & \qw & \qw & \qw & \qw & \qw & \qw & \qw & \targ & \qw & \qw & \qw & \qw & \qw & \qw & \qw & \qw & \qw & \qw & \qw & \qw & \qw & \qw & \qw & \qw & \qw & \qw & \qw & \qw & \qw & \qw & \qw\\
	 	\nghost{{q}_{2} :  } & \lstick{{q}_{2} :  } & \gate{\mathrm{H}} & \qw & \qw & \ctrl{5} & \qw & \qw & \qw & \qw & \qw & \qw & \qw & \targ & \qw & \qw & \qw & \qw & \qw & \qw & \ctrl{5} & \qw & \qw & \qw & \qw & \qw & \qw & \qw & \qw & \qw & \qw & \qw & \qw\\
	 	\nghost{{q}_{3} :  } & \lstick{{q}_{3} :  } & \qw & \qw & \qw & \qw & \qw & \qw & \qw & \qw & \qw & \qw & \qw & \qw & \qw & \targ & \qw & \qw & \qw & \qw & \qw & \qw & \qw & \qw & \qw & \qw & \qw & \qw & \qw & \qw & \qw & \qw & \qw\\
	 	\nghost{{q}_{4} :  } & \lstick{{q}_{4} :  } & \qw & \qw & \qw & \qw & \qw & \targ & \qw & \qw & \qw & \qw & \qw & \qw & \qw & \qw & \qw & \qw & \qw & \qw & \qw & \qw & \qw & \qw & \qw & \targ & \qw & \qw & \qw & \qw & \qw & \qw & \qw\\
	 	\nghost{{q}_{5} :  } & \lstick{{q}_{5} :  } & \gate{\mathrm{H}} & \qw & \ctrl{4} & \qw & \qw & \qw & \ctrl{3} & \qw & \qw & \qw & \ctrl{4} & \qw & \qw & \qw & \targ & \qw & \qw & \targ & \qw & \qw & \qw & \targ & \ctrl{4} & \qw & \ctrl{3} & \ctrl{4} & \targ & \ctrl{3} & \qw & \qw & \qw\\
	 	\nghost{{q}_{6} :  } & \lstick{{q}_{6} :  } & \qw & \qw & \qw & \qw & \targ & \qw & \qw & \qw & \ctrl{1} & \qw & \qw & \qw & \qw & \qw & \qw & \ctrl{2} & \targ & \qw & \qw & \ctrl{1} & \ctrl{2} & \qw & \qw & \qw & \qw & \qw & \qw & \qw & \ctrl{2} & \qw & \qw\\
	 	\nghost{{q}_{7} :  } & \lstick{{q}_{7} :  } & \qw & \qw & \qw & \targ & \qw & \qw & \qw & \qw & \targ & \qw & \qw & \ctrl{-5} & \qw & \ctrl{-4} & \qw & \qw & \ctrl{-1} & \qw & \targ & \targ & \qw & \qw & \qw & \qw & \qw & \qw & \qw & \qw & \qw & \qw & \qw\\
	 	\nghost{{q}_{8} :  } & \lstick{{q}_{8} :  } & \qw & \targ & \qw & \qw & \ctrl{-2} & \qw & \targ & \qw & \qw & \targ & \qw & \qw & \ctrl{-8} & \qw & \ctrl{-3} & \targ & \qw & \ctrl{-3} & \qw & \qw & \targ & \ctrl{-3} & \qw & \qw & \targ & \qw & \ctrl{-3} & \targ & \targ & \qw & \qw\\
	 	\nghost{{q}_{9} :  } & \lstick{{q}_{9} :  } & \qw & \qw & \targ & \qw & \qw & \ctrl{-5} & \qw & \ctrl{-8} & \qw & \qw & \targ & \qw & \qw & \qw & \qw & \qw & \qw & \qw & \qw & \qw & \qw & \qw & \targ & \ctrl{-5} & \qw & \targ & \qw & \qw & \qw & \qw & \qw\\
\\ }}

\subsection{The $|+\rangle_L$ state of the $[[9,1,3]]$ Shor code on 2D grid connectivity (Google Sycamore)} 

Here, we show the result for the fault-tolerant preparation of the $|+\rangle_L$ state for the $[[9,1,3]]$ Shor code on a $4\times3$ grid. For the qubit placement, we follow the row-major order starting with $0$ at the top left of the qubit and $11$ at the bottom right of the qubit. The last three qubits are always given as flag qubits. The RL agent can decide whether to use them or not.

The qubit placement for the circuit below is: 1,0,2,5,3,8,6,9,11,4,7,10. One flag qubit is not used.

\noindent \scalebox{1.0}{
\Qcircuit @C=0.2em @R=0.2em @!R { \\
	 	\nghost{{q}_{0} :  } & \lstick{{q}_{0} :  } & \gate{\mathrm{H}} & \ctrl{2} & \ctrl{1} & \qw & \qw & \qw & \qw & \qw & \qw & \qw & \qw & \qw & \qw\\
	 	\nghost{{q}_{1} :  } & \lstick{{q}_{1} :  } & \qw & \qw & \targ & \qw & \ctrl{3} & \qw & \qw & \qw & \qw & \qw & \qw & \qw & \qw\\
	 	\nghost{{q}_{2} :  } & \lstick{{q}_{2} :  } & \qw & \targ & \ctrl{1} & \qw & \qw & \qw & \qw & \qw & \qw & \qw & \qw & \qw & \qw\\
	 	\nghost{{q}_{3} :  } & \lstick{{q}_{3} :  } & \gate{\mathrm{H}} & \ctrl{2} & \targ & \ctrl{2} & \qw & \qw & \qw & \qw & \qw & \qw & \ctrl{6} & \qw & \qw\\
	 	\nghost{{q}_{4} :  } & \lstick{{q}_{4} :  } & \qw & \qw & \qw & \qw & \targ & \qw & \qw & \qw & \targ & \ctrl{5} & \qw & \qw & \qw\\
	 	\nghost{{q}_{5} :  } & \lstick{{q}_{5} :  } & \qw & \targ & \ctrl{3} & \targ & \targ & \qw & \qw & \qw & \qw & \qw & \qw & \qw & \qw\\
	 	\nghost{{q}_{6} :  } & \lstick{{q}_{6} :  } & \qw & \qw & \qw & \qw & \qw & \qw & \qw & \targ & \ctrl{-2} & \qw & \qw & \qw & \qw\\
	 	\nghost{{q}_{7} :  } & \lstick{{q}_{7} :  } & \qw & \qw & \qw & \qw & \qw & \qw & \targ & \ctrl{-1} & \qw & \qw & \qw & \qw & \qw\\
	 	\nghost{{q}_{8} :  } & \lstick{{q}_{8} :  } & \qw & \qw & \targ & \qw & \ctrl{-3} & \ctrl{3} & \qw & \ctrl{3} & \qw & \qw & \qw & \qw & \qw\\
	 	\nghost{{q}_{9} :  } & \lstick{{q}_{9} :  } & \qw & \qw & \qw & \qw & \qw & \qw & \qw & \qw & \qw & \targ & \targ & \qw & \qw\\
	 	\nghost{{q}_{10} :  } & \lstick{{q}_{10} :  } & \qw & \qw & \qw & \qw & \qw & \qw & \qw & \qw & \qw & \qw & \qw & \qw & \qw\\
	 	\nghost{{q}_{11} :  } & \lstick{{q}_{11} :  } & \qw & \qw & \qw & \qw & \qw & \targ & \ctrl{-4} & \targ & \qw & \qw & \qw & \qw & \qw\\
\\ }}
 
The qubit placement for the circuit below is: 6,7,10,8,1,3,5,11,9,4,0,2. One flag qubit is not used. In this case, one data qubit has the flag qubit as its neighbor, so the agent needs to use the flag qubit as a bridge to that data qubit. 

\noindent \scalebox{1.0}{
\Qcircuit @C=0.2em @R=0.2em @!R { \\
	 	\nghost{{q}_{0} :  } & \lstick{{q}_{0} :  } & \gate{\mathrm{H}} & \ctrl{8} & \qw & \qw & \qw & \qw & \qw & \targ & \ctrl{1} & \qw & \qw & \qw & \qw & \qw & \qw & \qw & \qw\\
	 	\nghost{{q}_{1} :  } & \lstick{{q}_{1} :  } & \qw & \qw & \qw & \qw & \qw & \qw & \qw & \qw & \targ & \qw & \qw & \ctrl{2} & \ctrl{1} & \ctrl{8} & \qw & \qw & \qw\\
	 	\nghost{{q}_{2} :  } & \lstick{{q}_{2} :  } & \qw & \qw & \qw & \targ & \qw & \ctrl{5} & \qw & \qw & \targ & \qw & \qw & \qw & \targ & \qw & \qw & \qw & \qw\\
	 	\nghost{{q}_{3} :  } & \lstick{{q}_{3} :  } & \qw & \qw & \qw & \qw & \qw & \qw & \qw & \qw & \qw & \targ & \ctrl{3} & \targ & \qw & \qw & \qw & \qw & \qw\\
	 	\nghost{{q}_{4} :  } & \lstick{{q}_{4} :  } & \gate{\mathrm{H}} & \qw & \ctrl{6} & \qw & \qw & \qw & \ctrl{6} & \qw & \qw & \qw & \qw & \qw & \qw & \qw & \ctrl{5} & \qw & \qw\\
	 	\nghost{{q}_{5} :  } & \lstick{{q}_{5} :  } & \qw & \qw & \qw & \qw & \targ & \qw & \qw & \ctrl{-5} & \qw & \qw & \qw & \qw & \qw & \qw & \qw & \qw & \qw\\
	 	\nghost{{q}_{6} :  } & \lstick{{q}_{6} :  } & \qw & \qw & \qw & \qw & \qw & \qw & \qw & \qw & \qw & \qw & \targ & \ctrl{3} & \qw & \qw & \qw & \qw & \qw\\
	 	\nghost{{q}_{7} :  } & \lstick{{q}_{7} :  } & \qw & \qw & \qw & \qw & \qw & \targ & \qw & \qw & \ctrl{-5} & \ctrl{-4} & \qw & \qw & \qw & \qw & \qw & \qw & \qw\\
	 	\nghost{{q}_{8} :  } & \lstick{{q}_{8} :  } & \qw & \targ & \qw & \ctrl{-6} & \qw & \qw & \qw & \qw & \qw & \qw & \qw & \qw & \qw & \qw & \qw & \qw & \qw\\
	 	\nghost{{q}_{9} :  } & \lstick{{q}_{9} :  } & \qw & \qw & \qw & \qw & \qw & \qw & \qw & \qw & \qw & \qw & \qw & \targ & \qw & \targ & \targ & \qw & \qw\\
	 	\nghost{{q}_{10} :  } & \lstick{{q}_{10} :  } & \qw & \qw & \targ & \qw & \ctrl{-5} & \qw & \targ & \qw & \qw & \qw & \qw & \qw & \qw & \qw & \qw & \qw & \qw\\
	 	\nghost{{q}_{11} :  } & \lstick{{q}_{11} :  } & \qw & \qw & \qw & \qw & \qw & \qw & \qw & \qw & \qw & \qw & \qw & \qw & \qw & \qw & \qw & \qw & \qw\\
\\ }}

\subsection{The $|-\rangle_L$ state of the $[[5,1,3]]$ perfect code on IBMQ Tokyo}

The qubit placement for the circuit below on the IBMQ Tokyo~\cite{tokyo} connectivity: 6,1,7,5,11,2,10.

\noindent \scalebox{1.0}{
\Qcircuit @C=0.2em @R=0.2em @!R { \\
\nghost{{q}_{0} :  }	 & \lstick{{q}_{0} :  }	 & \gate{\mathrm{H}}	 & \ctrl{3}	 & \ctrl{1}	 & \targ	 & \ctrl{4}	 & \ctrl{1}	 & \qw	 & \targ	 & \gate{\mathrm{H}}	 & \qw	 & \qw	 & \control\qw	 & \qw	 & \control\qw	 & \control\qw	 & \control\qw	 & \qw	 & \qw	 & \qw	 & \qw\\
\nghost{{q}_{1} :  }	 & \lstick{{q}_{1} :  }	 & \qw	 & \qw	 & \targ	 & \qw	 & \qw	 & \targ	 & \gate{\mathrm{H}}	 & \ctrl{-1}	 & \qw	 & \qw	 & \targ	 & \qw	 & \qw	 & \qw	 & \qw	 & \qw	 & \qw	 & \qw	 & \qw	 & \qw\\
\nghost{{q}_{2} :  }	 & \lstick{{q}_{2} :  }	 & \gate{\mathrm{H}}	 & \qw	 & \qw	 & \ctrl{-2}	 & \qw	 & \qw	 & \qw	 & \qw	 & \qw	 & \qw	 & \qw	 & \qw	 & \qw	 & \qw	 & \qw	 & \qw	 & \control\qw	 & \qw	 & \qw	 & \qw\\
\nghost{{q}_{3} :  }	 & \lstick{{q}_{3} :  }	 & \qw	 & \targ	 & \gate{\mathrm{H}}	 & \qw	 & \qw	 & \qw	 & \qw	 & \qw	 & \qw	 & \qw	 & \qw	 & \qw	 & \control\qw	 & \qw	 & \qw	 & \qw	 & \qw	 & \qw	 & \qw	 & \qw\\
\nghost{{q}_{4} :  }	 & \lstick{{q}_{4} :  }	 & \qw	 & \qw	 & \qw	 & \qw	 & \targ	 & \qw	 & \qw	 & \qw	 & \qw	 & \targ	 & \qw	 & \qw	 & \qw	 & \qw	 & \qw	 & \qw	 & \qw	 & \qw	 & \qw	 & \qw\\
\nghost{{q}_{5} :  }	 & \lstick{{q}_{5} :  }	 & \qw	 & \qw	 & \qw	 & \qw	 & \qw	 & \qw	 & \qw	 & \qw	 & \gate{\mathrm{H}}	 & \ctrl{-1}	 & \qw	 & \ctrl{-5}	 & \ctrl{-2}	 & \ctrl{-5}	 & \qw	 & \ctrl{-5}	 & \qw	 & \gate{\mathrm{H}}	 & \qw	 & \qw\\
\nghost{{q}_{6} :  }	 & \lstick{{q}_{6} :  }	 & \qw	 & \qw	 & \qw	 & \qw	 & \qw	 & \qw	 & \qw	 & \qw	 & \gate{\mathrm{H}}	 & \qw	 & \ctrl{-5}	 & \qw	 & \qw	 & \qw	 & \ctrl{-6}	 & \qw	 & \ctrl{-4}	 & \gate{\mathrm{H}}	 & \qw	 & \qw\\
\\ }}

The qubit placement for the circuit below on the IBMQ Tokyo~\cite{tokyo} connectivity: 2,7,1,12,13,6,8. 
\noindent \scalebox{1.0}{
\Qcircuit @C=0.2em @R=0.2em @!R { \\
\nghost{{q}_{0} :  }	 & \lstick{{q}_{0} :  }	 & \gate{\mathrm{H}}	 & \ctrl{1}	 & \gate{\mathrm{H}}	 & \qw	 & \qw	 & \ctrl{1}	 & \qw	 & \qw 	 & \qw	 & \qw	 & \qw	 & \qw	 & \control\qw	 & \qw	 & \qw	 & \qw	 & \qw	 & \qw	 & \qw	 & \qw	 & \qw\\
\nghost{{q}_{1} :  }	 & \lstick{{q}_{1} :  }	 & \qw	 & \targ	 & \ctrl{3}	 & \targ	 & \ctrl{2}	 & \targ	 & \targ	 & \gate{\mathrm{H}}	 & \qw	 & \qw	 & \targ	 & \qw	 & \qw	 & \targ	 & \targ	 & \targ	 & \qw	 & \qw	 & \qw	 & \qw	 & \qw\\
\nghost{{q}_{2} :  }	 & \lstick{{q}_{2} :  }	 & \gate{\mathrm{H}}	 & \qw	 & \qw	 & \ctrl{-1}	 & \qw	 & \qw	 & \qw	 & \qw	 & \control\qw	 & \qw	 & \qw	 & \qw	 & \qw	 & \qw	 & \qw	 & \qw	 & \qw	 & \qw	 & \qw	 & \qw	 & \qw\\
\nghost{{q}_{3} :  }	 & \lstick{{q}_{3} :  }	 & \qw	 & \qw	 & \qw	 & \qw	 & \targ	 & \gate{\mathrm{H}}	 & \qw	 & \qw	 & \qw	 & \control\qw	 & \qw	 & \targ	 & \qw	 & \qw	 & \qw	 & \qw	 & \qw	 & \qw	 & \qw	 & \qw	 & \qw\\
\nghost{{q}_{4} :  }	 & \lstick{{q}_{4} :  }	 & \qw	 & \qw	 & \targ	 & \qw	 & \qw	 & \qw	 & \ctrl{-3}	 & \qw	 & \qw	 & \qw	 & \qw	 & \qw	 & \qw	 & \qw	 & \qw	 & \qw	 & \targ	 & \control\qw	 & \qw	 & \qw	 & \qw\\
\nghost{{q}_{5} :  }	 & \lstick{{q}_{5} :  }	 & \qw	 & \qw	 & \qw	 & \qw	 & \qw	 & \qw	 & \qw	 & \gate{\mathrm{H}}	 & \ctrl{-3}	 & \qw	 & \ctrl{-4}	 & \qw	 & \ctrl{-5}	 & \ctrl{-4}	 & \qw	 & \ctrl{-4}	 & \qw	 & \qw	 & \gate{\mathrm{H}}	 & \qw	 & \qw\\
\nghost{{q}_{6} :  }	 & \lstick{{q}_{6} :  }	 & \qw	 & \qw	 & \qw	 & \qw	 & \qw	 & \qw	 & \qw	 & \gate{\mathrm{H}}	 & \qw	 & \ctrl{-3}	 & \qw	 & \ctrl{-3}	 & \qw	 & \qw	 & \ctrl{-5}	 & \qw	 & \ctrl{-2}	 & \ctrl{-2}	 & \gate{\mathrm{H}}	 & \qw	 & \qw\\
\\ }}

\section{Varying integrated fault-tolerant logical state preparation task weight rewards}
\label{app:ftlsp-varymiu}

\begin{figure}[htb]
	\includegraphics[width=.5\textwidth]{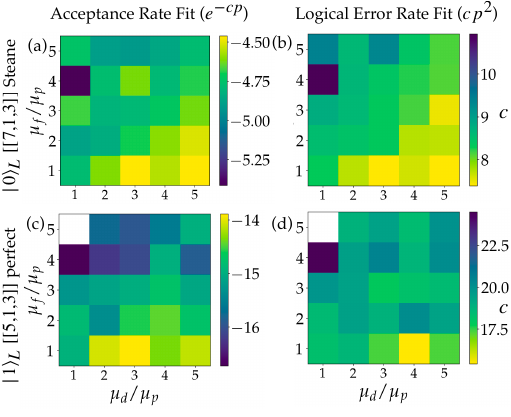}
	\caption{\label{fig:vary-miu-iftlsp} Varying the weights of the reward function for integrated fault-tolerant logical state preparation (IFT-LSP). We vary $\mu_f / \mu_p$ and $\mu_d / \mu_p$ ranging from $1$ to $5$ with an interval of $1$. The heatmap shows the average fitting coefficients for the acceptance rate (in (a) and (c), the higher the better) and the logical error rate (in (b) and (d), the lower the better). We evaluate for integrated fault-tolerant logical state preparation of the $|0\rangle_L$ state of the $[[7,1,3]]$ Steane code  (in (a) and (b)) and the $|1\rangle_L$ state of the $[[5,1,3]]$ perfect code (in (c) and (d)). The white color means that no agent has converged on that parameter.}
\end{figure}

Here, we vary the weights for the flag reward $\mu_f$, the complementary distance reward $\mu_d$, and the product state reward $\mu_p$ of the reward function that is defined in Eq.~(\ref{eq:vcs-reward}) for the integrated fault-tolerant logical state preparation task (IFT-LSP). We then evaluate how it affects the acceptance and the logical error rates. 

Effectively, only the weight ratios matter, since scaling the reward function generally does not affect the performance of the reinforcement learning training. We vary the ratios $\mu_f / \mu_p$ and $\mu_d / \mu_p$ and prepare the fault-tolerant logical state at each point. We then compute the acceptance and logical error rates, and fit them with exponential and quadratic functions, respectively. We then compare the average coefficients over $10$ different circuits.

In Fig.~\ref{fig:vary-miu-iftlsp}, we see that the best strategy for the integrated fault-tolerant logical state preparation task is to prioritize the weight of the complementary distance reward $\mu_d$. This is expected since the RL training starts from scratch, so $\mu_d$ must be prioritized.

\end{document}